\documentclass[]{aastex63}

\usepackage{graphicx}
\usepackage{amsmath}
\usepackage{amssymb}
\usepackage{romannum}
\usepackage{textcomp}
\usepackage{upgreek}
\submitjournal{PSJ}
\shorttitle{Exploring Retrievals with Solar System Observations}
\shortauthors{Robinson \& Salvador}
\graphicspath{{./}{}}
\begin{document}

%
\pagenumbering{arabic}
%

\title{Exploring and Validating Exoplanet Atmospheric Retrievals with Solar System Analog Observations}

\correspondingauthor{Tyler D. Robinson}
\email{tdrobin@arizona.edu}

%
\author[0000-0002-3196-414X]{Tyler D. Robinson}
\affiliation{Lunar \& Planetary Laboratory, University of Arizona, Tucson, AZ 85721 USA}
\affiliation{Department of Astronomy and Planetary Science, Northern Arizona University, Flagstaff, AZ 86011, USA}
\affiliation{Habitability, Atmospheres, and Biosignatures Laboratory, University of Arizona, Tucson, AZ 85721, USA}
\affiliation{NASA Nexus for Exoplanet System Science Virtual Planetary Laboratory, University of Washington, Box 351580, Seattle, WA 98195, USA}

\author[0000-0001-8106-6164]{Arnaud Salvador}
\affiliation{Lunar \& Planetary Laboratory, University of Arizona, Tucson, AZ 85721 USA}
\affiliation{Department of Astronomy and Planetary Science, Northern Arizona University, Flagstaff, AZ 86011, USA}
\affiliation{Habitability, Atmospheres, and Biosignatures Laboratory, University of Arizona, Tucson, AZ 85721, USA}
%

\begin{abstract}
    Solar System observations that serve as analogs for exoplanet remote sensing data can provide important opportunities to validate ideas and models related to exoplanet environments. Critically, and unlike true exoplanet observations, Solar System analog data benefit from available high-quality ground- or orbiter-derived ``truth'' constraints that enable strong validations of exoplanet data interpretation tools. In this work, we first present a versatile atmospheric retrieval suite, capable of application to reflected light, thermal emission, and transmission observations {spanning a broad range of wavelengths and thermochemical conditions}. The tool\,---\,dubbed \texttt{rfast}\,---\,is designed, in part, to enable exoplanet mission concept feasibility studies. Following model validation, the retrieval tool is applied to a range of Solar System analog observations for exoplanet environments. Retrieval studies using Earth reflected light observations from NASA's \textit{EPOXI} mission provide a key proof-of-concept for under-development exo-Earth direct imaging concept missions. Inverse modeling applied to an infrared spectrum of Earth from the \textit{Mars Global Surveyor} Thermal Emission Spectrometer achieves good constraints on atmospheric gases, including many biosignature gases. Finally, retrieval analysis applied to a transit spectrum of Titan derived from the \textit{Cassini} Visual and Infrared Mapping Spectrometer provides a proof-of-concept for interpreting more feature-rich transiting exoplanet observations from NASA's \textit{James Webb Space Telescope} (\textit{JWST}). In the future, Solar System analog observations for exoplanets could be used to verify exoplanet models and parameterizations, and future exoplanet analog observations of any Solar System worlds from planetary science missions should be encouraged.
\end{abstract}
%

%

%
%

%
\section{Introduction} \label{sec:intro}
%

Atmospheric remote sensing has proven essential to interpreting exoplanet observations \citep{madhusudhan2018,fortneyetal2021}, with applications ranging from some of the first-ever potential constraints on the composition of exoplanet atmospheres \citep{tinettietal2007,pontetal2008,swainetal2008,singetal2009,beanetal2010} through to the more-modern atmospheric retrieval (or inverse modeling) approaches developed in a wide range of studies  \citep{irwinetal2008,madhusudhan&seager2009,lineetal2012,benneke&seager2012,lineetal2013a,lineetal2014a,kreidbergetal2014a,knutsonetal2014b,stevensonetal2014,morleyetal2016,barstowetal2017,macdonald&madhusudhan2017,tsiarasetal2019,molliereetal2019,bennekeetal2019,zhangetal2019,kitzmannetal2020,colonetal2020,minetal2020,mansfieldetal2022}. In the very near future, atmospheric retrieval tools are likely to see widespread application to exoplanet observations from NASA's \textit{JWST} \citep{cowanetal2015,greeneetal2016,krissansentottonetal2018b,nixon&madhusudhan2022}. Inverse modeling tools are also likely to prove key when interpreting exoplanet observations from near-future exoplanet-themed missions, such as NASA's \textit{Nancy Grace Roman Space Telescope} \citep[\textit{Roman}; ][]{akesonetal2019,kasdinetal2020,marleyetal2014,lupuetal2016,nayaketal2017} and ESA's \textit{Ariel} mission \citep{tinettietal2016,barstowetal2022}.

Recommendations from the recent Decadal Survey on Astronomy and Astrophysics 2020\footnote{\url{https://doi.org/10.17226/26141}} indicate that studies of exoplanet environments should continue to expand in both quantity and quality, engaging complementary ground- and space-based resources. When looking forward to future exoplanet exploration strategies, atmospheric inverse modeling will play at least two major roles. First, and most obviously, atmospheric retrieval tools will be needed to interpret data from any near- or far-future observing facilities and characterize exoplanet properties. Second, and maybe less obviously, inverse models can provide the connection between proposed instrument/telescope performance and the expected constraints on the parameters that describe an exoplanet environment. In fact, atmospheric retrieval is actively being used to refine designs of the Habitable Exoplanet Observatory \citep[HabEx;][]{gaudietal2018}, the Large UltraViolet-Optical-InfraRed Surveyor \citep[LUVOIR;][]{robergeetal2018}, and the Origins Space Telescope \citep{battersbyetal2018} mission concepts \citep[or their successors;][]{fengetal2018,smithetal2020,tremblayetal2020,damiano&hu2021}.

Given the large number of exoplanet-themed missions and instruments, either operational or on the horizon, it may be easy to become focused on environments that are many parsecs away from Earth. Such an outlook can miss important opportunities that Solar System worlds present for guiding exoplanet science \citep[][]{robergeetal2017,keithly&savransky2021}. For example, Solar System planets and moons can serve as models for the predicted appearance of analogous exoplanet targets \citep{tinettietal2005,tinettietal2006a,stam2008,kaltenegger&traub2009,zuggeretal2010,robinsonetal2011,fujiietal2014,robinsonetal2014a,dalbaetal2015,mayorgaetal2016,lustigyaegeretal2018,macdonald&cowan2019,kaneetal2019,mayorgaetal2020,mayorgaetal2021}.

From the perspective of exoplanet atmospheric remote sensing, Solar System worlds and observations can yield opportunities to both validate retrieval model results and capabilities and to test the simplifying parameterizations necessarily adopted in these tools. {A recent review of connections between Solar System planetary science and exoplanet science \citep{kaneetal2021} highlighted measurables from Solar System worlds as a ``pathway forward'' for more-correct interpretations of exoplanet data.} However, relatively few works have examined such applications of Solar System observations. \citet{marleyetal2014} \citep[and a more-formal companion study,][]{lupuetal2016}, in exploratory work relevant to (what is now) \textit{Roman}, performed retrievals on visible-wavelength observations of Jupiter, Saturn, and Uranus \citep[from][]{karkoschka1998} and demonstrated that methane abundances could be reliably inferred using a forward model that included two distinct cloud decks. The assumption of grey cloud properties was found to be generally acceptable, although haze absorption at wavelengths shorter than those considered in the study would be an important consideration for future mission concepts. \citet{heng&li2021} used high-quality phase curves of Jupiter from the \textit{Cassini} Imaging Science Subsystem \citep{porcoetal2004,lilimingetal2018} to infer properties of Jovian clouds with potential implications for \textit{JWST}. Finally, \citet{tribbettetal2021} performed retrievals on effective transit spectra of Titan\,---\,generated from stellar occultations observed by the \textit{Cassini} Ultraviolet Imaging Spectrograph \citep{espositoetal2004,koskinenetal2011}\,---\,and showed that simple parameterizations of haze extinction failed to detect the presence of known haze layers/over-densities.

The work that follows explores a variety of novel retrieval studies for Solar System worlds treated as exoplanet analogs, which is a strongly under-explored area of study. Section~\ref{sec:methods} develops a versatile and efficient inverse modeling suite whose origins stem from exoplanet direct imaging mission concept studies. Validations against existing tools are showcased in Section~\ref{sec:valid}. Section~\ref{sec:results} first demonstrates an application to direct imaging studies of Earth-like exoplanets and subsequently demonstrates retrieval results as applied to disk-integrated reflected light observations of Earth, disk-integrated infrared observations of Earth, and near-infrared transit spectra of Titan. Key findings from these retrieval studies are discussed in the context of near- and further-future exoplanet-themed missions in Section~\ref{sec:discuss} while take-away findings are enumerated in Section~\ref{sec:conc}.

%
\section{Methods} \label{sec:methods}
%

Atmospheric retrieval (or inference) requires a suite of interconnected tools: a parameter space sampling tool, a radiative transfer ``forward'' model, and an instrument model. When retrieving on a noisy observation, the sampling tool uses information about goodness of fit to statistically explore a posterior distribution for a collection of atmospheric and planetary parameters (i.e., ``state'' vectors). For a particular instance of a state vector, the radiative transfer model, {in general}, predicts a high resolution spectrum that is subsequently spectrally degraded to match the resolution of the observed spectrum via the instrument model. A likelihood comparison between the observed spectrum and the degraded prediction then enables the sampling algorithm to further explore posterior space. For mission design purposes\,---\,where true observations do not yet exist\,---\,it is often the case that faux/synthetic observations must be generated using a forward model and an instrument model. Adopting this faux observation into the retrieval framework then enables exploration of how changes to key mission/instrument parameters map to changes in expected constraints on planetary environments.

Material below describes the components of a computationally efficient, open source, and user friendly generalized atmospheric retrieval package, called \texttt{rfast}. Core elements of the \texttt{rfast} package were developed to support studies for the HabEx and LUVOIR mission concepts. Direct imaging capabilities were inspired by a rocky exoplanet retrieval tool described in \citet{fengetal2018}, although many changes and upgrades have been introduced: opting for a fully Python-based implementation rather than merged Python/Fortran, allowing for a wider variety of atmospheric gases (and, thus, planet types) with both pressure- and temperature-dependent opacities, and enabling users to straightforwardly toggle on/off which atmospheric and planetary parameters should be included in the retrieval analysis and what priors should be adopted for these parameters. Other notable model capabilities include treatments for vertically-varying gas and temperature profiles and the ability to divide a synthetic spectrum into bands with distinct wavelength coverages, noise levels, and spectral resolutions. Finally, to enable both mission concept and feasibility studies beyond reflected light direct imaging scenarios, the \texttt{rfast} suite also includes options for spectroscopic studies in thermal emission, combined reflected light and thermal emission, and transit transmission.

\subsection{Reflected Light Forward Model} \label{subsec:reflect}

The \texttt{rfast} tool includes treatments of reflected light spectroscopy and photometry where the planet is either treated as a single plane-parallel scene (one-dimensional) or as a pixelated globe (three-dimensional). Radiative transfer for each pixel in the three dimensional treatment makes a local plane-parallel assumption and thereby allows for simulated observations that depend on planetary phase angle. While the single scene option does not allow for phase-dependent studies, the reduction from three dimensional to one dimensional geometry retains suitability for broad mission concept studies while also offering large computational efficiency improvements (i.e.,~usually at least an order of magnitude improvement in runtime).

Radiative transfer in the single scene option follows a diffuse two-stream flux adding treatment developed in \citet{robinson&crisp2018}. For each model layer, layer reflectivity ($r_k$) and transmissivity ($t_k$) terms are computed by integrating the hemispheric mean two-stream radiative transfer equations over optical depth assuming a diffuse illumination source, yielding,
\begin{equation}
    r_k = \frac{A_{\infty}\left(1 - e^{-2 a_k b_k \Delta \tau_k} \right)}{1 - A_{\infty}^2e^{-2 a_k b_k \Delta \tau_k}} ~~~ \left(0 \leq \omega_k < 1\right) \ ,
\end{equation}
\begin{equation}
    t_k = \frac{\left(1 - A_{\infty}^2 \right)e^{-a_k b_k \Delta \tau_k}}{1 - A_{\infty}^2e^{-2 a_k b_k \Delta \tau_k}}  ~~~ \left(0 \leq \omega_k < 1\right) \ ,
\end{equation}
with $a_k = \sqrt{1-\omega_k}$, $b_k = \frac{3}{2}\sqrt{1-w_kg_k}$, and $A_{\infty} = (\frac{2}{3}b_k - a_k)/(a_k + \frac{2}{3}b_k)$. Here, $\Delta \tau_k$ is the layer extinction optical depth, $\omega_k$ is the layer single scattering albedo, $g_k$ is the layer scattering asymmetry parameter, and $A_{\infty}$ is the layer reflectivity as the optical depth tends to infinity, all of which can generally depend on wavelength. A single scattering albedo of unity represents a special case where,
\begin{equation}
    r_k = \frac{ \frac{3}{4}(1-g_k)\Delta\tau_k }{1 + \frac{3}{4}(1-g_k)\Delta\tau_k} ~~~ (\omega_k = 1) \ ,
\end{equation}
\begin{equation}
    t_k = \frac{1}{1 + \frac{3}{4}(1-g_k)\Delta\tau_k}  ~~~ (\omega_k = 1) \ .
\end{equation}
The reflectivity of the inhomogeneous atmospheric column extending upward from the surface is then determined via a recursive relation, 
\begin{equation}
    R_k = r_k + \frac{t_k^2 R_{k+1}}{1 - r_{k}R_{k+1}} \ ,
\end{equation}
where $R_k$ is the reflectivity of the atmospheric column extending from the surface to the top of the $k$-th layer (i.e., so that $R_1$ can be taken as the planetary reflectivity in the single scene model). The lower boundary condition for a model with $N$ atmospheric levels is applied as $R_N = A_{\rm s}$, where $A_{\rm s}$ is the wavelength-dependent surface albedo.

The phase-dependent option adds a treatment for the direct solar beam that follows \citet{hapke1981}. In each pixel, radiation scattered in a given layer that does not remain in the direct beam enters the diffuse field and is included in layer flux upwelling and downwelling source terms ($s_j^+$ and $s_j^-$, respectively). The fraction of the diffuse flux that enters either the upwelling or downwelling source terms is determined via integrating the scattering phase function over the upwelling and downwelling directions given the pixel illumination geometry. Given the layer source terms and the layer reflectivity and transmissivity (which only apply to the diffuse field), the diffuse flux traveling upward at the top of the atmospheric pixel is determined using Equations (4), (5), and (7--9) in \citet{robinson&crisp2018}. Combining the diffuse flux with the emergent direct beam intensity yields the total emergent intensity from a plane-parallel pixel.

Summation over the pixelated disk is accomplished using Gauss-Chebyshev integration, as detailed in \citet{horak&little1965}. For a model with spatial degree $M$, the Gauss points and weights ($x_{{\rm G},i}$ and $w_{{\rm G},i}$, respectively) are based on the roots of the Legendre polynomials of degree $M$, while the Chebyshev points and weights ($x_{{\rm C},j}$ and $w_{{\rm C},j}$) are based on the Chebyshev polynomials of the first kind \citep[see Section 2 of][]{webberetal2015}. The cosine of the solar and observer zenith angles for a given pixel are then given, respectively, by,
\begin{equation}
    \mu_{\rm s,ij} = \sin\left( \cos^{-1} x_{{\rm C},j} \right) \cos \left( \zeta_i - \alpha \right) \ ,
\end{equation}
\begin{equation}
    \mu_{\rm o,ij} = \sin\left( \cos^{-1} x_{{\rm C},j} \right) \cos \left( \zeta_i \right) \ ,
\end{equation}
where $\alpha$ is the planetary phase angle and,
\begin{equation}
    \sin \zeta_i = \frac{\cos\alpha+1}{2} \left(x_{{\rm G},i} - \frac{\cos\alpha - 1}{\cos\alpha + 1}\right) \ .
\end{equation}
Given this formalism, the flux emerging from the spatially integrated disk is,
\begin{equation}
    F_{\lambda}\left( \alpha \right) = \frac{\cos\alpha + 1}{2} \sum_{i=1}^{M} \sum_{j=1}^{M} w_{{\rm G},i}w_{{\rm C},j} I_{\lambda}\left(\tau=0; \alpha, \mu_{\rm s,ij}, \mu_{\rm o,ij} \right) \ ,
    \label{eqn:gcsum}
\end{equation}
where $I_{\lambda}\left(\tau=0; \alpha, \mu_{\rm s,ij}, \mu_{\rm o,ij} \right)$ is the wavelength-dependent specific intensity emerging from the $ij$-th pixel at the given planetary phase angle. The number of pixels on the illuminated disk scales as $M^2$ and spatially homogeneous models can halve the number of Chebyshev points due to a symmetry about the illumination equator. Adopting a normal-incidence top-of-atmosphere specific stellar flux of unity causes Equation~\ref{eqn:gcsum} to yield the wavelength-dependent planetary geometric albedo ($A_{\rm g}$) at full phase ($\alpha = 0$) and the product of the geometric albedo and the planetary phase function ($\Phi\left(\alpha\right)$) at other phase angles.

\subsection{Emitted Light Forward Model} \label{subsec:emit}

Treatments for emitted light spectra are similar to the single scene reflected light model. Layer reflectivity and transmissivity are computed as in the reflected light case, and it is useful to define a layer absorptivity,
\begin{equation}
    a_k = 1 - r_k - t_k \ .
\end{equation}
Following \citet{robinson&crisp2018}, layer thermal flux source terms in the upwelling and downwelling directions are given by,
\begin{equation}
    s_k^+ = \pi \left[ a_k B_{\lambda}\left( T_k \right) - \delta B_{\lambda,k} \right] \ ,
\end{equation}
\begin{equation}
    s_k^- = \pi \left[ a_k B_{\lambda}\left( T_{k+1} \right) + \delta B_{\lambda,k} \right] \ ,
\end{equation}
where $B_{\lambda}$ is the Planck function, $T_k$ is the temperature at level $k$ (incrementing downward), and,
\begin{equation}
    \delta B_{\lambda,k} = \left(1 - a_k + \frac{a_k}{1-a_{k}} \right) \left[ B_{\lambda}\left( T_{k} \right) - B_{\lambda}\left( T_{k+1} \right) \right] \ ,
\end{equation}
which is a correction that ensures the net thermal flux across a layer tends towards the radiation diffusion limit for large optical depths. Lower boundary conditions use a surface emissivity and a Planck-like surface emission term while the upper boundary condition is zero incident downwelling thermal radiation. Upwelling thermal flux at the top of the planetary atmosphere is determined using the flux adding expressions in Equations (4), (5), and (7--9) of \citet{robinson&crisp2018}.

\subsection{Transit Spectroscopy Forward Model} \label{subsec:transit}

Transit spectra are computed in the geometric limit (i.e., assuming straight line ray trajectories) using the one-dimensional path distribution approach developed by \citet{robinson2017} \citep[see also][]{macdonald&lewis2021}. For an atmospheric layer centered at radial distance $r_k$ with width $\Delta r_k$, and for a ray incident on the atmosphere with impact parameter $b$, the geometric path distribution is given by,
\begin{equation}
  \mathcal{P}_{b}\!\left(r_k\right) = 
  \begin{cases}
    0 \ , & b \geq r_k +\Delta r_k/2 \\
    \frac{2}{\Delta r_k} \sqrt{ \left( r_k + \Delta r_k/2 \right)^2 - b^2} \ , &  r_k - \Delta r_k/2 < b < r_k + \Delta r_k/2 \\
    \frac{2}{\Delta r_k} \left[\sqrt{ \left( r_k + \Delta r_k/2 \right)^2 - b^2} - \sqrt{ \left( r_k - \Delta r_k/2 \right)^2 - b^2} \right] \ , & b \leq r_k - \Delta r_k/2 \ .  \\ 
  \end{cases}
\end{equation}
For an atmospheric model with $N$ levels and assuming that the grid of impact parameters corresponds to layer midpoints, the geometric path distribution is a matrix of size $N-1 \times N - 1$. Given a vector of layer differential optical depths, $\boldsymbol{\Delta \tau}$, the wavelength-dependent slant path transmissivity, $\boldsymbol{t}$, for each impact parameter can be computed using matrix algebra with,
\begin{equation}
  \boldsymbol{t} = \boldsymbol{1} - \boldsymbol{a} 
                                          = {\rm{EXP}}\! \left( - \boldsymbol{\Delta \tau} \cdot \boldsymbol{\mathcal{P}} \right) \ ,
\end{equation}
where we have defined the absorptivity vector, $\boldsymbol{a}$. If we define a vector of annulus areas as $A_{k} = 2 \pi b_{k} \Delta b_{k}$, then the wavelength dependent transit spectrum can be written as,
\begin{equation}
  \left( \frac{R_{\rm{p},\lambda}}{R_{\rm{s}}} \right)^{\! \! 2} =  \frac{1}{R_{\rm{s}}^2} \left( R_{\rm{p}}^2  + 
                                                                                       \frac{1}{\pi} \boldsymbol{a} \cdot \boldsymbol{A} \right) \ ,
\end{equation}
where $R_{\rm p}$ is a reference planetary radius (e.g., the solid body radius or the radius {at a specified atmospheric pressure}) and $R_{\rm s}$ is the host stellar radius.

Transit spectra in the \texttt{rfast} tool are generally treated in the pure absorption limit, so that the optical depths adopted in the expressions above are extinction optical depths. For particles, aerosol forward scattering\,---\,which can reduce slant path optical depths\,---\,is treated using the analytic formalism of \citet{robinsonetal2017}. A refractive floor to the transit spectrum follows existing analytic treatments \citep{sidis&sari2010,betremieux&kaltenegger2014,robinsonetal2017}.

\subsection{Other Model Considerations}

The relationship between pressure and altitude\,---\,which is especially important for transit spectroscopy\,---\,is determined by solving the hydrostatic equation given the atmospheric thermal and chemical state. Assuming that gravitational acceleration is proportional to $\left( R_{\rm p} + z \right)^2$ (where $z$ is altitude above the planetary radius) and that temperature varies linearly with pressure through a layer yields the recursion,
\begin{equation}
    \frac{R_{\rm p}}{1 + z_k/R_{\rm p}} = \frac{R_{\rm p}}{1 + z_{k+1}/R_{\rm p}} - h_k
\end{equation}
with,
\begin{equation}
    h_k = \frac{k_{\rm B}}{g_{\rm s} m_k} \left[ \left(T_{k+1} - \frac{T_k - T_{k+1}}{p_k - p_{k+1}} \right)\ln\frac{p_{k+1}}{p_k} - T_{k} + T_{k+1}\right] \ ,
\end{equation}
where $k_{\rm B}$ is Boltzmann's constant, $g_{\rm s}$ is the acceleration due to gravity at $R_{\rm p}$ (e.g., at the planetary surface), $m$ is the layer mean molecular mass, and $T$ and $p$ are the level-dependent temperature and pressure, respectively. The acceleration due to gravity at any altitude is simply,
\begin{equation}
    g\left( z \right) = \frac{g_{\rm s}}{\left( 1 + z/R_{\rm p} \right)^2} \ .
\end{equation}

Unless otherwise noted, molecular opacities are derived from the HITRAN database \citep{gordonetal2022} using the Line-By-Line ABSorption Coefficients tool \citep[\texttt{LBLABC};][]{meadows&crisp1996}. The \texttt{rfast} radiative transfer tools can also interface to the \citet{freedmanetal2008} opacities database \citep[see also][]{freedmanetal2014}. {Full line-resolving opacities are placed onto a wavenumber grid at 1\,cm$^{-1}$ resolution (0.1\,cm$^{-1}$-resolved opacities are also available for high-resolution applications) and then further degraded in resolution when forward or inverse modeling to at least an order of magnitude finer resolving power than the relevant observational data. For each incorporated molecule, opacities span 0.1--100\,$\upmu$m so that the highest resolving power that can be accommodated (assuming no over-sampling) at optical/near-infrared/thermal wavelengths is roughly 10,000/2,000/100.} {As the core radiative transfer solvers for \texttt{rfast} are indifferent to thermochemical conditions, the input opacities are then the only determinant of the types of worlds that can be simulated using the \texttt{rfast} suite. Cold and clement worlds are emphasized below, so the adopted opacities span only 50--700\,K. Nevertheless, \texttt{LBLABC}-generated opacities have been shown to compare well to other tools even under hot Jupiter-like conditions \citep{robinson2017}.}

A primary design consideration for the \texttt{rfast} tool is rapid exploration of retrieval scenarios. Software is nearly entirely written using linear algebra techniques, thereby taking advantage of vectorized computational approaches. Exceptions occur for aspects of atmospheric recursion relations and integration over atmospheric pixels in the three-dimensional reflected light option. As the number of atmospheric levels or planetary pixels are generally at least 1--2 orders of magnitude smaller than the number of spectral points, these exceptions do not impart any significant model inefficiencies. On a single processor, the \texttt{rfast} tool can generate a spectrum with 10k spectral points for a model atmosphere with 50 vertical levels and eight absorbing gas species (including opacity interpolation over both pressure and temperature) in 400\,ms for the single scene reflectance option, 1\,s for the three-dimensional phase-dependent reflectance option (with $M=3$), 600\,ms for the thermal emission option, and 300\,ms for the transit spectroscopy option.

The \texttt{rfast} model currently adopts the widely-used and versatile \texttt{emcee} Markov chain Monte Carlo sampler \citep{foremanmackeyetal2013} when employed as a retrieval tool. Functions for computing the likelihood, prior probability, and posterior probability could straightforwardly be adapted for use with analogous samplers, and efficiencies may be gained by adopting a multi-nested sampling routine \citep{buchneretal2014}. The \texttt{rfast} framework allows for retrieving on more than 20 atmospheric, planetary, and orbital parameters: {atmospheric surface pressure, atmospheric temperature, surface albedo, atmospheric mean molar weight, planetary radius, planetary mass, surface gravity, cloud top pressure, cloud vertical extent, cloud optical thickness, fractional cloudiness, orbital distance, planetary phase angle, as well as gas mixing ratios for argon, molecular nitrogen, molecular oxygen, water vapor, carbon dioxide, ozone, carbon monoxide, nitrous oxide, methane, helium, and molecular hydrogen}. Users may adopt uninformed or Gaussian priors in either log or linear space. Additionally, gas abundance retrievals may be performed with the center-log ratio approach, which has been adapted in exoplanet applications to prevent biased priors for a background gas in \citet{benneke&seager2012} \citep[see also][]{damiano&hu2021,pietteetal2022}.

%
\section{Model Validations} \label{sec:valid}
%

Theoretical aspects of the \texttt{rfast} forward model have already seen applications in various Solar System and exoplanet studies, although implementations there were Fortran-based \citep{robinsonetal2011,robinson2017,robinson&crisp2018}. Nevertheless, the novel applications within the \texttt{rfast} framework warrant validation. Importantly, aspects of core radiative transfer engines as well as overall forward modeling capabilities require verification. Finally, the retrieval capabilities of the \texttt{rfast} suite can be verified against a key initial investigation into atmospheric inference for directly imaged Earth-like exoplanets \citep{fengetal2018}.

\subsection{Isochromatic Core Radiative Transfer Model Validations} \label{subsec:corert}

\begin{figure}
    \centering
    \includegraphics[scale=0.4]{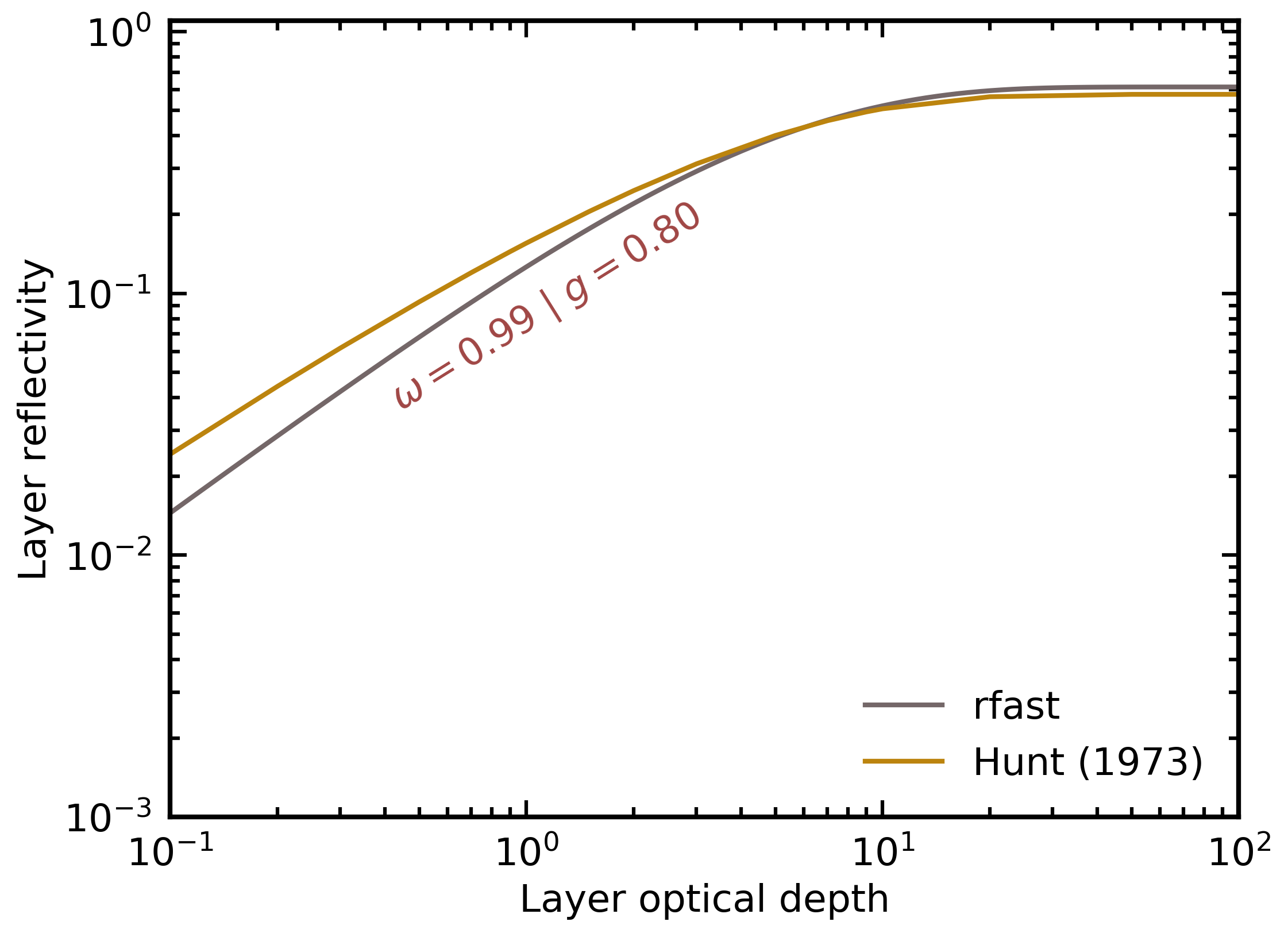}
    \includegraphics[scale=0.4]{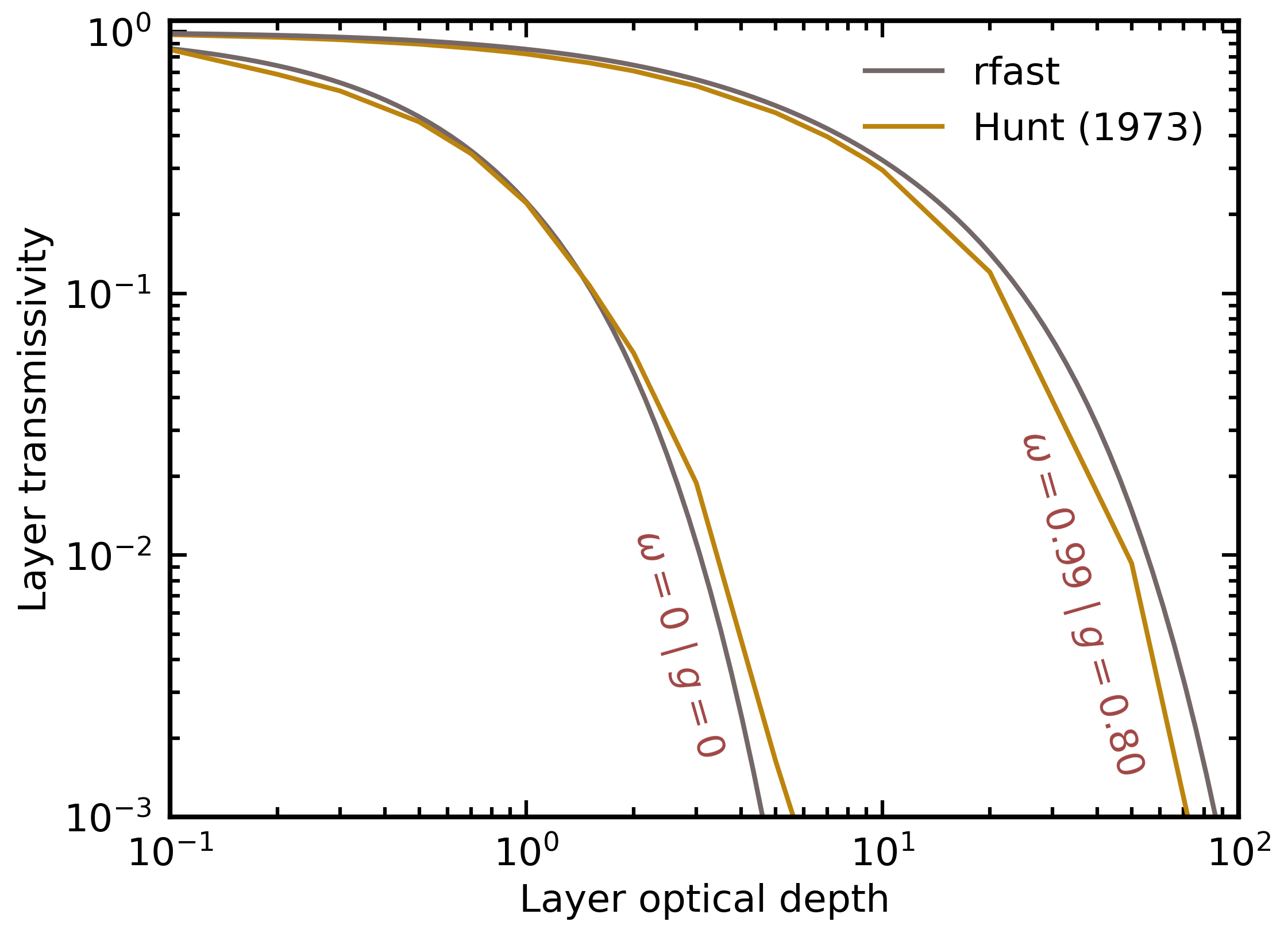}
    \includegraphics[scale=0.4]{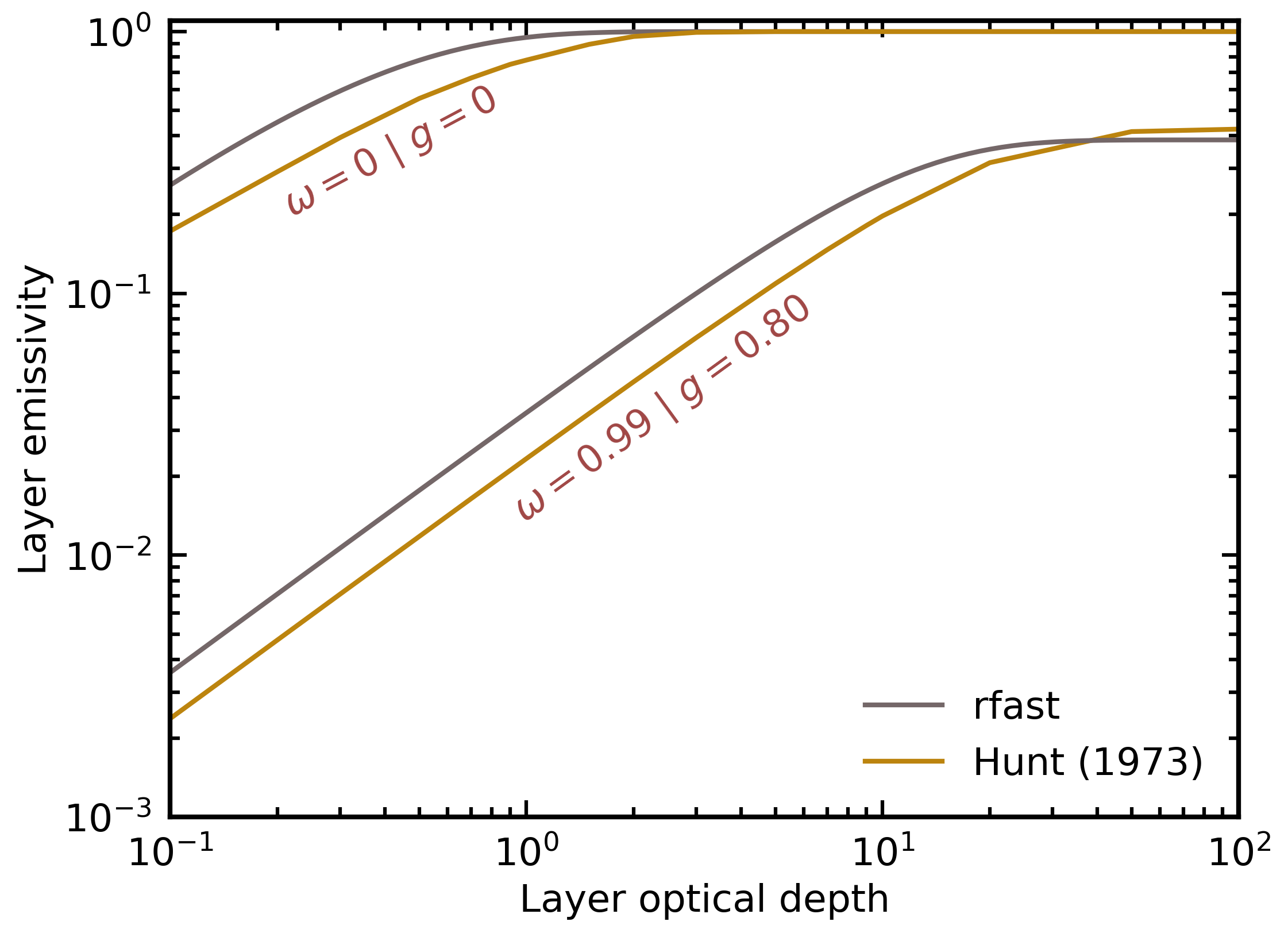}
    \caption{Layer reflectivity (top left), transmissivity (top right), and emissivity (bottom) as a function of layer extinction optical depth in the \texttt{rfast} model (grey lines) versus detailed calculations from \citet{hunt1973} (yellow lines). Limiting cases of pure absorption and forward scattering are considered, where layers have no reflectivity in the pure absorption case.}
    \label{fig:hunt}
\end{figure}

A standard check of the radiative properties of a single homogeneous atmospheric layer (i.e., a layer with uniform optical properties throughout) is to compare against the detailed numerical solutions of \citet{hunt1973}. In this earlier work, the flux reflectivity, transmissivity, and emissivity (analogous to $r_k$, $t_k$, and $a_k$ in this present work) were studied for layers of various optical thickness and constant single scattering albedo and asymmetry parameter. Figure~\ref{fig:hunt} compares results from the \citet{hunt1973} study to those from the \texttt{rfast} two-stream treatment for two limiting cases\,---\,pure absorption and forward scattering. While systematic biases are apparent, these are equivalent to other two-stream approaches \citep{toonetal1989}. {More specifically, reflectivity and transmissivity biases are comparable in magnitude and direction to those reported in \citet{toonetal1989}. Emissivity biases can be large (greater than 10\%) for cases with small optical depths (i.e., optical depths below a few tenths), and two-stream models presented in \citet{toonetal1989} also struggle in these conditions}. Note that these layer properties underpin the single scene reflectance and thermal emission options in \texttt{rfast} as well as the treatment of multiply-scattered radiation in the phase-dependent option.

\begin{figure}
    \centering
    \includegraphics[scale=0.4]{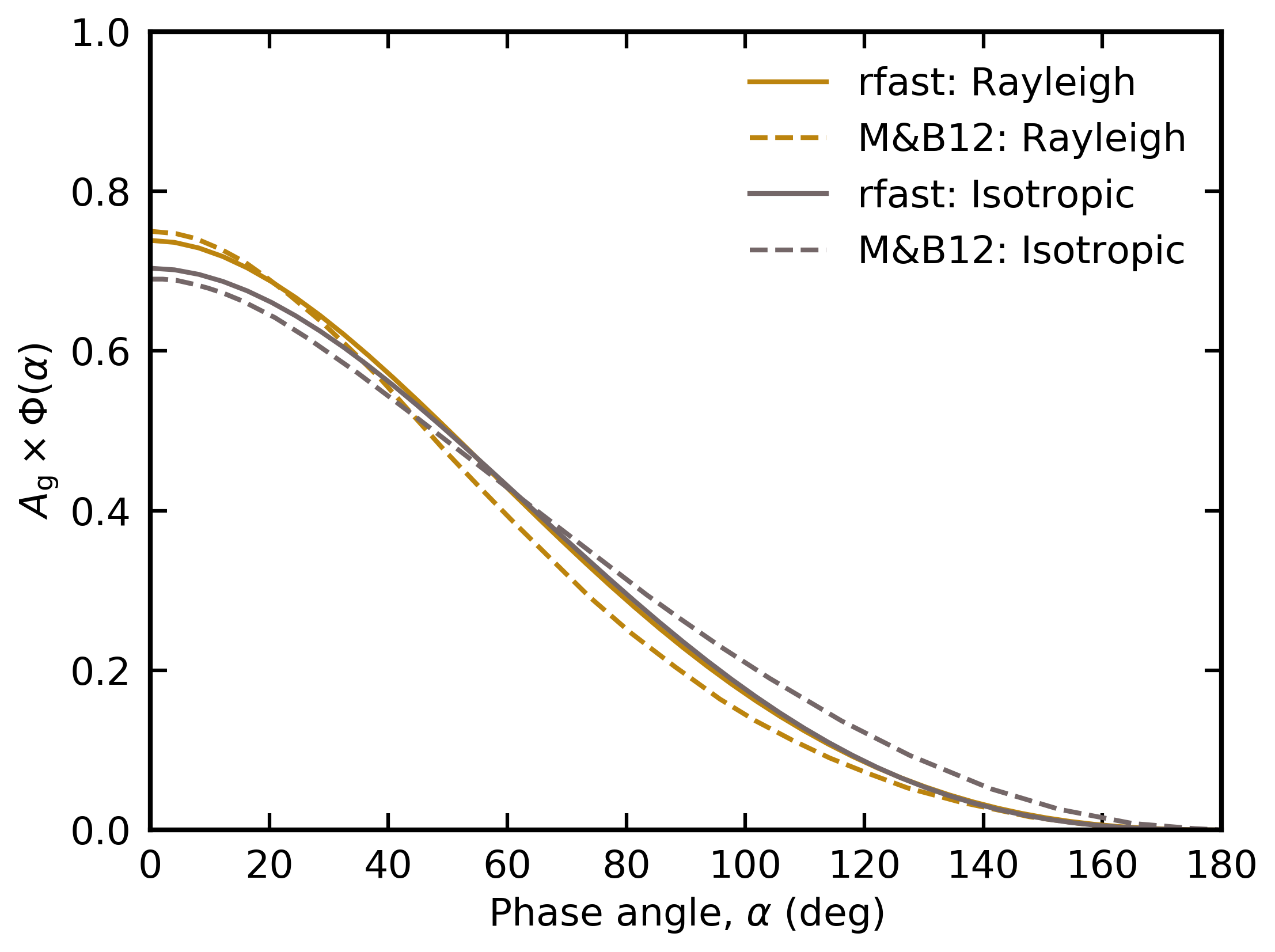}
    \caption{Phase-dependent planetary reflectivity for infinitely deep Rayleigh and isotropic scattering atmospheres from \citet{madhusudhan&burrows2012} as compared to \texttt{rfast}.} 
    \label{fig:mb}
\end{figure}

\begin{figure}
    \centering
    \includegraphics[scale=0.4]{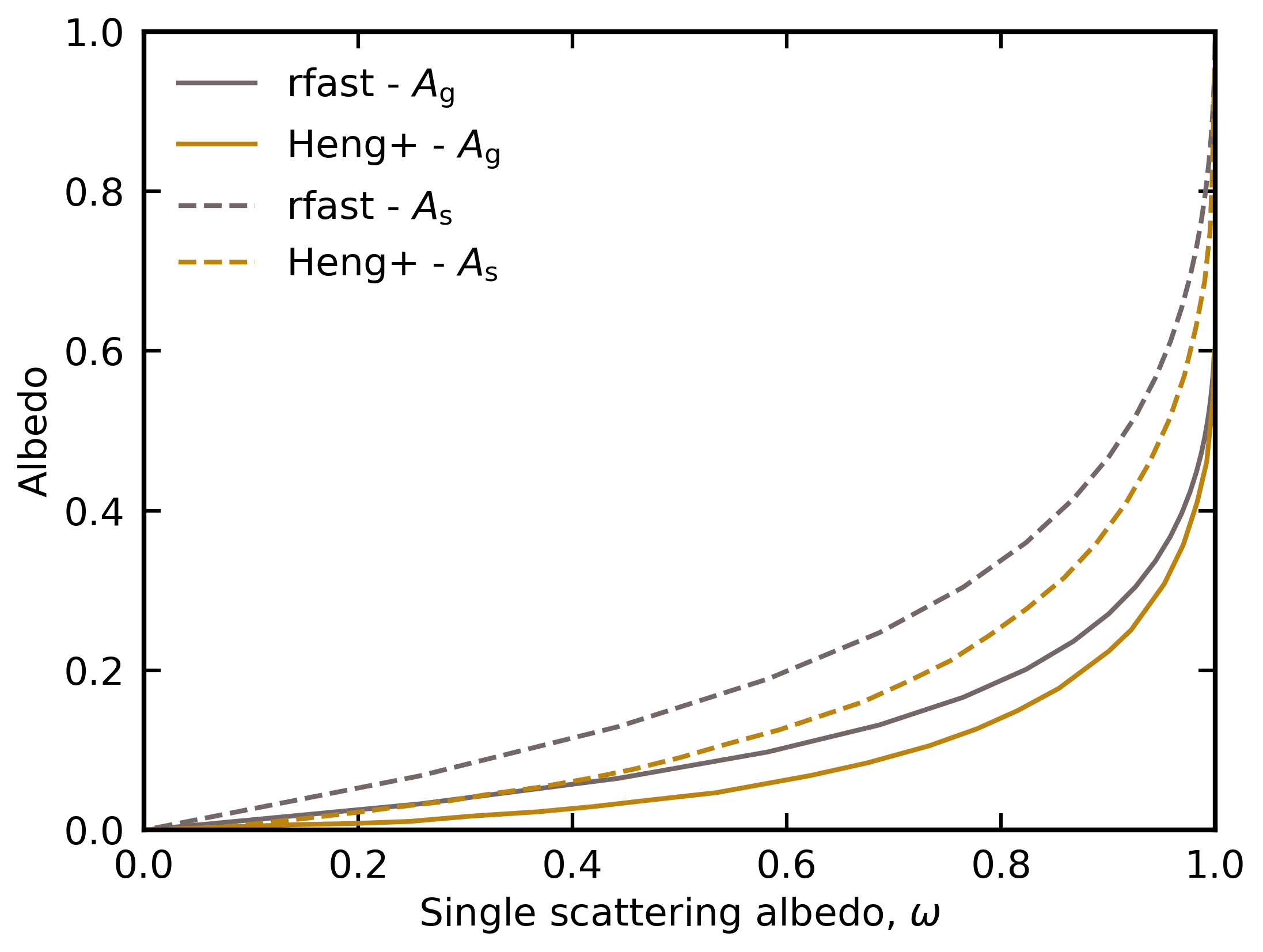}
    \caption{Planetary geometric ($A_{\rm g}$; solid lines) and spherical ($A_{\rm s}$; dashed lines) albedo as a function of single scattering albedo from \citet{hengetal2021} and \texttt{rfast} for an infinitely deep atmosphere whose medium obeys a Henyey-Greenstein scattering phase function of asymmetry parameter 0.508 \citep[selected to reproduce population-level results for hot Jupiters by][]{hengetal2021}.}
    \label{fig:heng}
\end{figure}

The theoretical study of the phase-dependent reflection of planetary bodies with homogeneous atmospheres has a long history \citep{horak1950,sobolev1975}. More recent studies provide straightforward opportunities to validate phase-dependent treatments within \texttt{rfast}\,---\,both \citet{madhusudhan&burrows2012} and \citet{hengetal2021} present analytic or near-analytic results for light reflection from planetary bodies with homogeneous atmospheres (or surfaces) that have different scattering properties. Figure~\ref{fig:mb} compares phase-dependent reflectance values (i.e., the product of the geometric albedo and planetary phase function) from the \citet{madhusudhan&burrows2012} work to those from \texttt{rfast} for (optically) infinitely deep Rayleigh and isotropically scattering cases. {A Lambertian surface case (where the phase-dependent reflectance has an analytic solution) is well-reproduced by \texttt{rfast} so is not shown}. Similarly, Figure~\ref{fig:heng} compares results from \citet{hengetal2021} and \texttt{rfast} for the planetary geometric albedo and spherical albedo ($A_{\rm s}$) as a function of single scattering albedo for an infinitely deep atmosphere whose medium has a Henyey-Greenstein phase function of asymmetry parameter $g=0.508$ \citep{henyey&greenstein1941}. (Note that the planetary spherical albedo is given by the integral of $A_{\rm g} \phi \left( \alpha \right) \sin \alpha$ over all phase angles.) All phase-dependent validations adopt $M=10$ for Gauss-Chebyshev integration, which was shown to provide better than 1\% precision. Discrepancies between \texttt{rfast} and the more-sophisticated calculations do occur for geometric albedo and reflectivity calculations (which are most relevant to reflected light observations) at the level of 10\% (or more, under some circumstances), {which stems from the simplifying assumption of hemispheric mean radiative transfer in the multiply-scattered radiation field}. {Future work could incorporate more-accurate radiative transfer solvers that still maintain high computational efficiency \citep[e.g.,][]{spurr&natraj2011}}. As described in Section~\ref{sec:discuss}, this precludes some phase-dependent retrievals on observational data, but does not preclude retrievals performed on phase-dependent synthetic observations generated with \texttt{rfast}.


\subsection{Spectral Validations} \label{subsec:spectral}

The primary utility of the \texttt{rfast} tool is in generating spectra, so spectral validations are of central importance. Comparisons/validations in reflected or emitted light presented here are against the plane-parallel, line-by-line, multiple scattering Spectral Mapping Atmospheric Radiative Transfer (\texttt{SMART}) model \citep[developed by D.~Crisp;][]{meadows&crisp1996} while transit spectrum comparisons are against the \texttt{scaTran} addition to \texttt{SMART} \citep{robinson2017}. Phase-dependent \texttt{SMART} results come from a disk integration technique developed in \citet{robinsonetal2011}. All comparisons adopt a standard Earth atmospheric model \citep{mcclatcheyetal1972} and include gas opacity from N$_2$, O$_2$, H$_2$O, CO$_2$, O$_3$, CO, CH$_4$, and N$_2$O.

\begin{figure}
    \centering
    \includegraphics[scale=0.4]{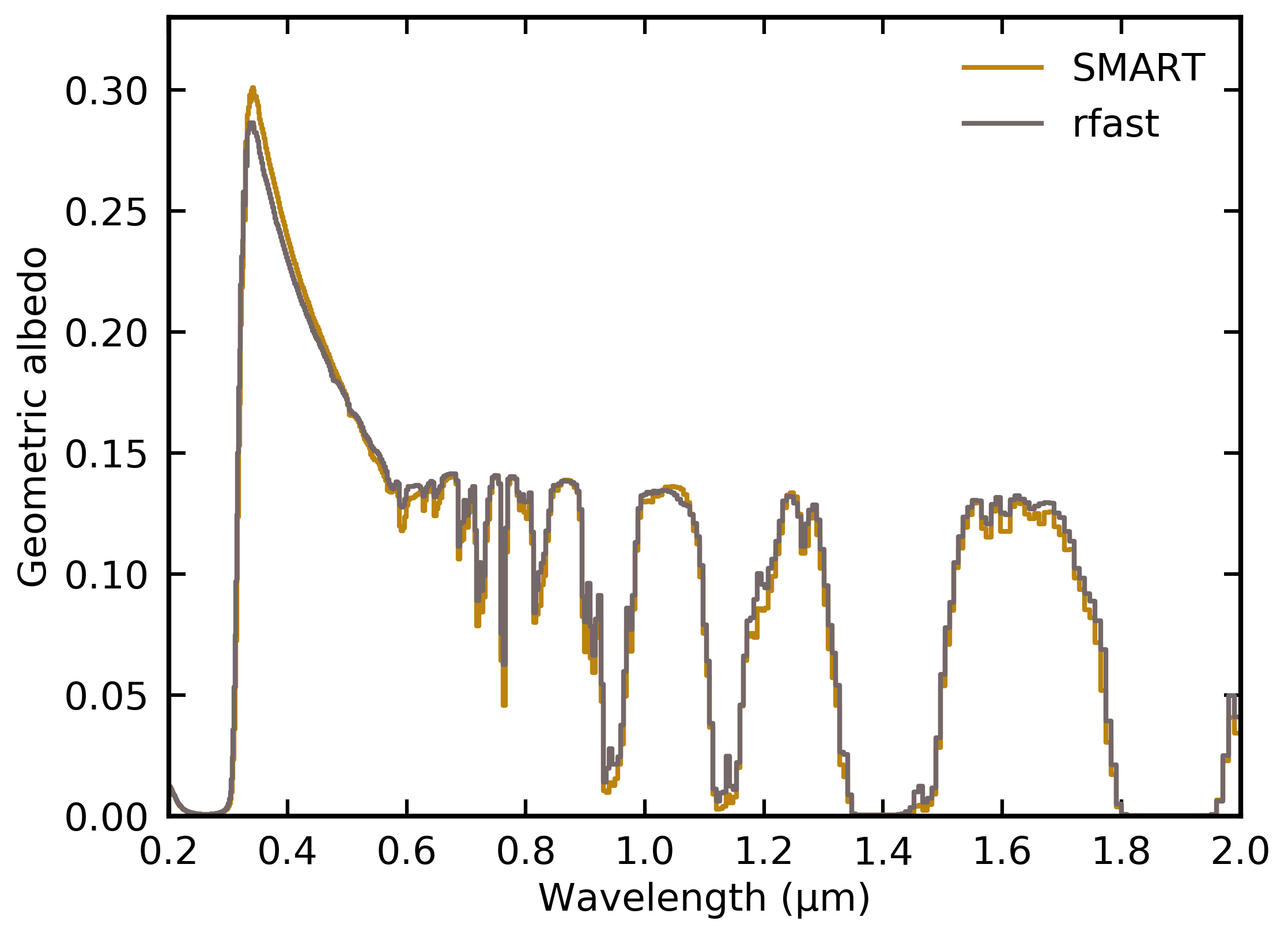}
    \includegraphics[scale=0.4]{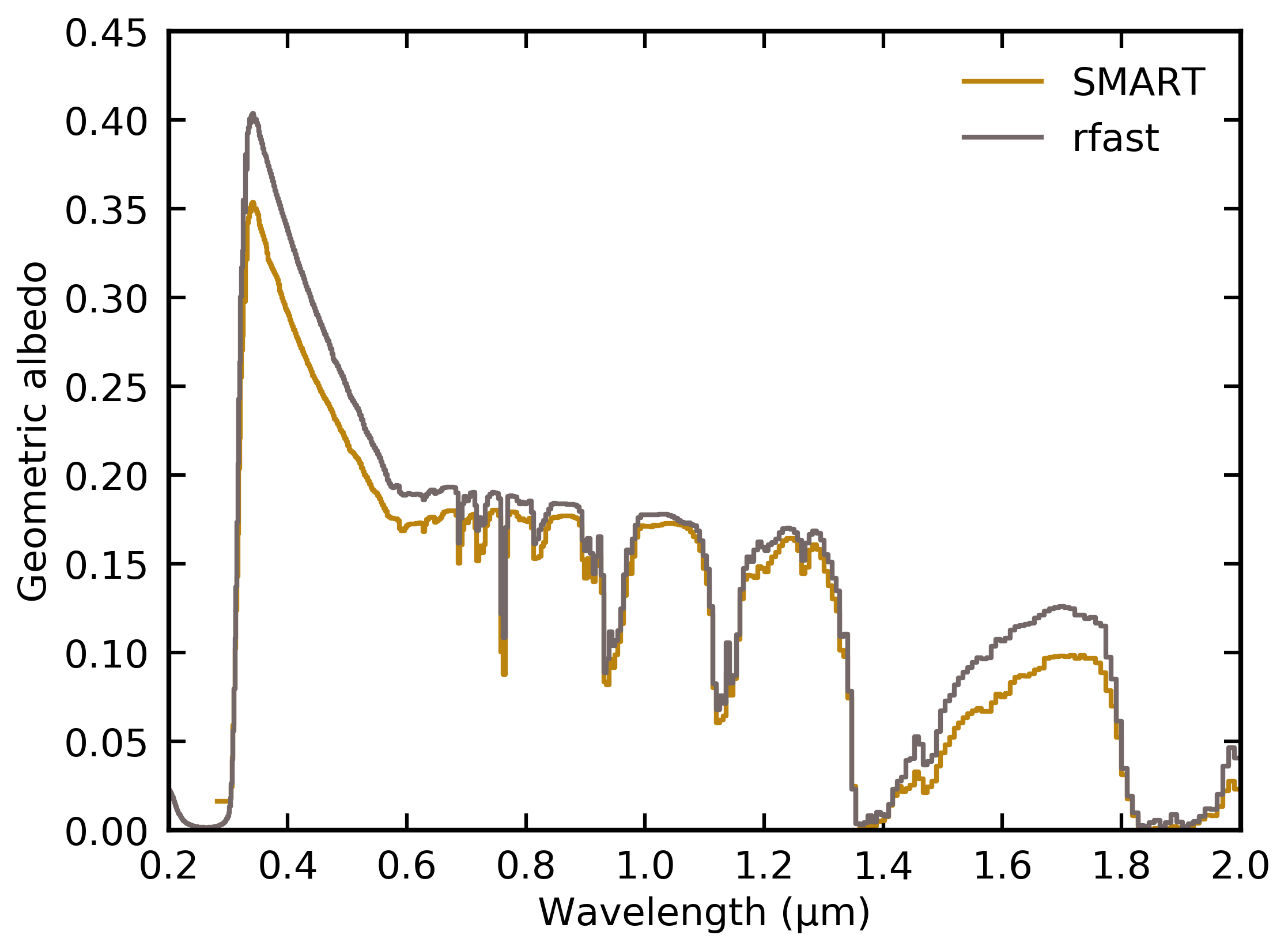}
    \caption{Reflection spectra for a partially-clouded (50\% cloud coverage) Earth-like case from the \texttt{rfast} (grey) and \texttt{SMART} (yellow) models. Left figure is for a single-scene, plane-parallel setup. Right figure is for a three-dimensional treatment where plane-parallel calculations are run for pixels on the planetary disk and then spatially integrated.} 
    \label{fig:smart_refl}
\end{figure}

Figure~\ref{fig:smart_refl} compares results from plane-parallel \texttt{SMART} and single-scene \texttt{rfast}, both at a resolving power ($\lambda$/$\Delta \lambda$) of 200. Simulations {with both \texttt{rfast} and \texttt{SMART}} include 50\% coverage of water ice and liquid clouds with realistic wavelength-dependent scattering properties and a Henyey-Greenstein phase function. The \texttt{SMART} simulation adopts a solar zenith angle of 60\textdegree, and all results adopt a Lambert-like scaling factor of 2/3 to convert from scene albedo to geometric albedo. Agreement between the two models is strong, especially considering the large difference in model complexity. Figure~\ref{fig:smart_refl} also shows an analogous comparison between three-dimensional \texttt{SMART} and \texttt{rfast} cases, both shown at full phase. Discrepancies between the pair of three-dimensional treatments are larger than single-scene cases, owing to the rather simple treatment of diffuse scattering in the \texttt{rfast} model. {Marked differences are seen in the Rayleigh scattering continuum and in the continuum near 1.6\,$\upmu$m. Issues in the Rayleigh continuum stem from the scattering at these wavelengths coming from a combination of cloud optical properties and Rayleigh scattering. The more-simple \texttt{rfast} treatment of diffuse scattering also struggles near 1.6\,$\upmu$m where ice clouds become more absorptive while liquid water clouds remain reflective, which is consistent with overestimates of geometric albedo shown in Figure~\ref{fig:heng}}.

\begin{figure}
    \centering
    \includegraphics[scale=0.4]{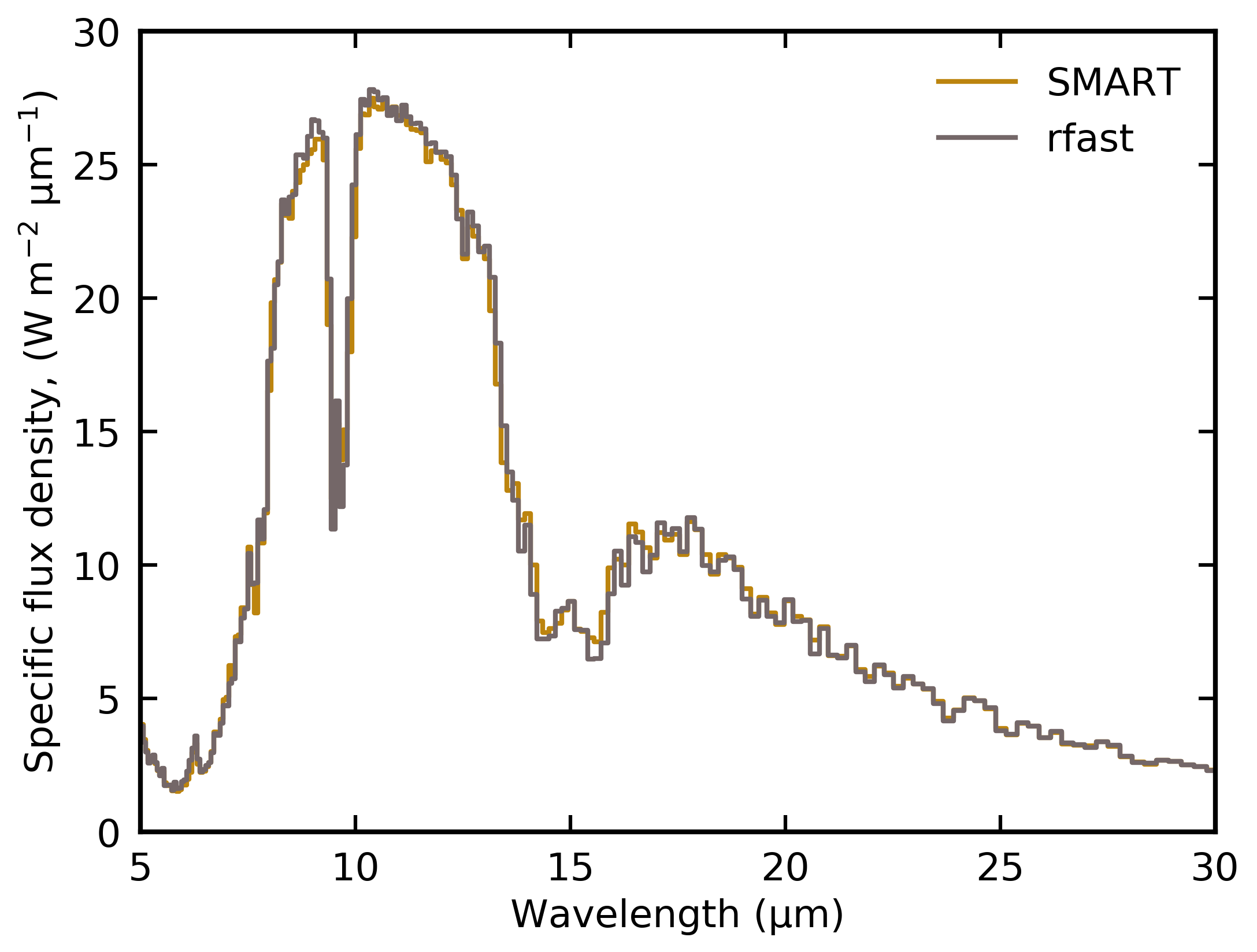}
    \caption{Thermal emission spectra for a clearsky Earth-like case from the \texttt{rfast} (grey) and \texttt{SMART} (yellow) models.} 
    \label{fig:smart_thrm}
\end{figure}

\begin{figure}
    \centering
    \includegraphics[scale=0.4]{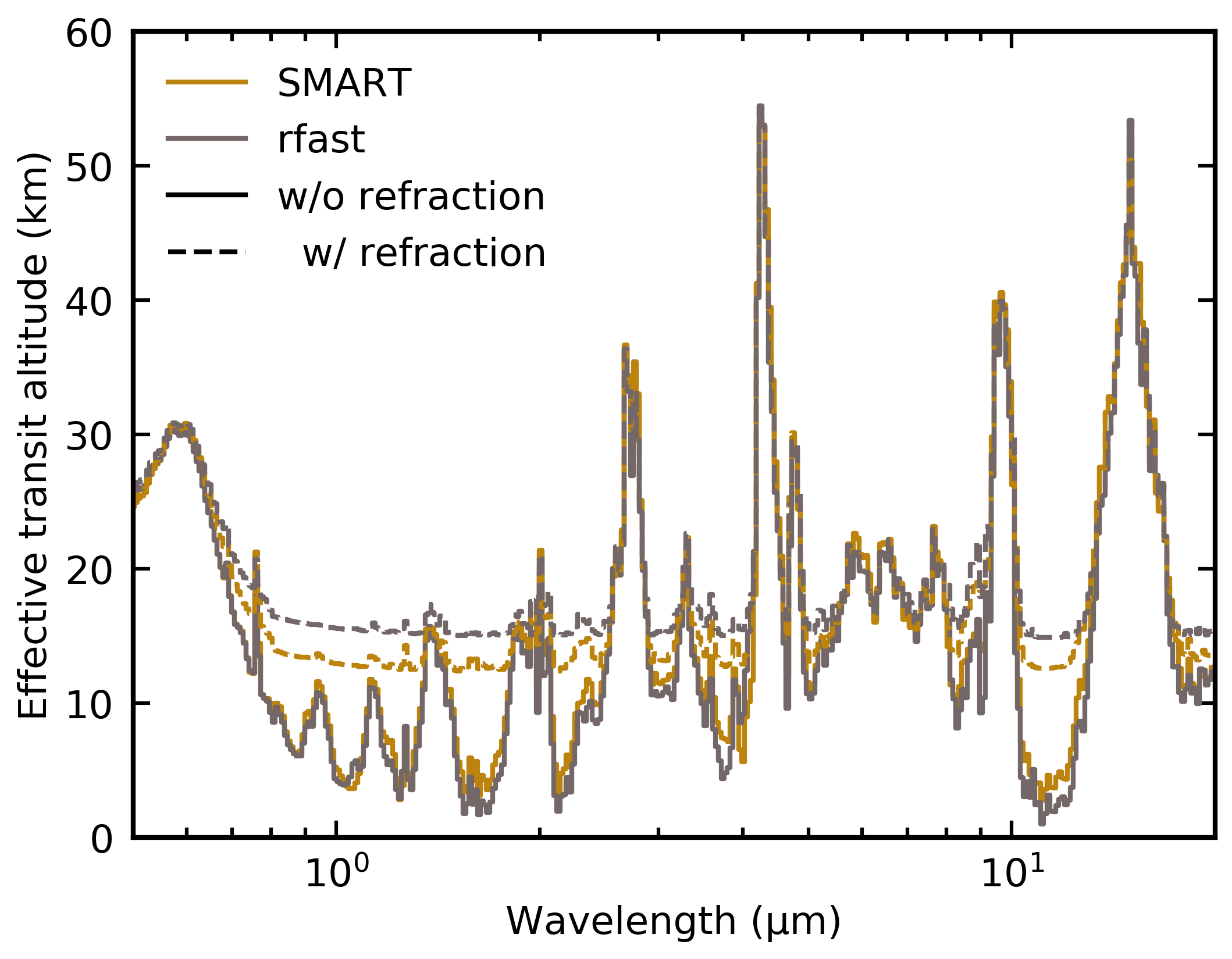}
    \caption{Transit spectra for a clearsky Earth-like case from the \texttt{rfast} (grey) and \texttt{SMART} (yellow) models. Scenarios that both include (dashed) and exclude (solid) refraction are shown.} 
    \label{fig:smart_trns}
\end{figure}

Figures~\ref{fig:smart_thrm} and \ref{fig:smart_trns} show comparisons between \texttt{rfast} and \texttt{SMART} for clearsky thermal emission and transit spectroscopy cases, respectively. Thermal spectra are at a resolving power of 100 and show excellent agreement. Transit spectroscopy cases are also at a resolving power of 100 and plot effective transit altitude ($z_{\rm eff}$), which is defined by separating the solid-body contribution to a transit spectrum from the atmospheric component,
\begin{equation}
    \left( \frac{R_{\rm p} + z_{\rm eff}}{R_{\rm s}} \right)^{\! \! 2} = \left( \frac{R_{\rm{p},\lambda}}{R_{\rm{s}}} \right)^{\! \! 2} \ .
\end{equation}
The transit spectrum comparison shows scenarios that both include and exclude refraction effects for an Earth-Sun twin, where the \texttt{scaTran} addition to the \texttt{SMART} model incorporates refraction effects through ray tracing. A discrepancy (at less than one atmospheric pressure scale height) in the location of the transit floor for the \texttt{rfast} model arises as the incorporated analytic treatment of refraction can only be derived assuming an isothermal atmosphere. As the refractive bending is sensitive to atmospheric number densities, ray tracing through an atmosphere with a thermal structure profile yields a different (and more accurate) result than the analytic treatment. {Fortunately, earlier modeling results show that refractive effects will be quite limited for the types of close-in exoplanets typically studied with transit spectroscopy \citep{betremieux&kaltenegger2014,misraetal2014,betremieux&swain2017,robinsonetal2017}}.

{To summarize, the underlying radiative transfer routines within the \texttt{rfast} suite work well for transit applications (especially when refraction can be ignored), as the transmissivity calculations are in-line with more-sophisticated tools. Single-scene and thermal emission applications of \texttt{rfast} agree well with high-fidelity models, but biases can arise at levels typically less than 5--10\%. Three-dimensional calculations of planetary reflectivity with \texttt{rfast} have larger disagreements when compared to high-fidelity models, and applications of the \texttt{rfast} inverse model to phase-dependent observations should be done with this limitation in mind. For comparison purposes, the high-fidelity, fully line-resolving, cloud-free column \texttt{SMART} simulations for single-scene reflectance, three-dimensional reflectance, thermal emission, and transit required single-core runtimes of 55\,min, 430\,min, 4.3\,min, and 13.3\,min, respectively, while the analogous \texttt{rfast} spectra required runtimes of 0.66\,s, 6.8\,s, 0.19\,s, and 0.36\,s, respectively. Thus, runtimes differ by a factor of 1,300--5,000.}

\subsection{Retrieval Validations} \label{subsec:feng}

\citet{fengetal2018} presented atmospheric retrieval results for simulated reflected light high contrast imaging observations of Earth-like exoplanets. Driven by initial ideas for the HabEx and LUVOIR concept missions, the \citet{fengetal2018} results focused on visible wavelengths (0.4--1.0\,$\upmu$m). Retrievals included 11 inferred parameters: planetary surface pressure ($p_{\rm s}$), planetary radius, planetary surface gravity, a grey surface albedo ($A_{\rm s}$), cloud top pressure ($p_{\rm c}$), cloud thickness (i.e., pressure extent; $\Delta p_{\rm c}$), cloud extinction optical depth ($\tau_{\rm c}$), cloud coverage fraction on the planetary disk ($f_{\rm c}$), and gas mixing ratios for water, ozone, and molecular oxygen ($f_{\rm H2O}$, $f_{\rm O3}$, and $f_{\rm O2}$, respectively; {assumed to have constant vertical profiles}). Except for planetary radius, all parameters were retrieved in log-space and molecular nitrogen was taken as the background gas. Uninformed priors were adopted for all parameters.

\begin{figure}
    \centering
    \includegraphics[scale=0.4]{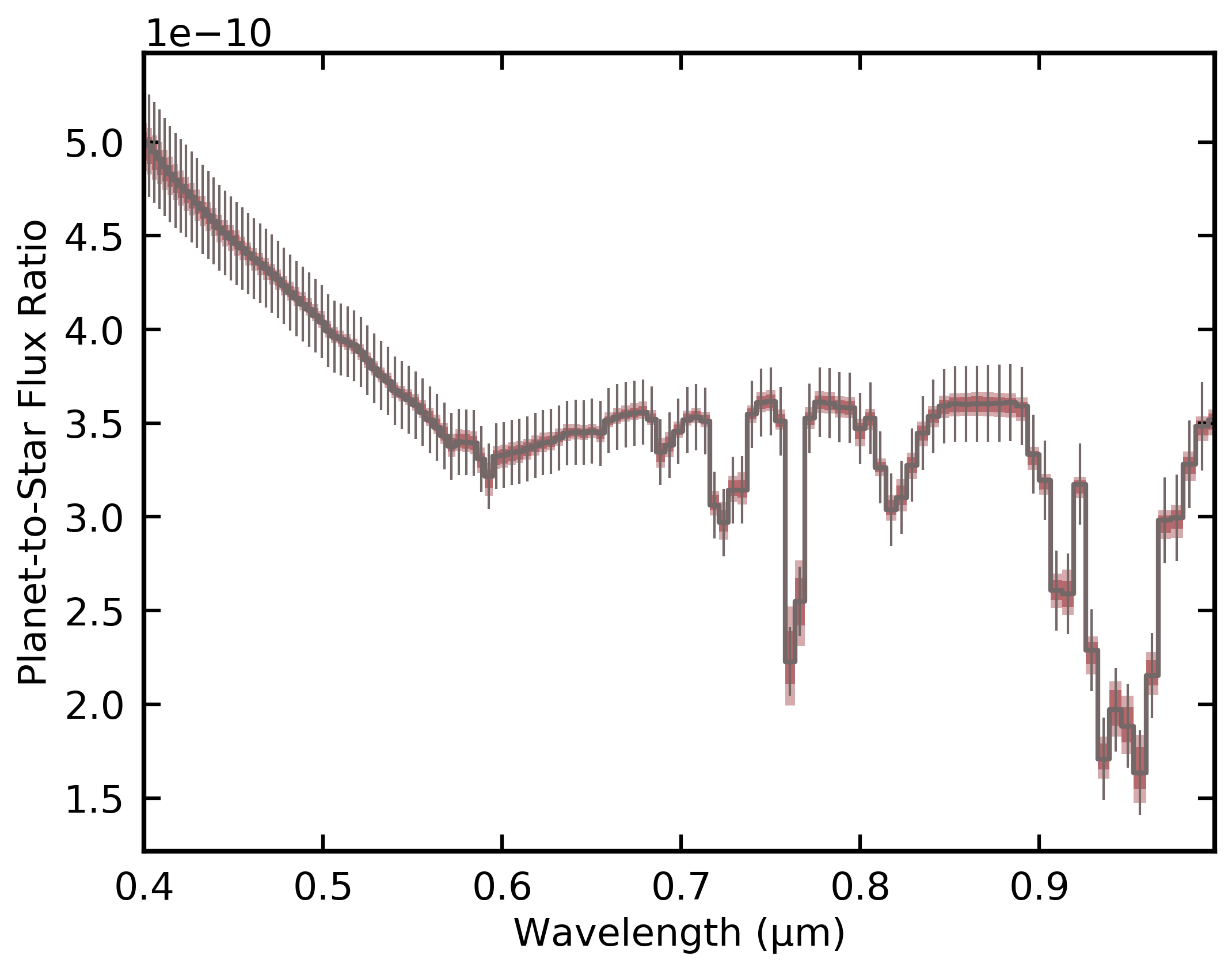}
    \caption{Simulated visible-wavelength reflected light observation of an exo-Earth at resolving power of 140 and a V-band SNR of 20 \citep[points and associated error-bars, after][]{fengetal2018}. Forward model spread from retrieval analysis as applied to the simulated observation is shown as darker and lighter swaths for 16--84 and 5--95 percentiles.} 
    \label{fig:feng_swaths}
\end{figure}

\begin{figure}
    \centering
    \includegraphics[scale=0.25]{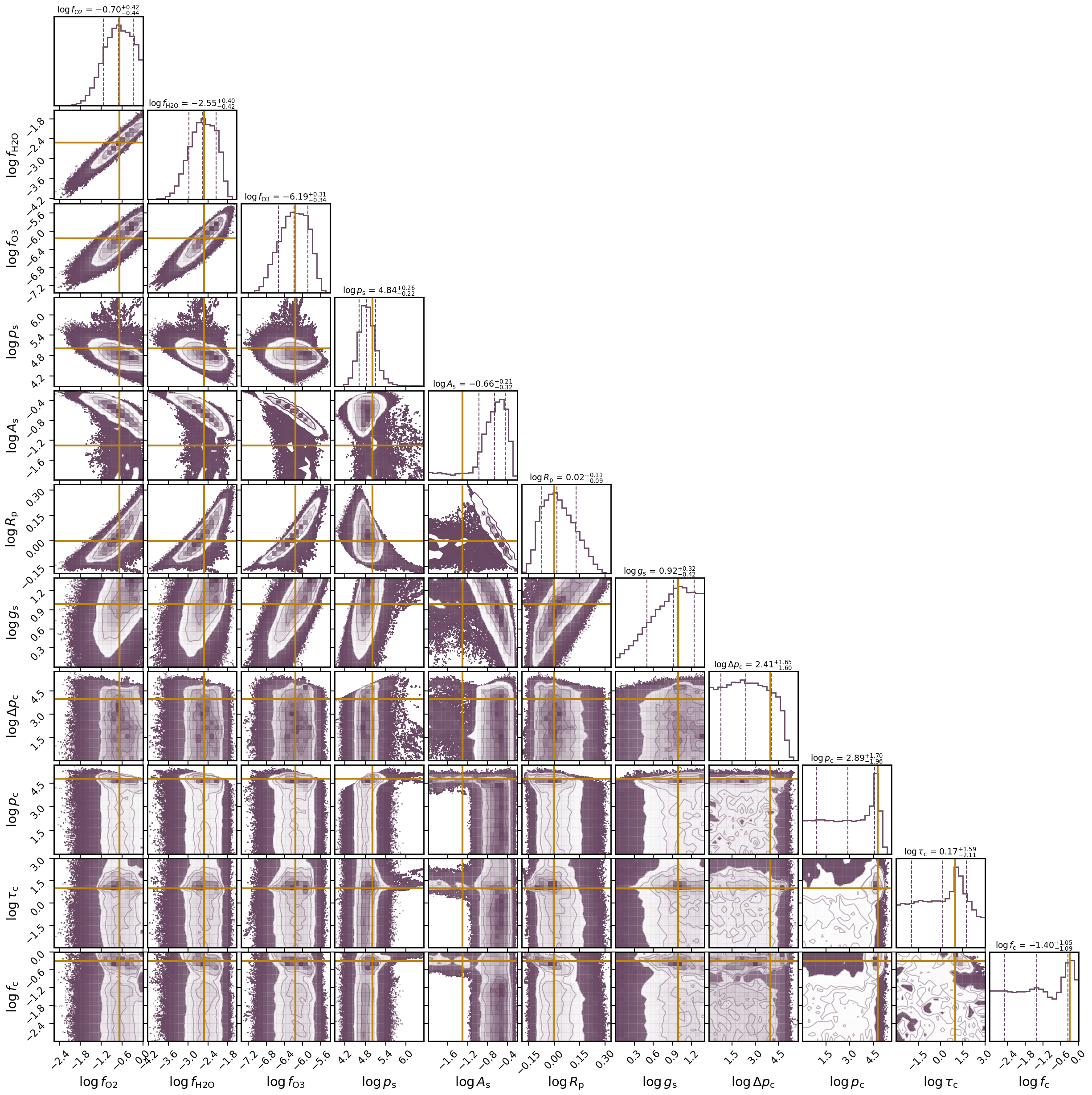}
    \caption{Visualized posterior distribution for eleven parameters from the \texttt{rfast} tool when retrieving on the simulated observations in Figure~\ref{fig:feng_swaths}. One-dimensional marginal distributions are shown along the diagonal with associated 16/50/84-percentile values indicated above. Values used to generate the faux observation (i.e., ``truth'' values) are indicated as orange vertical and horizontal lines.} 
    \label{fig:feng_corner}
\end{figure}

Figures~\ref{fig:feng_swaths} and \ref{fig:feng_corner} show retrieval results from the \texttt{rfast} tool that are analogous to a scenario in the \citet{fengetal2018} work where the resolving power was fixed at 140 and the V-band SNR was taken as 20 (see simulated data with error bars in Figure~\ref{fig:feng_swaths}). {The simulated observations did not have uncertainties randomly applied to maintain consistency with \citet{fengetal2018}, who demonstrated that a statistical sampling of retrievals with randomized spectral errors yielded similar inference results to a case where errors are non-randomized and simply centered on the noise-free simulation}. {Note that the \citet{fengetal2018} model is three-dimensional and uses 100 Gauss-Chebyshev integration points over the illuminated disk while the \texttt{rfast} retrieval was executed using the single-scene option}. Figure~\ref{fig:feng_corner} visualizes the full posterior distribution using the Python \texttt{corner} package \citep{foremanmackey2016} with one-dimensional marginal distributions for each parameter shown along the diagonal. Figure~\ref{fig:feng_swaths} shows forward model swaths at the 16--84 and 5--95 percentiles (i.e., 1- and 2-sigma for a Gaussian distribution).

For validation purposes, Table~\ref{tab:feng} compares inferred parameter values at the 16/50/84 percentiles from the \citet{fengetal2018} study to those from \texttt{rfast}. Note that planetary radius was retrieved in log space with the \texttt{rfast} tool and then re-sampled to linear space for comparison to the \citet{fengetal2018} results. Agreement between constraints is generally strong with 16--84 percentile spreads for {nearly all cloud-unrelated parameters} in \texttt{rfast} falling within {40\%} of the \citet{fengetal2018} values, {as indicated by the spread comparison column which differences the 16--84th percentile ranges from the two tools relative to the \citet{fengetal2018} spread}. Key exceptions occur for parameters related to clouds ($p_{\rm c}$, $\Delta p_{\rm c}$, $\tau_{\rm c}$, and $f_{\rm c}$) where \texttt{rfast} finds markedly weaker constraints. Detailed investigation reveals that the \citet{fengetal2018} retrievals did not sufficiently progress Markov chain Monte Carlo simulations to map out the posterior distributions for poorly-constrained parameters\,---\,100k walker steps are taken in the \texttt{rfast} retrieval versus 10k--20k for the \citet{fengetal2018} retrievals. {An equivalent test with the three-dimensional version of \texttt{rfast} further confirmed these results}. Importantly, this comparison shows that, for low-SNR simulated data and retrievals related to mission concept studies, three-dimensional spectral models are {likely} not required {for a first-order understanding of} the mapping from predicted data quality to parameter constraints. When comparing the \texttt{rfast} tool to the \citet{fengetal2018} model, {this results in a runtime savings that scales with the number of disk integration points in the three-dimensional model}.

\begin{table}
\begin{center}
\caption{Retrieval Model Comparison}
\label{tab:feng}
\begin{tabular}{llrrrr}
 Parameter               & Units           & Input    & Feng+18                  & \texttt{rfast~\,}        & Spread Comparison \\ \hline \hline
 $\log p_{\rm s}$        & log Pa          & $5.0$    &  $ 5.07^{+0.39}_{-0.31}$ &  $ 4.84^{+0.26}_{-0.22}$ & 0.31  \\
 $R_{\rm p}$             & $R_{\oplus}$    & $1.0$    &  $ 1.05^{+0.42}_{-0.27}$ &  $ 1.04^{+0.20}_{-0.31}$ & 0.26  \\ 
 $\log g_{\rm s}$        & log m\,s$^{-2}$ & $0.99$   &  $ 1.20^{+0.50}_{-0.64}$ &  $ 0.92^{+0.32}_{-0.42}$ & 0.35  \\
 $\log A_{\rm s}$        &                 & $-1.3$   &  $-0.79^{+0.34}_{-0.69}$ &  $-0.66^{+0.21}_{-0.32}$ & 0.49  \\
 $\log p_{\rm c}$        & log Pa          & $4.8$    &  $ 4.34^{+0.53}_{-0.85}$ &  $ 2.89^{+1.70}_{-1.96}$ & 1.65  \\
 $\log \Delta p_{\rm c}$ & log Pa          & $4.0$    &  $ 3.51^{+1.00}_{-0.98}$ &  $ 2.41^{+1.65}_{-1.60}$ & 0.64  \\
 $\log \tau_{\rm c}$     &                 & $1.0$    &  $ 0.79^{+0.87}_{-1.40}$ &  $ 0.17^{+1.59}_{-2.11}$ & 0.63  \\
 $\log f_{\rm c}$        &                 & $-0.3$   &  $-0.76^{+0.54}_{-1.26}$ &  $-1.40^{+1.05}_{-1.09}$ & 0.19  \\
 $\log f_{\rm H2O}$      &                 & $-2.5$   &  $-2.43^{+0.39}_{-0.56}$ &  $-2.55^{+0.40}_{-0.42}$ & 0.14  \\
 $\log f_{\rm O3}$       &                 & $-6.2$   &  $-6.03^{+0.34}_{-0.48}$ &  $-6.19^{+0.31}_{-0.34}$ & 0.21  \\
 $\log f_{\rm O2}$       &                 & $-0.68$  &  $-0.60^{+0.43}_{-0.59}$ &  $-0.70^{+0.42}_{-0.44}$ & 0.16
\end{tabular}
\end{center}
\end{table}

\section{Results} \label{sec:results}
%

In what follows, inferences from a variety of Solar System analog observations for exoplanets are explored using the \texttt{rfast} tool. Prior to these explorations, a {demonstration application of the \texttt{rfast} tool} is provided for a scenario near to its original design use\,---\,exoplanet direct imaging feasibility studies. Following this demonstration, the \texttt{rfast} tool is applied to reflected light observations of the {distant Earth} from NASA's \textit{EPOXI} mission \citep{livengoodetal2011}, providing a strong proof-of-concept for future exoplanet direct imaging missions. Next, retrieval analysis is used to understand information from a spacecraft-measured whole-disk infrared spectrum of Earth. Finally, an observationally-derived transit spectrum of Titan is studied using the \texttt{rfast} tool. As is common for exoplanet atmospheric retrievals, gas mixing ratios are assumed constant throughout the atmosphere (although the \texttt{rfast} tool can accommodate vertical structure in gas mixing ratios). The studies in this section are not intended to be exhaustive\,---\,myriad questions could be asked of these analog observations, likely motivating many stand-alone studies. Instead, the studies below are meant to be an example of how retrieval approaches can be understood and validated through application to worlds where detailed \textit{in situ} (or orbiter/spacecraft) data exist. Finally, note that any detailed discussion of results derived in this section are reserved for Section~\ref{sec:discuss}.

\subsection{Exo-Earth Reflected Light Direct Imaging} \label{subsec:luvex}

\begin{figure}
    \centering
    \includegraphics[scale=0.4]{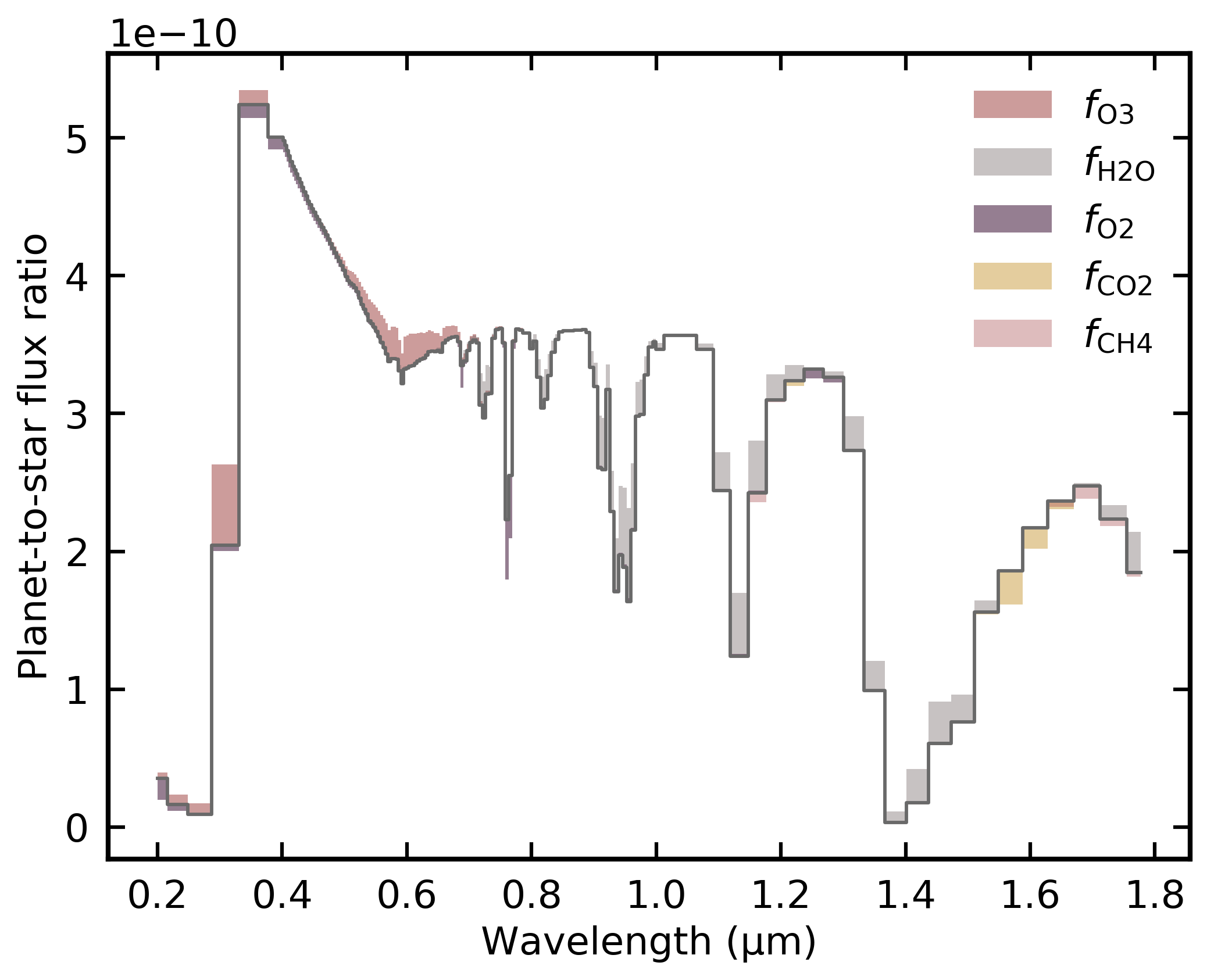}
    \caption{Simulated spectrum of an exo-Earth orbiting a solar twin at characteristic HabEx/LUVOIR spectral coverage and resolving power (ultraviolet [0.2--0.4\,$\upmu$m], visible [0.4--1.0\,$\upmu$m], and near-infrared [1.0--1.8\,$\upmu$m] at resolving power of 7, 140, and 70, respectively). Gas spectral impacts are indicated by showing sensitivity to mixing ratio changes for key species. For the sensitivity test, and relative to fiducial values, the water vapor and ozone mixing ratios were halved, the molecular oxygen mixing ratios was doubled, and the carbon dioxide and methane mixing ratios were increased by a factor of 5.} 
    \label{fig:luvex_sens}
\end{figure}

The \texttt{rfast} tool was originally designed to rapidly answer questions for development of exoplanet characterization-focused missions. As an example, Figure~\ref{fig:luvex_sens} shows a characteristic exo-Earth spectrum at resolving powers relevant to the HabEx and LUVOIR mission concepts (i.e., resolving powers of 7, 140, and 70 in the ultraviolet, optical, and near-infrared, respectively). Spectral impacts of species that are radiatively active in the depicted wavelength range are also indicated.

\begin{figure}
    \centering
    \includegraphics[scale=0.4]{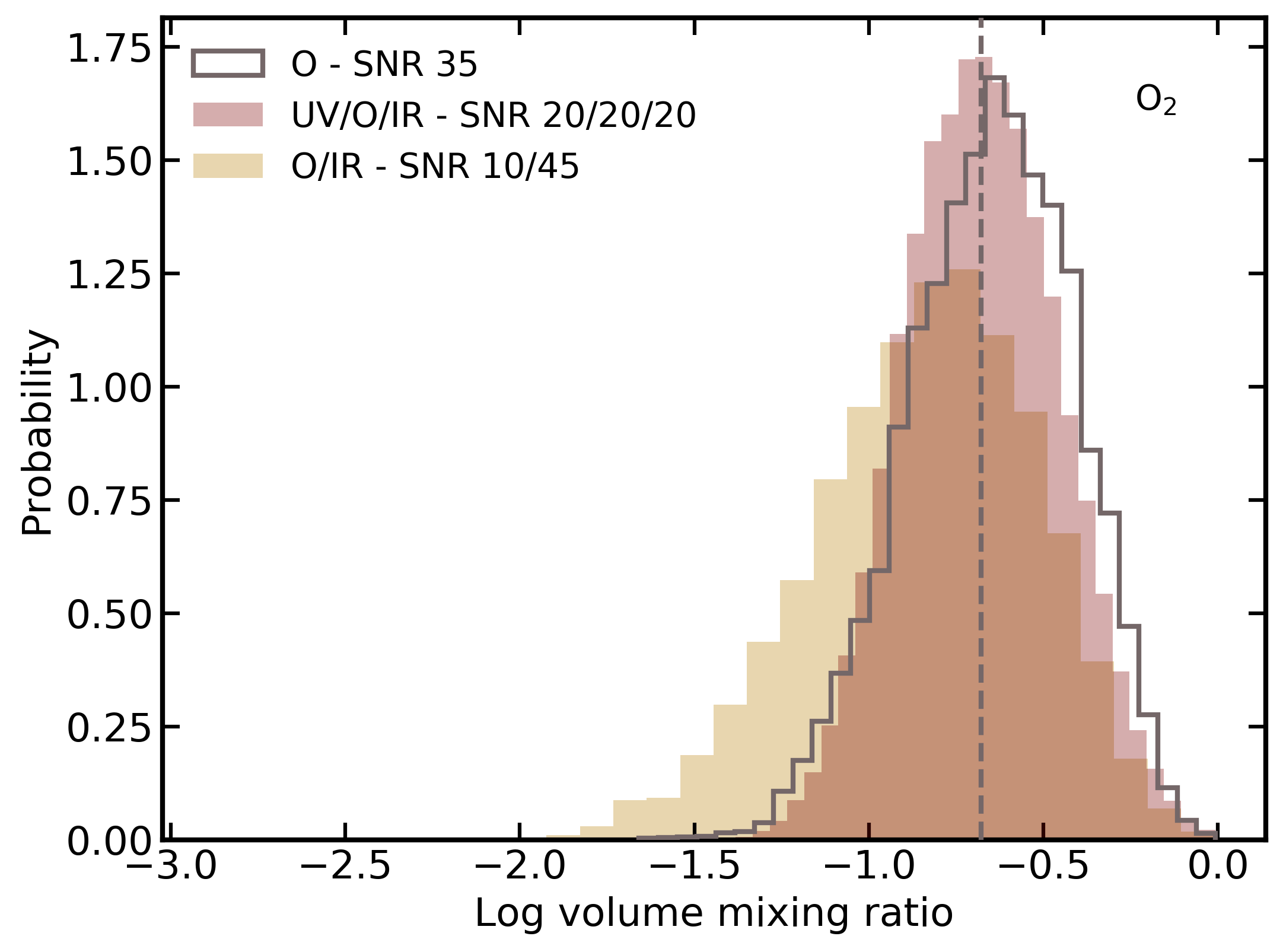}
    \includegraphics[scale=0.4]{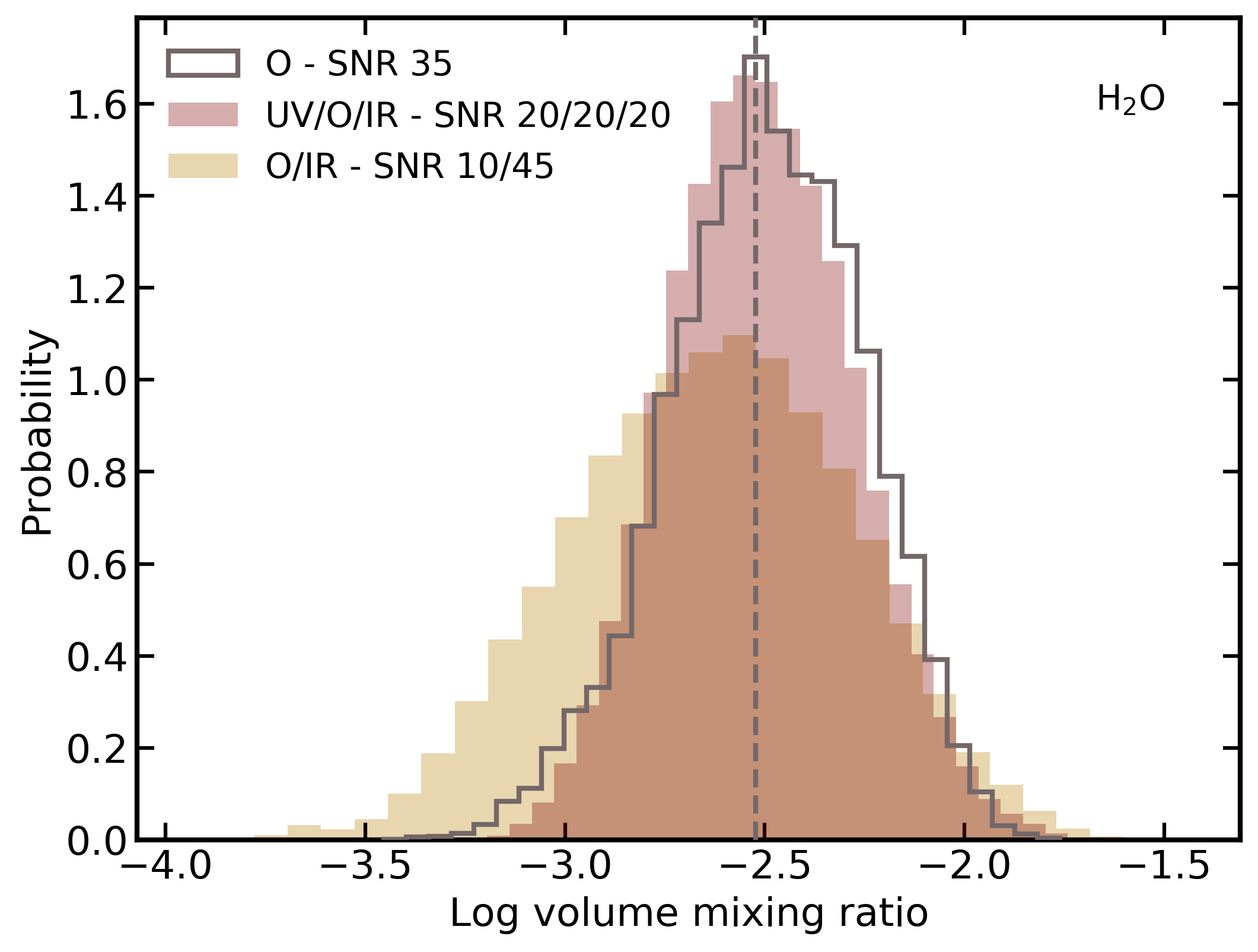}
    \includegraphics[scale=0.4]{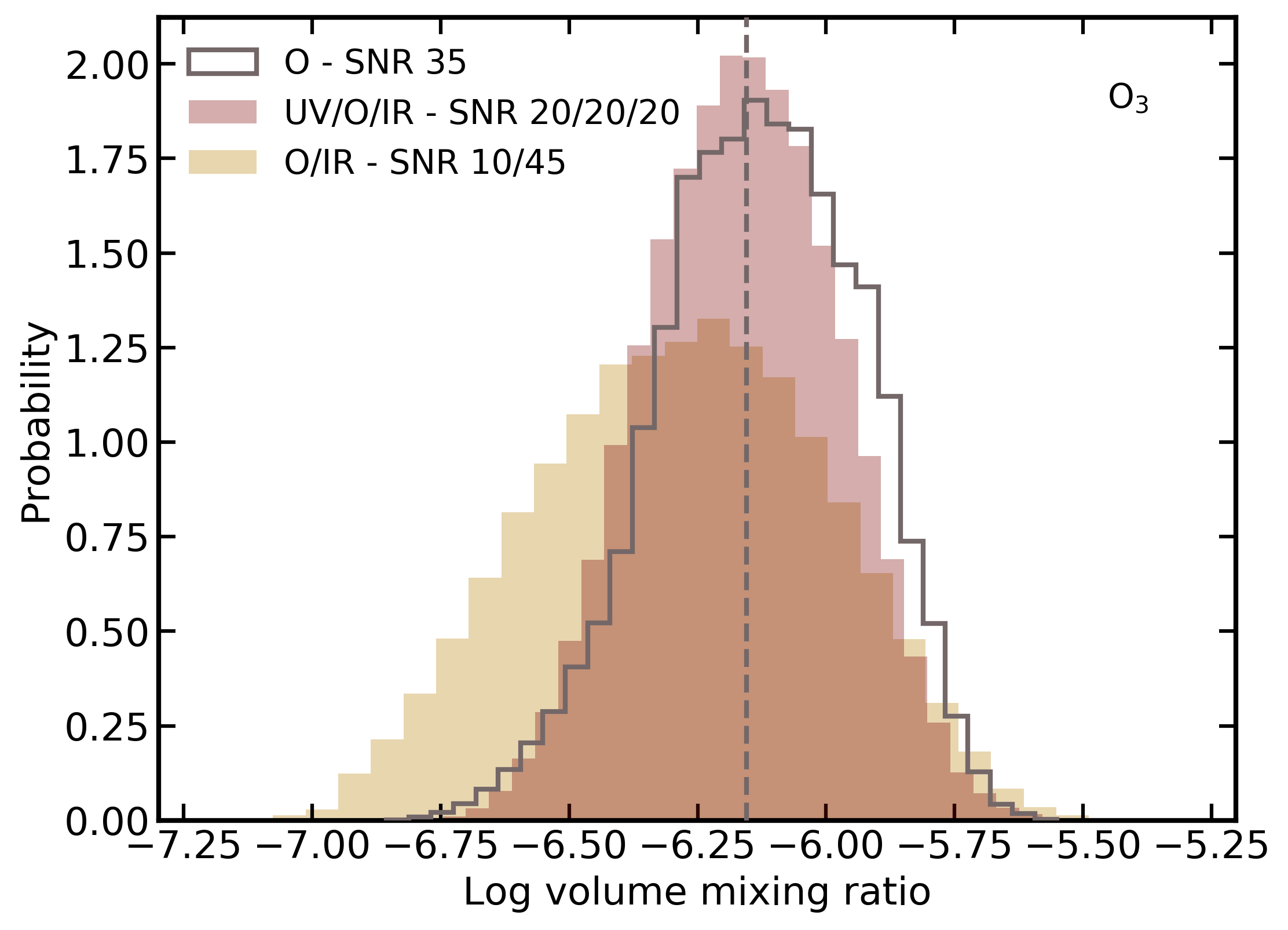}
    \includegraphics[scale=0.4]{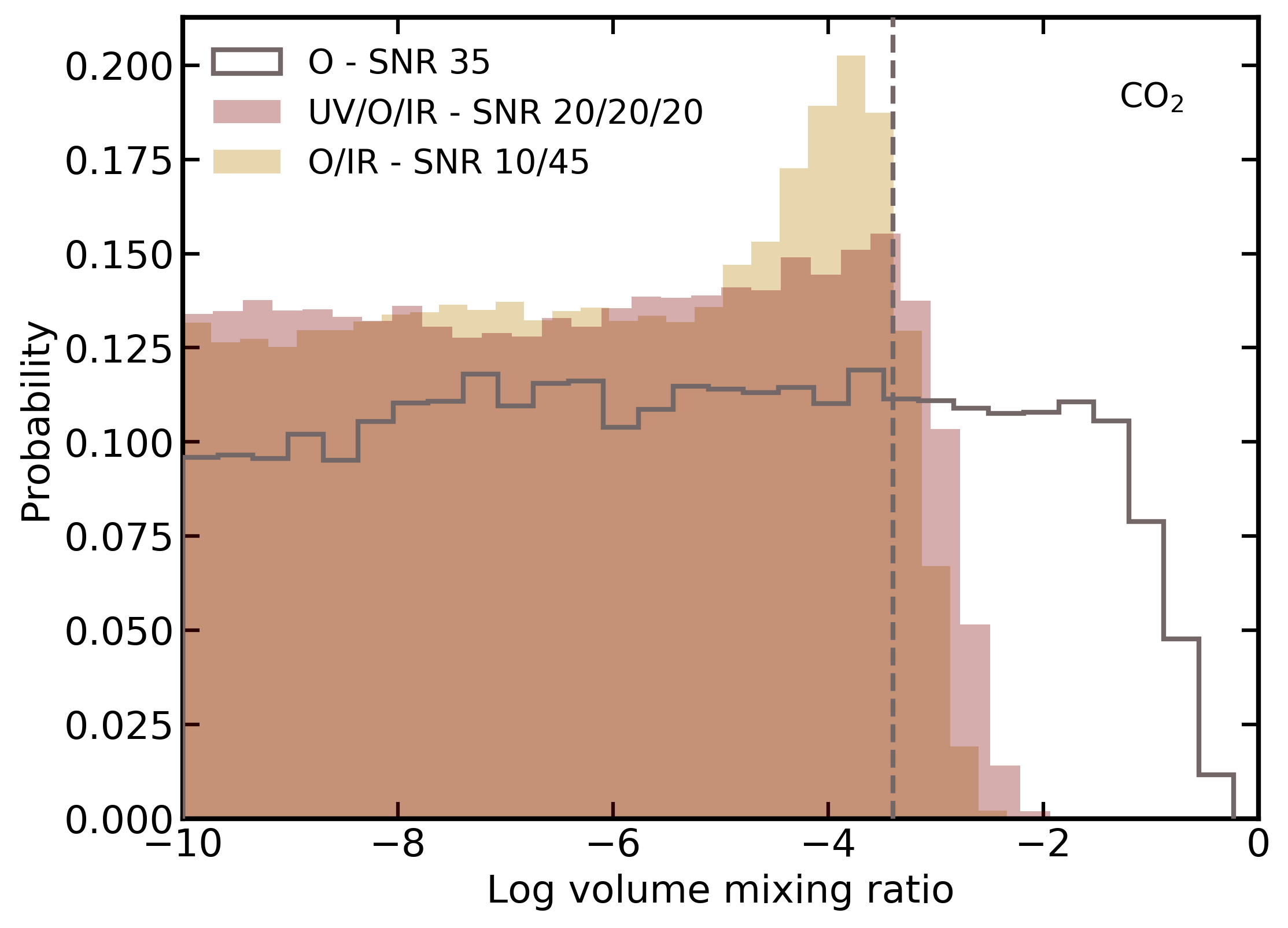}
    \includegraphics[scale=0.4]{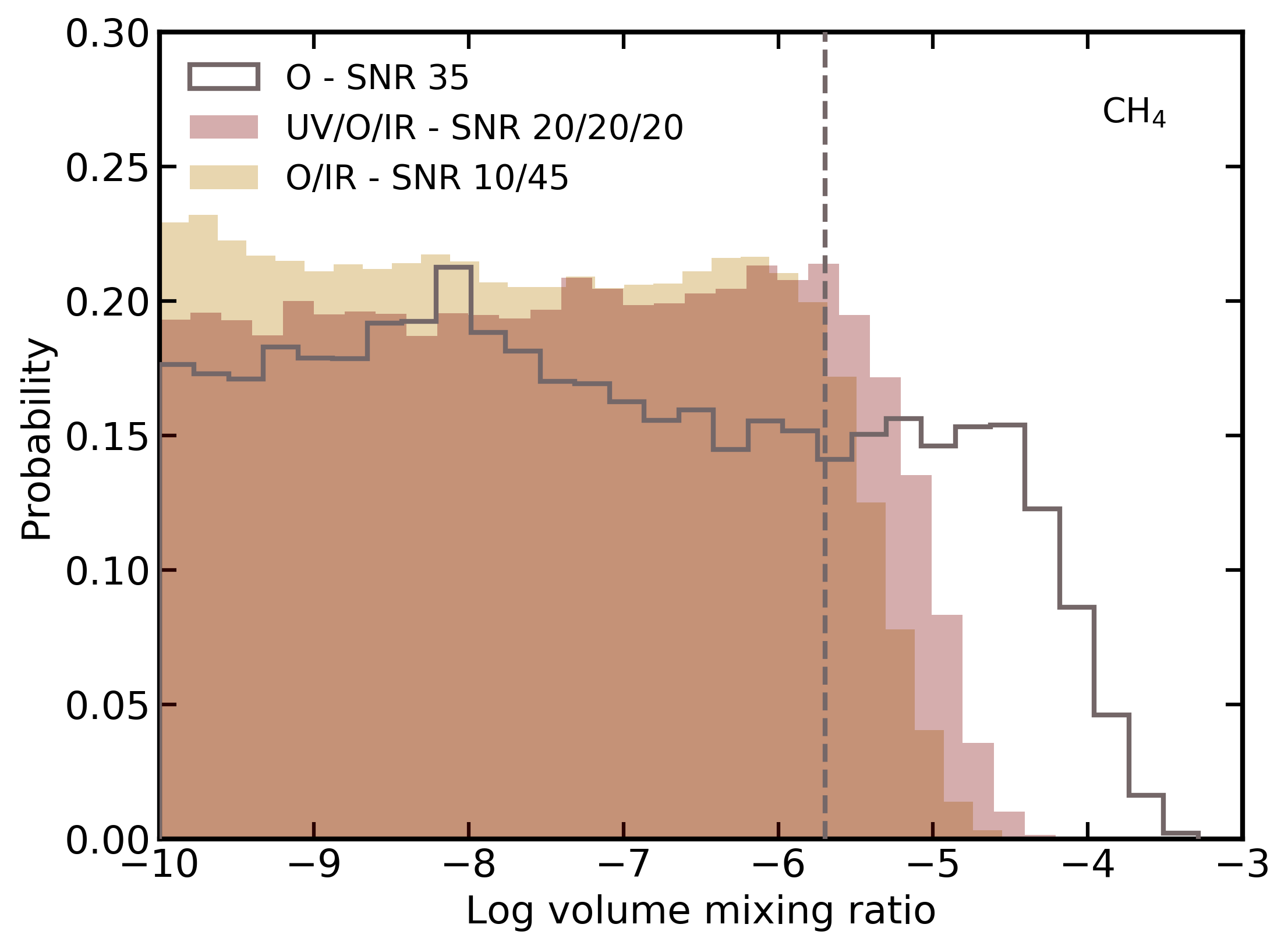}
    \caption{Constraints on key atmospheric species for three different HabEx/LUVOIR-like observing scenarios where spectral coverage and SNR are traded. One scenario has SNR of 20 across the ultraviolet, optical, and near-infrared bands, a second scenario has optical-only coverage at a enhanced SNR of 35, and a third has a reduced optical SNR of 10 and an enhanced near-infrared SNR of 45.} 
    \label{fig:luvex_gases}
\end{figure}

Both the HabEx and LUVOIR mission concepts did not include capabilities to perform observations that simultaneously spanned the full ultraviolet through near-infrared range. Thus, extended spectral coverage could be traded for increased exposure time (and, thus, SNR) in a given band. Figure~\ref{fig:luvex_gases} demonstrates how gas constraints are impacted by spectral coverage and band SNR. For the underlying retrievals, the simulated observation was derived from the baseline spectrum in Figure~\ref{fig:luvex_sens}, {the \texttt{rfast} tool was run in its single-scene mode for reflected light}, and the inferred parameters are the same {11 parameters} as in Section~\ref{subsec:feng} with the addition of mixing ratio inferences for CO$_2$ and CH$_4$: {surface pressure ($p_{\rm s}$), planetary radius ($R_{\rm p}$), surface gravity ($g_{\rm s}$), grey surface albedo ($A_{\rm s}$), cloud-top pressure ($p_{\rm c}$), cloud pressure extent ($\Delta p_{\rm c}$), cloud optical depth ($\tau_{\rm c}$), cloud covering fraction ($f_{\rm c}$), and mixing ratios for water vapor, ozone, and molecular oxygen.} One retrieval exercise was performed with the full spectral coverage (i.e., ultraviolet through near-infrared) and a V-band SNR of 20 (typical of what was proposed by the HabEx and LUVOIR concepts), another retrieval exercise used only the optical (0.4--1.0\,$\upmu$m) range and a V-band SNR of 35, and a third retrieval exercise omitted the ultraviolet band, adopted a reduced optical SNR of 10, and used an enhanced near-infrared SNR of 45. The feasibility of these observing scenarios would depend on a number of parameters, including target distance and the presence of any systematic noise floors, and the \texttt{rfast} tool is designed to enable exploration of any such relevant observing scenarios. Future work could intercompare observing scenarios for different types of worlds using the \texttt{rfast} suite.

\subsection{EPOXI Earth Retrievals} \label{subsec:epoxi}

\begin{figure}
    \centering
    \includegraphics[scale=0.4]{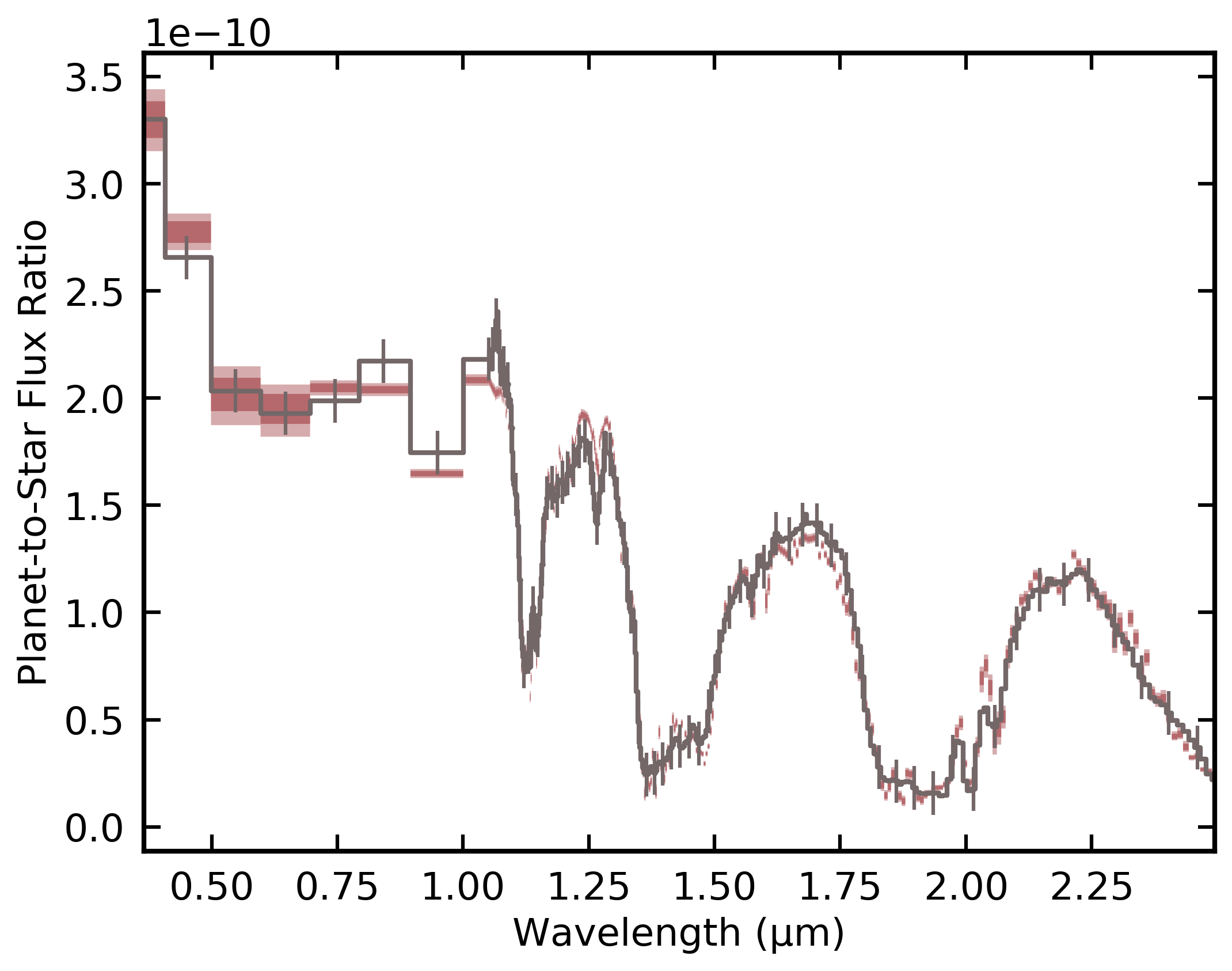}
    \caption{Photometric (below 1\,$\upmu$m) and spectroscopic (above 1\,$\upmu$m) reflected light observation of Earth from NASA's \textit{EPOXI} mission, rotationally averaged from the 18--19 Mar 2008 dataset \citep{livengoodetal2011}. Simulated wavelength-independent error bars yielding a V-band SNR of 20 are indicated, and only every fifth error bar is shown for spectroscopic data for clarity. Forward model spread from retrieval analysis is shown as darker and lighter swaths for 16--84 and 5--95 percentiles.}
    \label{fig:epoxi_data}
\end{figure}

A repurposed application of NASA's \textit{Deep Impact} flyby spacecraft, dubbed the \textit{EPOXI} mission, acquired whole-disk observations of Earth at distances of 0.18--0.34\,au on three separate occasions in northern spring 2008 \citep{livengoodetal2011}. These data are important testing grounds for ideas related to exo-Earth atmospheric inference for HabEx- or LUVOIR-like concepts as the observations span ultraviolet through near-infrared wavelengths. Specifically, ultraviolet and visible wavelengths (0.37--0.95\,$\upmu$m) are spanned by seven photometric bandpasses while near-infrared spectroscopy (1.1--4.54\,$\upmu$m) is acquired at variable resolving power ($\lambda/\Delta\lambda$ of 215--730). For retrieval studies presented here, data from the 18--19 March observing sequence (at a phase angle of 57.7\textdegree) was rotationally averaged and trimmed to emphasize wavelengths plainly dominated by reflected light (i.e., data longward of 2.5\,$\upmu$m were omitted). Additionally, data were scaled to full phase using a Lambertian phase function, which is an acceptable transformation for Earth at phase angles smaller than roughly 90\textdegree~\citep{robinsonetal2010}. Trimmed and scaled data are shown in Figure~\ref{fig:epoxi_data} with uncertainties that are wavelength-independent and yield a SNR of 20 at V-band (i.e., characteristic of predicted HabEx and LUVOIR uncertainties for exo-Earth targets).

Retrievals were performed on the \textit{EPOXI} data shown in Figure~\ref{fig:epoxi_data} {using the \texttt{rfast} single scene reflected light mode}. Thirteen (13) parameters are retrieved: the volume mixing ratios for O$_2$, H$_2$O, CO$_2$, O$_3$, and CH$_4$, as well as surface pressure ($p_{\rm s}$), an isothermal atmospheric temperature ($T_{\rm iso}$), a grey surface albedo ($A_{\rm s}$), planetary radius ($R_{\rm p}$), cloud pressure extent ($\Delta p_{\rm c}$), top-of-cloud pressure ($p_{\rm c}$), cloud optical thickness ($\tau_{\rm c}$), and cloud coverage fraction ($f_{\rm c}$). Blended water liquid/ice optical properties were assumed for cloud asymmetry parameter and single scattering albedo. Planetary mass was fixed at $1\,M_{\oplus}$, as could be the case for an exo-Earth with a mass constraint from radial velocity data. Figure~\ref{fig:epoxi_corner_snr20} (in Appendix) shows constraints from analyzing the \textit{EPOXI} spectrum at a SNR of 20, which are generally comparable to those from the SNR of 20 experiment in Section~\ref{subsec:luvex}. Notable differences include that both carbon dioxide and methane are confidently inferred in the \textit{EPOXI} retrievals, stemming from the presence of relatively strong carbon dioxide and methane features in the 1.8--2.5\,$\upmu$m range. Cloud fraction is constrained at roughly 30\% from the \textit{EPOXI} data and cases with opaque clouds that extend through the deep atmosphere with a cloud top pressure near the tropopause are preferred.

\begin{figure}
    \centering
    \includegraphics[scale=0.4]{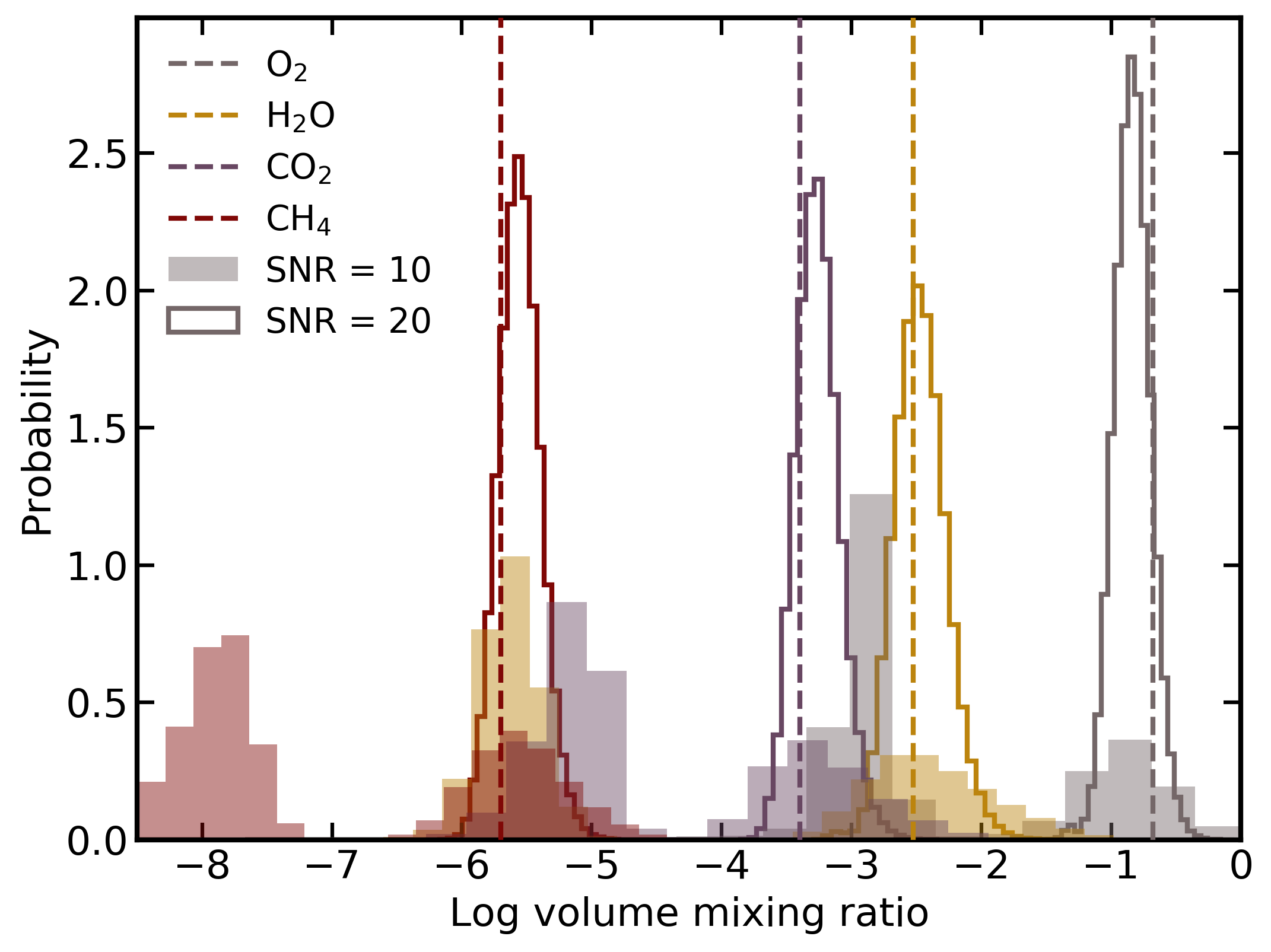}
    \caption{{Gas mixing ratio constraints from retrievals performed on \textit{EPOXI} observations of Earth degraded to V-band SNR of 10 (filled) and 20 (unfilled). Gases are represented with different colors and vertical lines indicate known ``truth'' values. At SNR of 10, gas posteriors include a peak near a realistic value and a second, unrealistic peak due to an inability to rule out solutions with a deep atmosphere}.}
    \label{fig:epoxi_gases}
\end{figure}

\begin{figure}
    \centering
    \includegraphics[scale=0.4]{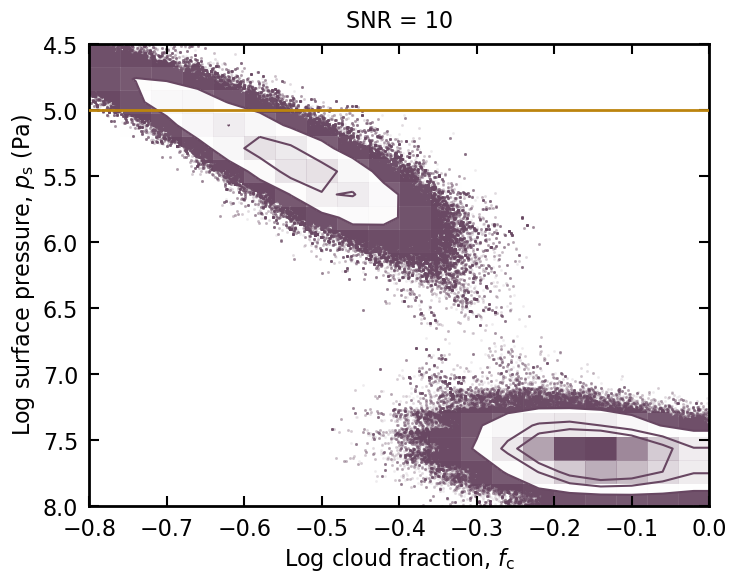}
    \caption{{Correlation between inferred cloud coverage fraction and surface pressure for retrievals performed on \textit{EPOXI} data at SNR of 10. Two classes of solutions emerge: one with realistic surface pressures (horizontal line) and cloud coverages and another with a deep atmosphere and near-global coverage of modest-thickness clouds}.}
    \label{fig:epoxi_snr10_corr}
\end{figure}

An additional experiment was run retrieving on the \textit{EPOXI} data where the wavelength-independent noise was then set to yield a SNR of 10 in V-band. Results are shown in Figure~\ref{fig:epoxi_corner_snr10} (in Appendix). Comparing the inferences from the SNR of 20 scenario to those from the SNR of 10 scenario provides insight into how constraints on atmospheric and planetary parameters degrade with decreasing SNR. Unsurprisingly, broader ranges of parameters (i.e., poorer constraints) are found to be consistent with the SNR of 10 \textit{EPOXI} data. More specifically, two distinct categories of atmospheric states are inferred as providing acceptable fits\,---\,an atmospheric state that is similar to the solutions found in the SNR of 20 experiment as well as an atmospheric model that has (1) near-total coverage of extended, diffuse clouds, (2) a deep atmosphere/surface boundary (i.e., large $p_{\rm s}$), and (3) gas mixing ratios that are reduced by orders of magnitude (to maintain roughly fixed column number density at these larger pressures). {Figure~\ref{fig:epoxi_gases} demonstrates the impact on gas mixing ratio constraints as the SNR is degraded from 20 to 10, and Figure~\ref{fig:epoxi_snr10_corr} shows the correlation between ``surface'' pressure and cloud fraction for the SNR of 10 scenario where an unrealistic set of solutions with an effectively infinitely deep atmosphere cannot be ruled out}. In such a case, additional prior information or longer exposure times would be required to better-refine constraints on the atmospheric state.

\subsection{Earth Infrared Retrievals} \label{subsec:mgs}
%

\begin{figure}
    \centering
    \includegraphics[scale=0.4]{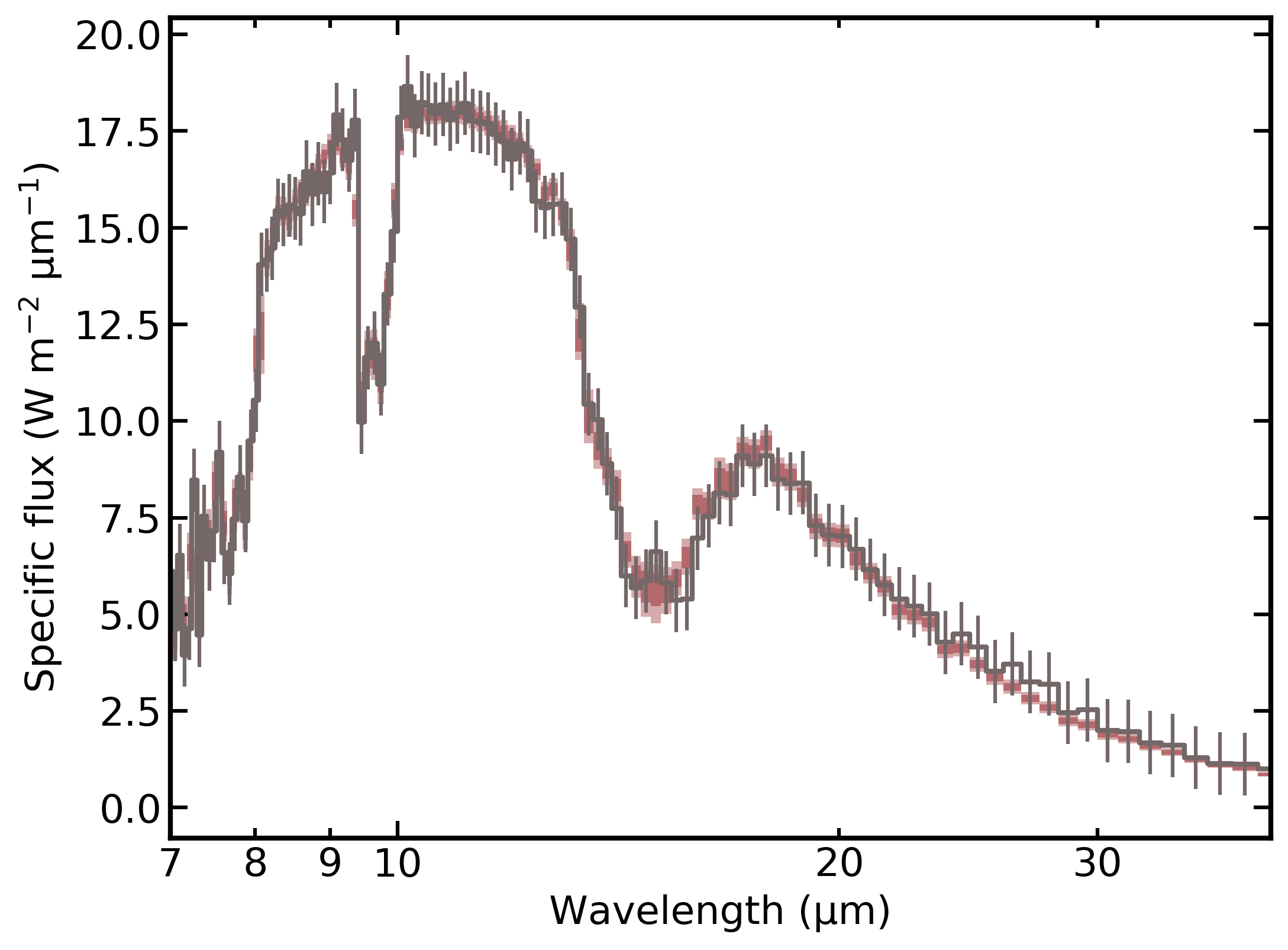}
    \caption{Earth thermal infrared spectrum from \textit{MGS}-TES \citep{christensen&pearl1997} trimmed to 7--40\,$\upmu$m, as described in text. Adopted error bars are constant in wavelength and yield SNR of 20 at 10\,$\upmu$m. Forward model spread from retrieval analysis is shown as darker and lighter swaths for 16--84 and 5--95 percentiles.}
    \label{fig:mgs_tes_data}
\end{figure}

The \textit{Mars Global Surveyor} Thermal Emission Spectrometer (\textit{MGS}-TES) captured a full-disk infrared spectrum of Earth on November 24, 1996, from a distance of $4.8\times10^6$\,km (0.032\,au) \citep{christensen&pearl1997}. The observing sequence was centered over the Pacific ocean at 18\textdegree\,N and 152\textdegree\,W, and spanned 6--50\,$\upmu$m at a constant resolution of 10\,cm$^{-1}$ (i.e., spanning resolving powers of 160--14 from the shortest to longest wavelengths). For retrieval studies presented here, data shortward of 7\,$\upmu$m and longward of 40\,$\upmu$m are omitted; flux densities at the shortest wavelengths are anomalously large \citep[see Figure 10 of][]{robinson&reinhard2020} (which strongly biased early retrieval studies explored for this work) and brightness temperatures at the longest wavelengths are unphysically large (i.e., exceeding 1,000\,K). The truncated spectrum is shown in Figure~\ref{fig:mgs_tes_data} where wavelength-independent uncertainties have been added to yield a SNR of 20 at 10\,$\upmu$m. Thus, adopted data quality is roughly consistent with under-study mid-infrared exo-Earth direct imaging mission concepts \citep{quanzetal2021}, where studies have investigated wavelength coverage of 3--20\,$\upmu$m, resolving powers of 20--100, and SNRs of 5--20 \citep{konradetal2021}.

A retrieval was performed on the data (with faux uncertainties) shown in Figure~\ref{fig:mgs_tes_data} using the \texttt{rfast} tool. Twelve (12) parameters were inferred, including: surface pressure ($p_{\rm s}$), planetary radius ($R_{\rm p}$), cloud fraction ($f_{\rm c}$), cloud-top pressure ($p_{\rm c}$), and the volume mixing ratios for water vapor, carbon dioxide, ozone, nitrous oxide, and methane. Planetary mass (or, equivalently, surface gravity) was assumed to be well-known, as would be the case, e.g., if prior precision radial velocity data were available \citep[recent, analogous retrieval results have shown that spectral inference does not serve to improve mass estimates beyond a mass prior;][]{aleietal2022}. Additionally, to maintain a simple model for clouds, the cloud pressure extent was taken as a single scale height. Finally, a baseline model assumes a thermal structure that follows a power-law from the surface through the ``troposphere,'' with,
\begin{equation}
    T \left( p \right) = T_{\rm s} \left( \frac{p}{p_{\rm s}} \right)^{\beta_{\rm t}} \ ,
\end{equation}
where the surface temperature, $T_{\rm s}$, and $p$-$T$ power-law index, $\beta_{\rm t}$, are fitted parameters. The ``stratosphere'' was taken as isothermal at a temperature $T_{\rm iso}$, which was also fitted. All priors were uninformed.

\begin{figure}
    \centering
    \includegraphics[scale=0.4]{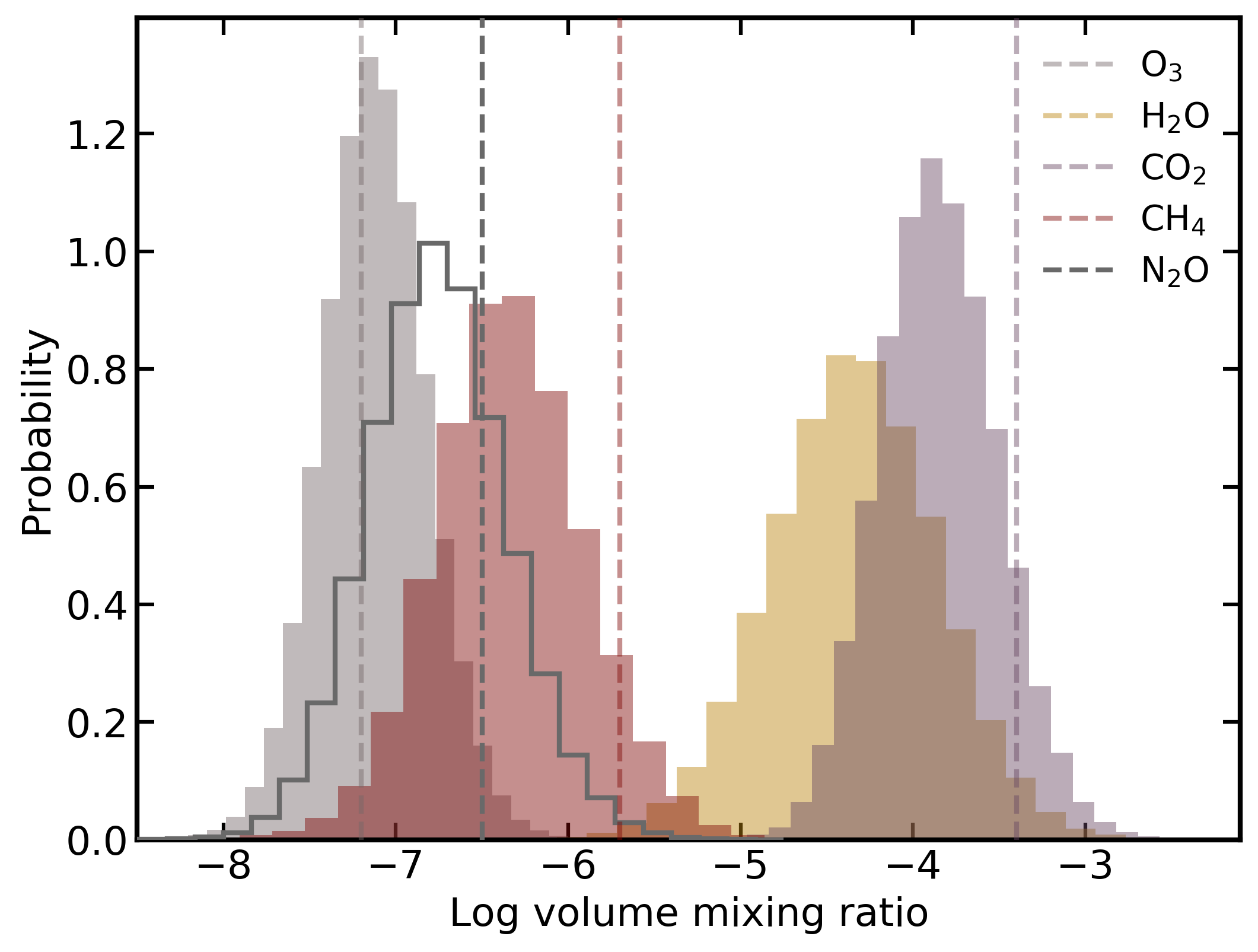}
    \caption{{Gas mixing ratio constraints from retrievals performed on the \textit{MGS}/TES observations of Earth shown in Figure~\ref{fig:mgs_tes_data}. Gases are represented with different colors and vertical lines indicate known ``truth'' values}.}
    \label{fig:mgs_tes_gases}
\end{figure}

Figure~\ref{fig:mgs_tes_corner} (in Appendix) shows retrieval results for the 12-dimensional fit to the \textit{MGS}-TES observations. {As shown in Figure~\ref{fig:mgs_tes_gases}}, gas mixing ratios are well-constrained and generally reasonable\,---\,characteristic values for the log of volume mixing ratios for carbon dioxide, nitrous oxide, and methane (in 1996) are -3.4, -6.5, and -5.7, respectively. Ozone mixing ratios in the deep atmosphere\,---\,as is mainly probed by the 9.6\,$\upmu$m ozone feature\,---\,span -7.3 to -6.3 (in log space). The inferred water vapor volume mixing ratio distribution (with characteristic values below about 0.01\%) appears to be biased low, potentially pointing to remaining systematic issues affecting the 6.3\,$\upmu$m water vapor band. Surface pressures are biased high by roughly a factor of $5$ and the planetary radius constraint\,---\,which would be strong by most exoplanet standards\,---\,is biased towards smaller radii (by about 10\%). Cloud-top pressure is largely unconstrained and only near-total cloud coverage fractions are ruled out. The inferred surface temperature is generally above the water freezing temperature, and preferred thermal structures have decreasing temperatures with pressures. Figure~\ref{fig:mgs_tes_data} shows spectral forward model swaths at the 16--84 and 5--95 percentiles.

\begin{figure}
    \centering
    \includegraphics[scale=0.4]{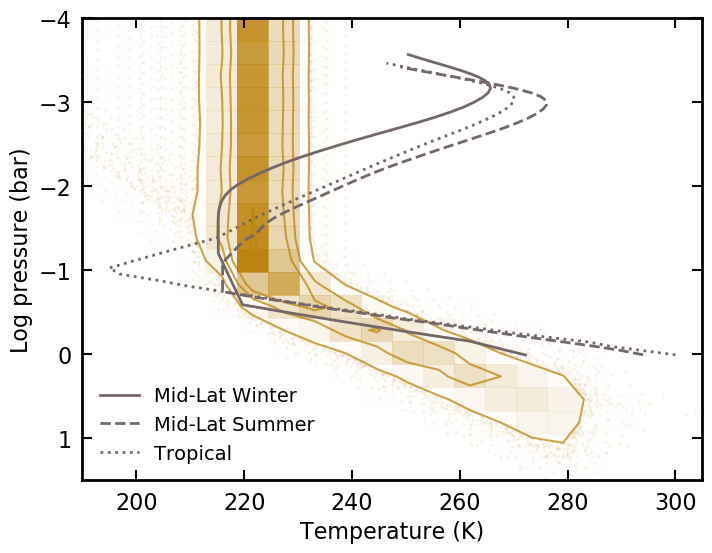}
    \includegraphics[scale=0.4]{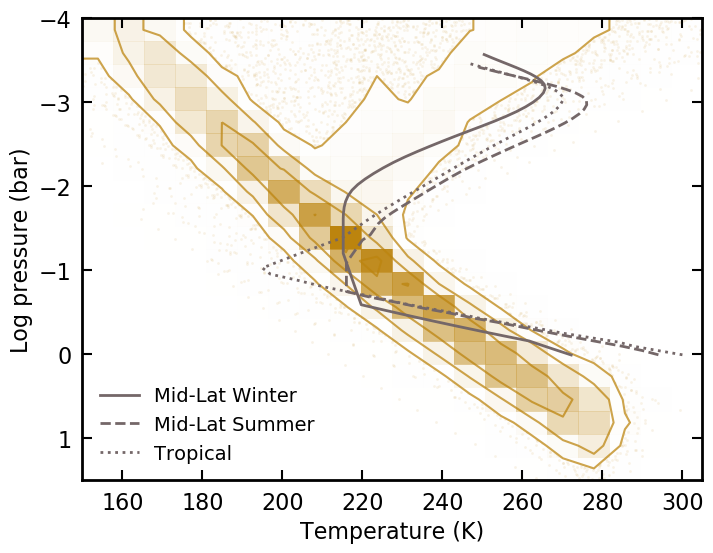}
    \caption{Two-dimensional histograms of thermal structure models randomly selected from retrievals on \textit{MGS}-TES Earth observations. Standard Earth thermal structures from \citet{mcclatcheyetal1972} are shown as grey lines. Left figure is from a retrieval with a three-parameters thermal structure model (power-law troposphere profile and isothermal stratosphere) while right figure is from a retrieval with a four-parameters thermal structure model (power laws in both the troposphere and stratosphere).} 
    \label{fig:mgs_tes_Tp}
\end{figure}

Figure~\ref{fig:mgs_tes_Tp} shows a two-dimensional histogram of inferred thermal structures for the 12-dimensional fit (and its three-parameter thermal structure model). Characteristic Earth thermal structure profiles (tropical and mid-latitudes) from \citet{mcclatcheyetal1972} are also shown. Model isothermal stratospheres show a strong preference for values near 220\,K, and agreement with realistic Earth $p$-$T$ data is acceptable through the upper troposphere. However, retrieved thermal structures generally have a less-steep $p$-$T$ relation through the troposphere (as compared to the Earth data) and extend to higher pressures (stemming from the biased-high surface pressure constraint). To explore a more realistic thermal structures\,---\,and given that the largest data-model discrepancies in Figure~\ref{fig:mgs_tes_data} occur in the core of the 15\,$\upmu$m carbon dioxide band (which is sensitive to a thermal inversion in the stratosphere of Earth)\,---\,a second retrieval was performed with a thermal structure model that introduced a stratospheric $p$-$T$ power law, thereby allowing for thermal inversions. Thermal structures from this 13-dimensional fit are shown in Figure~\ref{fig:mgs_tes_Tp}, which demonstrates only weak constraints on a stratospheric inversion. Reduced chi-squared values for the 12-dimensional and 13-dimensional fits are nearly identical (0.72 versus 0.71, respectively), which disfavors the model with an added treatment for stratospheric inversions.

\subsection{Titan Transit Retrievals} \label{subsec:vims}

\begin{figure}
    \centering
    \includegraphics[scale=0.4]{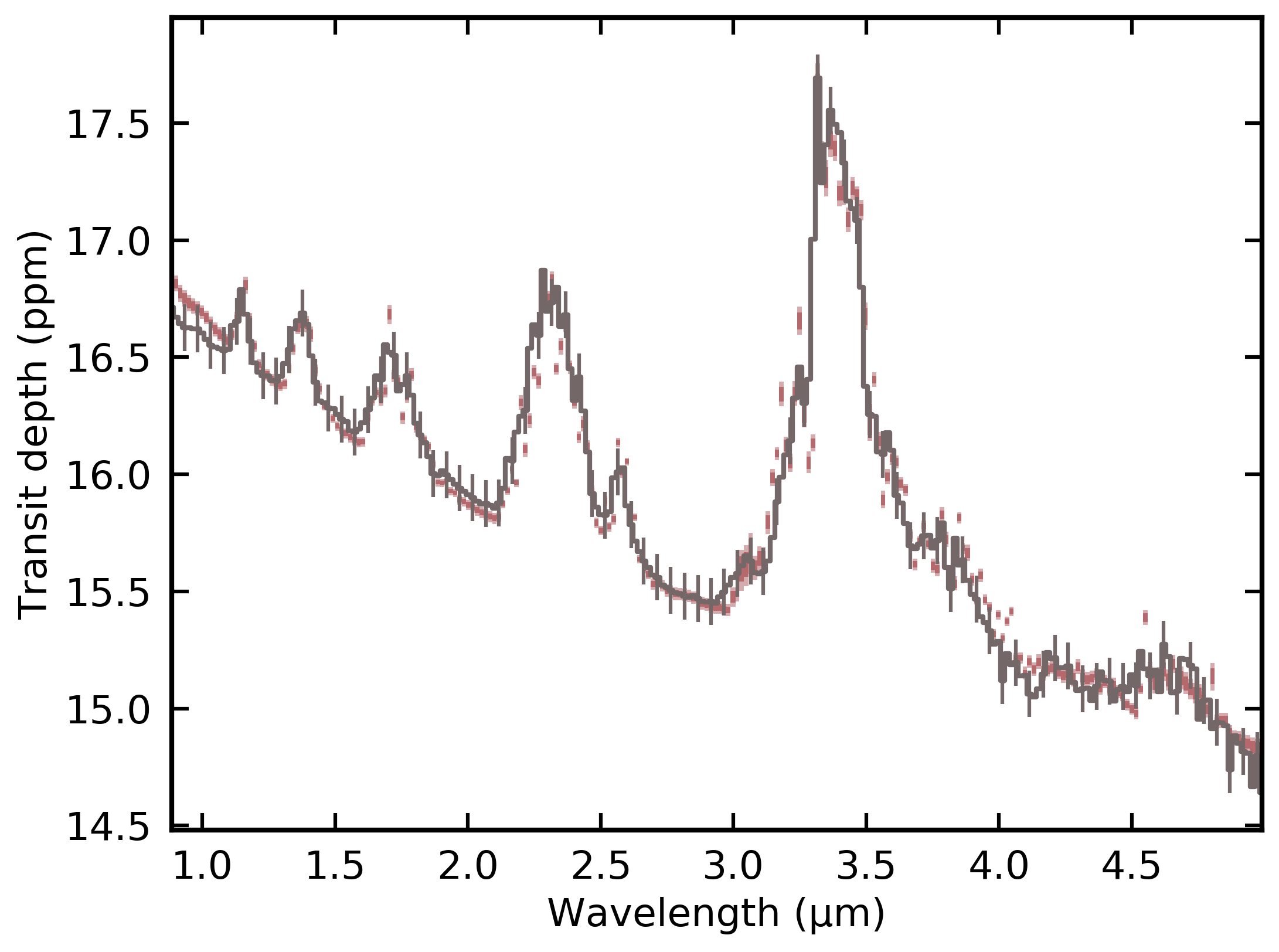}
    \caption{Titan transit spectrum derived from \textit{Cassini}-VIMS observations \citep[from][]{robinsonetal2014a}. Adopted error bars are 0.1\,ppm and are constant with wavelength, as justified in the text. Forward model spread from retrieval analysis is shown as darker and lighter swaths for 16--84 and 5--95 percentiles.}
    \label{fig:cassini_data}
\end{figure}

The Visual and Infrared Mapping Spectrometer \citep[VIMS;][]{brownetal2004} aboard NASA's \textit{Cassini} mission observed many solar occultations by Titan \citep{belluccietal2009,hayneetal2014,maltagliatietal2015}, and these can be translated into effective transit spectra \citep{robinsonetal2014a}. Figure~\ref{fig:cassini_data} shows a Titan transit spectrum derived from an occultation observation at 27\textdegree\,N, which was observed in September of 2011 and is an intermediate haze-extinction case provided by \citet{robinsonetal2014a}. Data span 0.88--5\,$\upmu$m and resolution increases with wavelength from 12\,nm to 18\,nm. {Refractive loss effects were removed from the underlying occultation data \citep[following][]{robinsonetal2014a} so that fits to the resulting Titan transit spectrum need not consider refraction.}

The \texttt{rfast} model was used to perform atmospheric retrievals on the transit spectrum shown in Figure~\ref{fig:cassini_data}. Faux error bars were assigned at the level of 0.1\,ppm, which results in a ratio of error bar size to spectral feature depth that is roughly comparable to those from \textit{JWST}-relevant clearsky, solar metallicity, warm Neptune cases investigated by \citet{greeneetal2016}. Based on previous work analyzing VIMS Titan occultation data \citep{maltagliatietal2015,coursetal2020}, fits included volume mixing ratios for carbon monoxide, methane, acetylene (C$_2$H$_2$), and propane (C$_3$H$_8$). The fitted planetary radius was applied at the 10\,mbar pressure level \citep{benneke&seager2012}. The spectral impact of haze was incorporated using a three-parameter model with (1) the vertical optical depth following an exponential with scale height, $H_{\rm h}$, (2) the wavelength-dependent opacity following a power law in wavelength with exponent $\beta_{\rm h}$, and (3) the optical depth at a wavelength of 1\,$\upmu$m and at 10\,mbar atmospheric pressure given by $\tau_{\rm h}$. To capture the known strong decrease in temperature from Titan's stratosphere to a cold tropopause, a three-parameter temperature model was adopted that follows,
\begin{equation}
    T\left( p \right) = \left( T_{\rm{st}} - T_0 \right)e^{-\frac{p}{p_{\rm T}}} + T_0 \,
\end{equation}
where $T_{\rm{st}}$ represents the hot stratopause temperature, $T_0$ represents a cold deep-atmosphere temperature, and $p_{\rm T}$ is the pressure scale for the decrease in stratospheric temperatures. 

Initial retrievals resulted in inferred values for $\beta_{\rm h}$ that were much smaller than previously-derived results \citep{hubbardetal1993,tomaskoetal2008,belluccietal2009,robinsonetal2014a}. Closer investigations revealed that these errant power values were driven by attempts to use haze opacity to fit continuum near 4.3\,$\upmu$m. Experiments revealed that this continuum could be better reproduced by including N$_2$-N$_2$ collision-induced absorption, so the nitrogen volume mixing ratio was added as a fitted parameter.

Several previous works have highlighted the role of a 3.4\,$\upmu$m C-H stretch feature in Titan occultation observations \citep{maltagliatietal2015,robinsonetal2014a,coursetal2020}. Optical depth data from interstellar medium observations were adopted to model the shape of this stretch feature \citep{pendleton&allamandola2002}. As the feature tracks condensed-phase hydrocarbons, the stretch feature optical depths were fitted by scaling the 1\,$\upmu$m haze optical depths by a factor $x_{\rm C-H}$, and multiplying by the interstellar medium optical depths to capture the wavelength-dependent feature shape.

\begin{figure}
    \centering
    \includegraphics[scale=0.4]{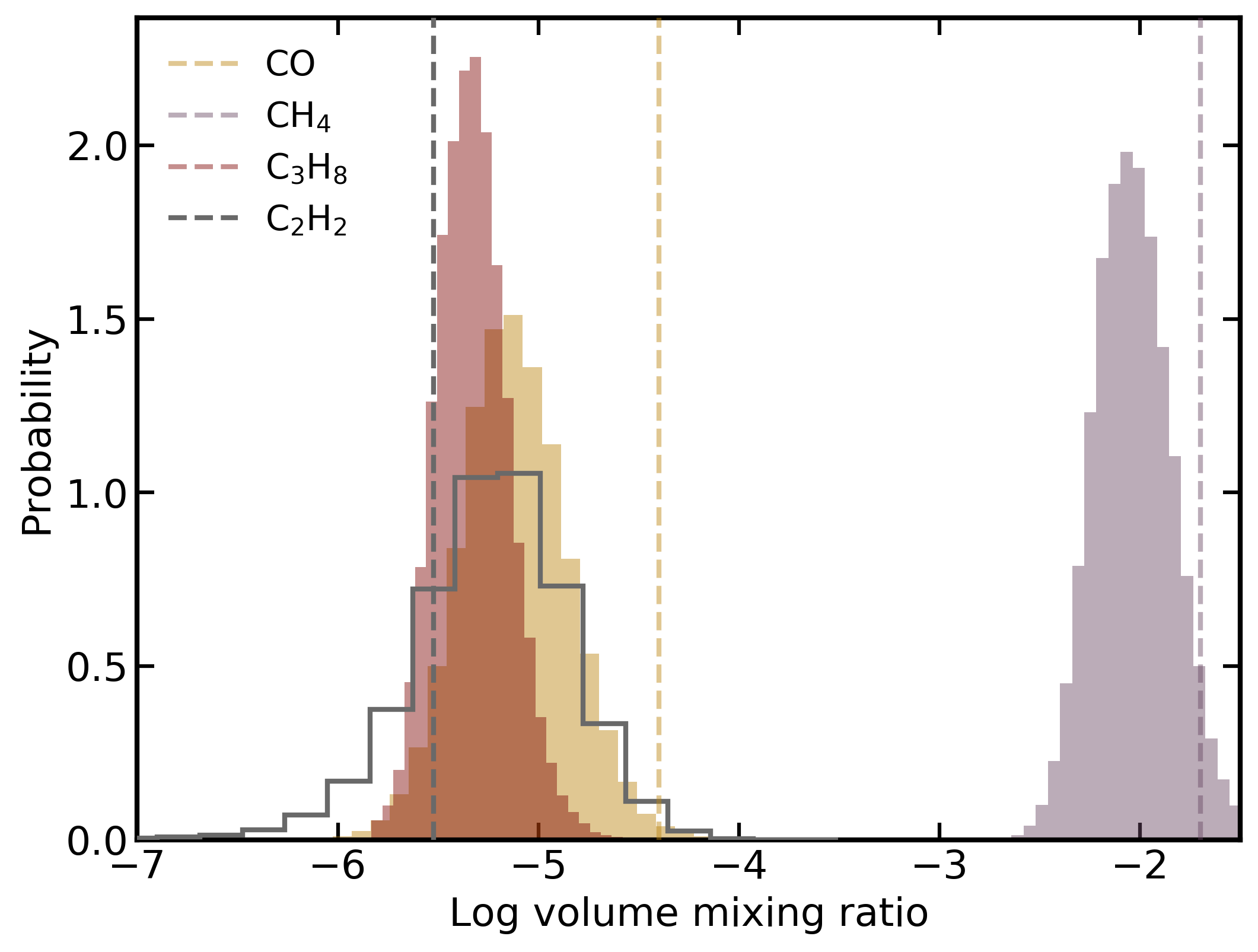}
    \caption{{Gas mixing ratio constraints from retrievals performed on the \textit{Cassini}/VIMS-derived transit spectrum for Titan shown in Figure~\ref{fig:cassini_data}. Gases are represented with different colors and vertical lines indicate known ``truth'' values, when available}.}
    \label{fig:cassini_vims_gases}
\end{figure}

Figure~\ref{fig:cassini_corner} (in Appendix) shows retrieval results for the 13-dimensional fit to the Titan transit spectrum derived from \textit{Cassini}-VIMS observations. When applicable or available, ``truth'' parameter values are indicated as solid vertical lines and are taken from separate analyses of the occultation observations \citep{maltagliatietal2015}. {Figure~\ref{fig:cassini_vims_gases} demonstrates constraints on all gases (except molecular nitrogen)}, where only the carbon monoxide constraint shows a substantial bias (i.e., is underestimated at roughly the 95\% confidence level). As this feature forms deeper in the atmosphere, this bias could result from our adopted thermal structure model and poor constraints on the deep-atmosphere temperatures in the transit spectrum. Upper-atmosphere temperatures are biased roughly 20\% warmer than a thermal structure inferred from \textit{Cassini} Composite Infrared Spectrometer (CIRS) data \citep{vinatieretal2015}, potentially pointing to minor issues with using HITRAN-derived opacities versus more specialized linelists \citep{campargueetal2013}. Finally, at least at the adopted artificial noise level, the presence of C-H stretch opacity is not well-constrained or even required. In fact, an analogous retrieval with the C-H stretch parameter removed yielded a slightly improved reduced chi-squared (1.49 versus 1.50).


%
\section{Discussion} \label{sec:discuss}
%

The \texttt{rfast} atmospheric retrieval suite developed above is designed to enable efficient explorations of the mapping between observational quality and constraints on atmospheric/planetary parameters. Comparisons to more-sophisticated radiative transfer results and tools in Figures~\ref{fig:hunt}--\ref{fig:smart_trns} provide strong validations of the core treatments of radiation within the \texttt{rfast} forward model. The most significant differences when compared to full-physics models occur for the three-dimensional approach to planetary reflectivity (Figure~\ref{fig:smart_refl}, right panel) where simplifications in the scattering treatment can lead to discrepancies at the level of several tens of percents. Thus, retrievals on real reflected light observations using the three-dimensional \texttt{rfast} forward model could lead to biased results. However, retrievals on synthetic observations created by the three-dimensional \texttt{rfast} forward model are still useful for informing mission designs as systematic effects will cancel out.

Retrieval comparisons between the single-scene \texttt{rfast} mode and the three-dimensional retrieval results presented in \citet{fengetal2018} are in good agreement, as shown in Table~\ref{tab:feng}. This indicates that computationally-expensive three-dimensional treatments are not necessarily needed for understanding the connection between SNR and atmospheric constraints, at least at moderate SNRs or when planetary phase is not an important consideration. Thus, future exoplanet direct imaging mission concept studies can save computational resources by exploring atmospheric retrievals with tools analogous to the single-scene approach described above. Section~\ref{subsec:luvex} demonstrates an application of the \texttt{rfast} single-scene reflectance mode to a concept relevant to the development of an exo-Earth direct imaging mission\,---\,trading exposure times in different bandpasses for atmospheric constraints. As shown in Figure~\ref{fig:luvex_gases}, the addition of near-infrared capabilities provides better upper limit constraints in methane abundances and markedly improved upper limit constraints on carbon dioxide. Omitting ultraviolet observations and reducing optical SNRs while enhancing near-infrared SNRs slightly weakens detections/constraints on molecular oxygen, water vapor, and ozone while weakly improving upper limit constraints on methane and carbon dioxide. A near-infrared SNR of 45 results in constraints on carbon dioxide that are not quite a true detection, which indicates that slightly higher near-infrared SNRs would be needed to detect carbon dioxide for a modern Earth analog (and lower SNRs would be needed to detect carbon dioxide for Earth-like worlds with enhanced atmospheric CO$_2$ abundances).

The bulk of the results presented above emphasize the utility of Solar System observations in exploring retrieval approaches for exoplanets. Unfortunately, exoplanet analog Solar System observations are rare \citep[see, e.g.,][]{robinson&reinhard2020}. However, the limited available data that do exist have potential applications that span exoplanet transit observations with \textit{JWST} to further-future exoplanet direct imaging missions.

\subsection{EPOXI Earth Discussion}

Retrievals on \textit{EPOXI} observations of the distant Earth in Section~\ref{subsec:epoxi} are a strong proof-of-concept for future exoplanet direct imaging missions. At SNR of 20, gas abundances are constrained to better than 0.5\,dex. As mentioned above, the methane and carbon dioxide constraints rely most heavily on bands beyond 1.8\,$\upmu$m, which is typically the longest wavelength adopted for exo-Earth direct imaging mission concepts (beyond this wavelength, telescope thermal emission becomes a leading noise term for non-cooled telescopes). The inferred surface pressure is biased high while the inferred planetary radius is biased low. The Rayleigh scattering feature in Earth's reflectance spectrum plays an important role in constraining both of these quantities, and it may be that a bias results from using a one-dimensional reflectance model to represent the complex, three-dimensional disk of Earth. Finally, opaque clouds are detected in the observations and cover roughly 30\% of the illuminated disk.

As contrasted to the SNR of 20 \textit{EPOXI} retrievals, the SNR of 10 results tell a cautionary tale. Specifically, the lower-SNR observations cannot rule out scenarios with near-total, deeper-atmosphere cloud coverage on the illuminated disk. In these cases, the surface pressure can be large (to be beneath the near planet-wide cloud deck at roughly 0.1\,bar), and gas mixing ratios become erroneously small to maintain constant column abundance. Thus, future exo-Earth direct imaging missions may need to obtain spectra at SNRs larger than 10 to enable reliable results from atmospheric retrieval analyses.

\subsection{MGS-TES Earth Discussion}

Section~\ref{subsec:mgs} explores thermal infrared retrievals on a disk-integrated Earth spectrum from \textit{MGS}-TES, where findings are relevant to the under-development Large Interferometer For Exoplanets (LIFE) mission concept \citep{quanzetal2021,quanzetal2021b,dannertetal2022}. At the adopted SNR of 20, concentrations of water vapor, carbon dioxide, ozone, nitrous oxide, and methane\,---\,the latter three of which are key biosignature gases\,---\,are detected and well-constrained, with 16/84-percentile ranges being smaller than an order of magnitude. Importantly, the inferred surface temperature is found to be above the freezing point of water at the 99.7\% confidence level (i.e., 3-$\sigma$), and warmer surface temperatures are permitted for scenarios with higher fractional cloudiness. There is no strong evidence for a stratospheric thermal inversion.

While surface pressure and planetary radius are also well-constrained, the inferred values for these parameters are biased high and low, respectively, at roughly the 95\% confidence level (i.e., 2-$\sigma$). As both constraints on surface pressure and planetary radius are sensitive to continuum levels and spectral regions are from band centers, it may be that systematic calibration uncertainties as well as continuum-based data scaling \citep[noted in][]{christensen&pearl1997} have led to slight biases in these parameters. A key message may then be that, for exoplanet atmospheric characterization in general, some parameters will be more sensitive to systematic calibration uncertainties than others. One additional contributing factor to these biases may be water vapor pressure-induced absorption in the 10\,$\upmu$m window. Models adopted here use a constant water vapor mixing ratio profile, and the 6.3\,$\upmu$m water vapor band constrains such a column-averaged quantity to be small (below 1\%). At such low mixing ratios pressure-induced absorption is not significant, and models could compensate by using larger surface pressures to broaden the 15\,$\upmu$m carbon dioxide band. It may be that adopting a water vapor profile shape appropriate for a condensing gas could lead to improved constraints, which was an approach used in \citet{vonparisetal2013}.

\begin{figure}
    \centering
    \includegraphics[scale=0.4]{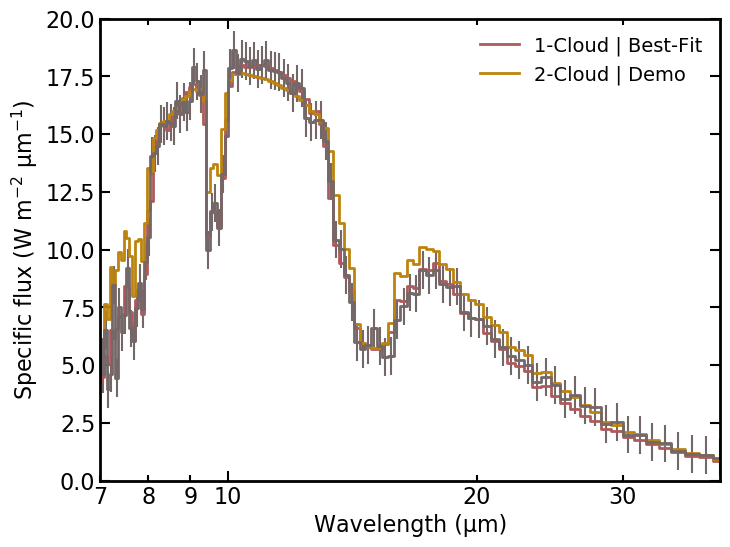}
    \caption{Comparison between the retrieved best-fit model from the single-cloud retrievals in Figure~\ref{fig:mgs_tes_corner} (red) and a demonstration with a two-clouds model and a realistic Earth thermal structure profile (yellow).}
    \label{fig:mgs_tes_two_cloud}
\end{figure}

A striking result from the \textit{MGS}-TES Earth retrievals is that the inferred thermal structures present an extremely low tropospheric lapse rate, with $d\ln T/d\ln p$ values near 0.06. By comparison, a typical Earth $d\ln T/d\ln p$ value is closer to 0.2 \citep{robinson&catling2014}. Figure~\ref{fig:mgs_tes_two_cloud} explores one potential explanation for this biasing. As thermal structure from the lower-stratosphere to the surface is constrained by spectral observations spanning the core to the wings of the 15\,$\upmu$m carbon dioxide band, allowing for clouds at multiple locations throughout the troposphere provides better control over fitting the band shape (e.g., high altitude clouds impact fluxes at all wavelengths that would probe the deep atmosphere while low clouds only impact wavelengths that probe the near-surface levels). Thus, and as a proof-of-concept, Figure~\ref{fig:mgs_tes_two_cloud} shows the best-fit model from the retrieval exercises presented above (which only adopted a single cloud type/location) as compared to an example model with Earth-like parameters (including an Earth-like thermal structure) and both a high and low altitude cloud, each covering about 30\% of the disk. While not definitive, results in Figure~\ref{fig:mgs_tes_two_cloud} show that adopting multiple cloud types could enable more accurate constraints on the tropospheric thermal structure, which is an important consideration for any future infrared emission-focused exoplanet direct imaging missions.

Comparisons can be made between the \textit{MGS}-TES Earth retrievals presented above and previous infrared retrieval efforts for Earth-like planets by \citet{vonparisetal2013} and \citet{konradetal2021}, although direct comparisons are difficult as these previous studies adopted synthetic observations from cloud-free models and used constant resolving power (as a reminder, the full \textit{MGS}-TES data have resolving power of 160--14 at wavelengths spanning 6--50\,$\upmu$m and were trimmed to use only 7--40\,$\upmu$m). The most relevant comparison point from \citet{vonparisetal2013} is an SNR 20 scenario with resolving power of 20 spanning 5--20\,$\upmu$m. Here, surface temperature and pressure constraints are similar, but gas mixing ratio constraints are weaker in \citet{vonparisetal2013}, likely owing to the limited number of spectral points spanning key gas absorption bands at such low resolving power. Also, constraints on the upper atmospheric temperatures are markedly weaker in results from \citet{vonparisetal2013}, which is likely driven by a modeling approach that applies a ``top of atmosphere'' temperature at an atmospheric pressure of $10^{-4}$\,bar where emission spectra have very little sensitivity to the thermal structure.

Work by \citet{konradetal2021} includes a relevant comparison point where the synthetic emission spectra for an Earth-like target are at a SNR of 20 and span 3--20\,$\upmu$m at a resolving power of 50. Overall, constraints on gas mixing ratios, surface pressure, surface temperature, and planetary radius are comparable. As the \citet{konradetal2021} study adopted cloud-free synthetic data and models, this leads to the conclusion that thermal direct imaging missions can achieve strong constraints on key atmospheric parameters for Earth-like worlds even when the target has patchy clouds.


\subsection{Cassini/VIMS Titan Discussion}

Transit retrievals explored in Section~\ref{subsec:vims} provide a strong proof-of-concept for using Solar System occultation observations as a validation point for exoplanet inference tools. While Solar System worlds with atmospheres may not be direct analogs for some exoplanet types, transit spectra and adopted error bars can be scaled to achieve a proper feature depth-to-uncertainty ratio. In general, Solar System analog transit spectra \citep{robinsonetal2014a,dalbaetal2015,macdonald&cowan2019} can be used to understand how constraints and model complexity relate to data quality and provide a timely connection to forthcoming \textit{JWST} observations. The Titan retrievals presented above show gas mixing ratio constraints that are generally consistent with orbiter retrievals. Temperature constraints are biased modestly warm. As some temperature information is gleaned, in part, from a somewhat limited number of more-strongly temperature sensitive lines, the biasing may be a result of application of the HITRAN linelist at conditions far from Earth-like. Titan transit retrievals also provide an important insight into how broad gas absorption features (e.g., N$_2$-N$_2$ collision-induced absorption near 4.3\,$\upmu$m) can impact haze/cloud inferences. A method for incorporating broad C-H stretch mode opacity near 3.4\,$\upmu$m was presented, although such a treatment was not needed at the data qualities adopted here.

\subsection{Broader Considerations}

Taken altogether, the Solar System retrievals presented here show promise while also demonstrating cautionary tales. On the positive side, the wide-ranging retrieval results explored above indicate that there is great utility in using Solar System analog observations to validate and refine approaches to exoplanet remote sensing. Caution is urged, however, as many of the findings above rely on using \textit{a prior} knowledge to ``know'' when a solution is adequate or inadequate. For example, biasing in the tropospheric thermal gradient inferred for Earth from the \texttt{MGS}-TES retrievals may not have been easily deduced to a truly external observer. Similarly, the addition of N$_2$-N$_2$ collision-induced absorption to the Titan-focused retrievals was introduced to better-reproduce the known wavelength-dependent slope in haze opacity.

Finally, the retrieval studies explored here\,---\,which span reflected light, thermal emission, and transit transmission\,---\,only scratch the surface of what can be learned from analog observations. Open questions remain, for example, regarding the complexity of cloud treatments warranted in reflected light retrievals, the extent to which thermal information can be extracted, and how well surface reflection signatures can be constrained. For thermal emission, an important next step is understanding how cloud treatments may (or may not) bias thermal structure inferences. Finally, Solar System transiting exoplanet analog studies provide an opportunity to test standard assumptions like constant profiles of either gas mixing ratios and/or atmospheric temperature. Any and all exoplanet-specific retrieval models could benefit from validation against Solar System observations.

%
\section{Conclusions} \label{sec:conc}
%

Solar System observations that can serve as analogs for exoplanet observations provide unique testing and validation opportunities for exoplanet science. While such analog observations are currently limited in number, future Solar System planetary science missions could make acquisition of exoplanet analog data more standard. The utility of Solar System analog observations for exoplanets was investigated here through a broad range of scenarios, and key findings/results are:

\begin{itemize}

    \item The new \texttt{rfast} retrieval suite compares well to more-sophisticated radiative transfer and inference models. This publicly-available tool was created with user-ease in mind and enables rapid and efficient explorations of how spectral data quality relates to constraints on key atmospheric and planetary parameters.

    \item Retrievals using the \texttt{rfast} model were applied to synthetic HabEx/LUVOIR-style exo-Earth observations to understand how bandpass SNR can be traded against wavelength coverage. Upper limit constraints on trace gases, such as methane and carbon dioxide\,---\,while not true detections\,---\,will still have utility for understanding exoplanet environments. The HabEx/LUVOIR-style retrievals showed that ultraviolet and/or visible data or data quality can be removed/reduced and near-infrared data qualities enhanced to achieve better constraints on difficult-to-detect gases like methane and carbon dioxide. Near-infrared SNRs of slightly greater than 45 may be required for carbon dioxide detections at the low levels present in modern Earth's atmosphere.
    
    \item Observations of Earth from NASA's \textit{EPOXI} mission can serve as a testing ground for exo-Earth reflected light direct imaging mission concepts. Retrievals for data limited to 0.3--2.5\,$\upmu$m at V-band SNR of 20 showed good constraints on gas mixing ratios and cloud parameters as well as a constraint on planet/column-averaged temperature. These are a strong proof-of-concept for a HabEx/LUVOIR-style missions, although carbon dioxide and methane constraints were enabled by features beyond 1.8\,$\upmu$m. Retrievals on the same data but at a V-band SNR of 10 cannot rule out scenarios with near planet-wide cloud coverage and a deep atmosphere underneath, and distinguish this from an Earth-like atmospheric state.
    
    \item Emitted-light retrievals were performed on observations of Earth from \textit{MGS}-TES instrument at a 10\,$\upmu$m SNR of 20. Realistic constraints on H$_2$O, CO$_2$, O$_3$, N$_2$O, and CH$_4$ were achieved, which, amongst other science cases, demonstrates feasibility of detecting key biosignature gases with an infrared exo-Earth direct imaging mission. Surface pressure and planetary radius constraints were biased high and low, respectively, potentially pointing to issues with calibration and/or the treatment of water vapor pressure-induced absorption. Surface temperature was constrained to be within the habitable range, although the inferred temperature gradient in the troposphere was unrealistically small. A cloud modeling treatment that allows for multiple cloud decks was shown to potentially remedy issues with tropospheric lapse rate.
    
    \item Transit retrievals on spectra derived from \textit{Cassini}-VIMS occultation observations of Titan are a strong proving ground for validation of concepts related to transiting exoplanet studies with \textit{JWST}, especially as the VIMS wavelength range (0.88--5\,$\upmu$m) overlaps with the ranges of several \textit{JWST} instruments. At a data quality similar to previous \textit{JWST}-relevant modeling studies \citep{greeneetal2016}, Titan transit spectra retrievals obtain gas mixing ratio constraints with uncertainties that are better than 0.5\,dex and that are roughly consistent with orbiter-derived results. Constrained temperatures are biased high by about 20\%, which may be due to linelist sensitivities. An approach to modeling the 3.4\,$\upmu$m C-H stretch mode feature is suggested, although the VIMS-derived Titan transit spectrum can be sufficiently fitted without this treatment at the adopted noise level of 0.1\,ppm.
    
\end{itemize}

\acknowledgements{TDR gratefully acknowledges support from NASA's Exoplanets Research Program (No.~80NSSC18K0349) and Exobiology Program (No.~80NSSC19K0473), the Nexus for Exoplanet System Science Virtual Planetary Laboratory (No.~80NSSC18K0829), and the Cottrell Scholar Program administered by the Research Corporation for Science Advancement. AS gratefully acknowledges support from NASA's Habitable Worlds Program (No.~80NSSC20K0226). {Both authors thank a pair of anonymous reviewers for constructive comments on an earlier version of this manuscript}.}

\software{
  \texttt{corner} \citep{foremanmackey2016},
  \texttt{emcee} \citep{foremanmackeyetal2013},
  \texttt{LBLABC} \citep{meadows&crisp1996},
  \texttt{rfast} \url{https://doi.org/10.5281/zenodo.7327817},
  \texttt{SMART} \citep{meadows&crisp1996}
  }

%
\appendix

\begin{figure}
    \centering
    \includegraphics[scale=0.25]{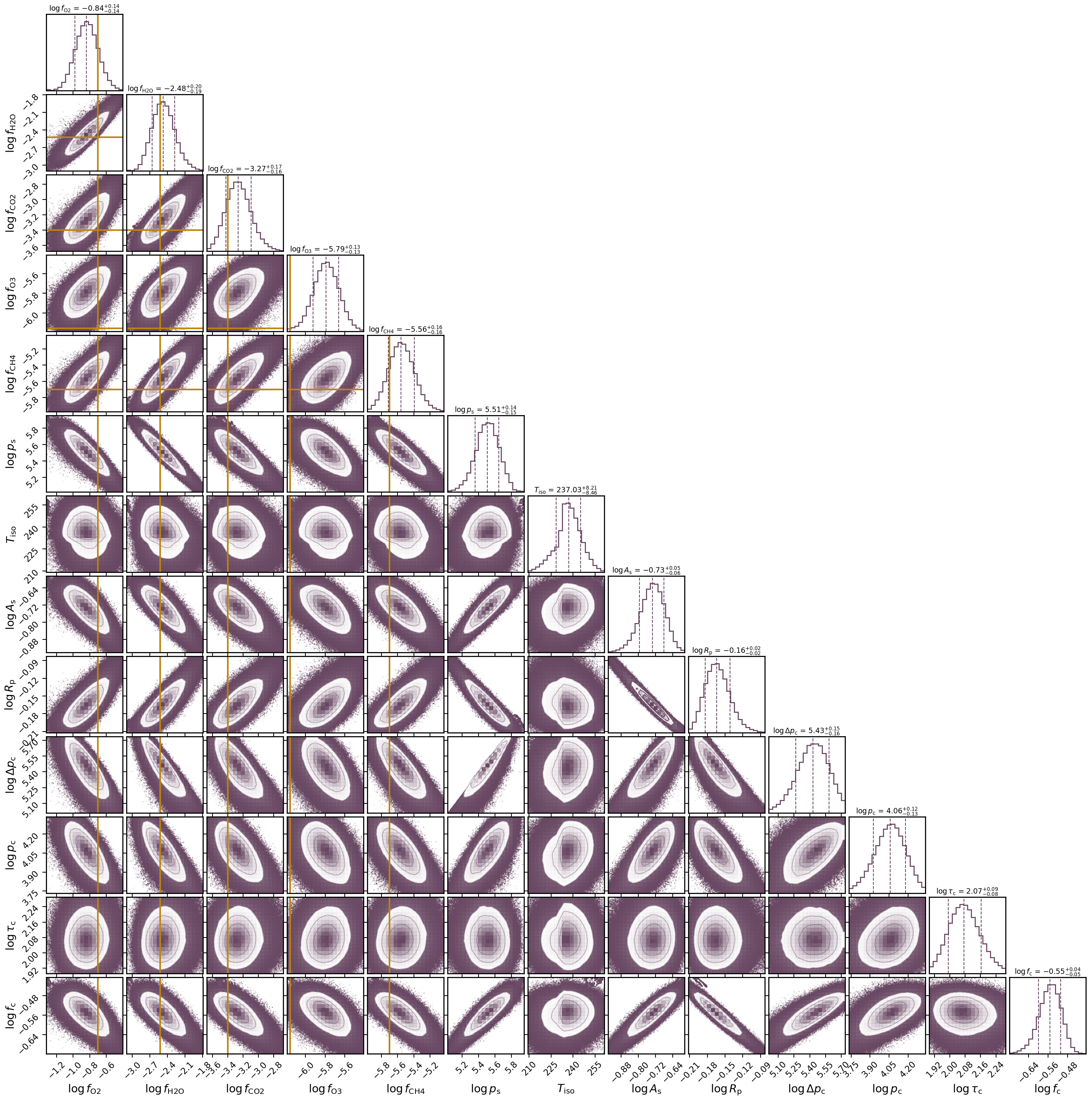}
    \caption{Visualized posterior distribution for a 13-parameters fit to the SNR of 20 \textit{EPOXI} Earth observations in Figure~\ref{fig:epoxi_data}. One-dimensional marginal distributions are shown along the diagonal with associated 16/50/84-percentile values indicated above and with vertical dashed lines. When relevant \textit{a priori} values are well-known, a vertical line indicates the characteristic ``truth'' value for Earth.} 
    \label{fig:epoxi_corner_snr20}
\end{figure}

\begin{figure}
    \centering
    \includegraphics[scale=0.25]{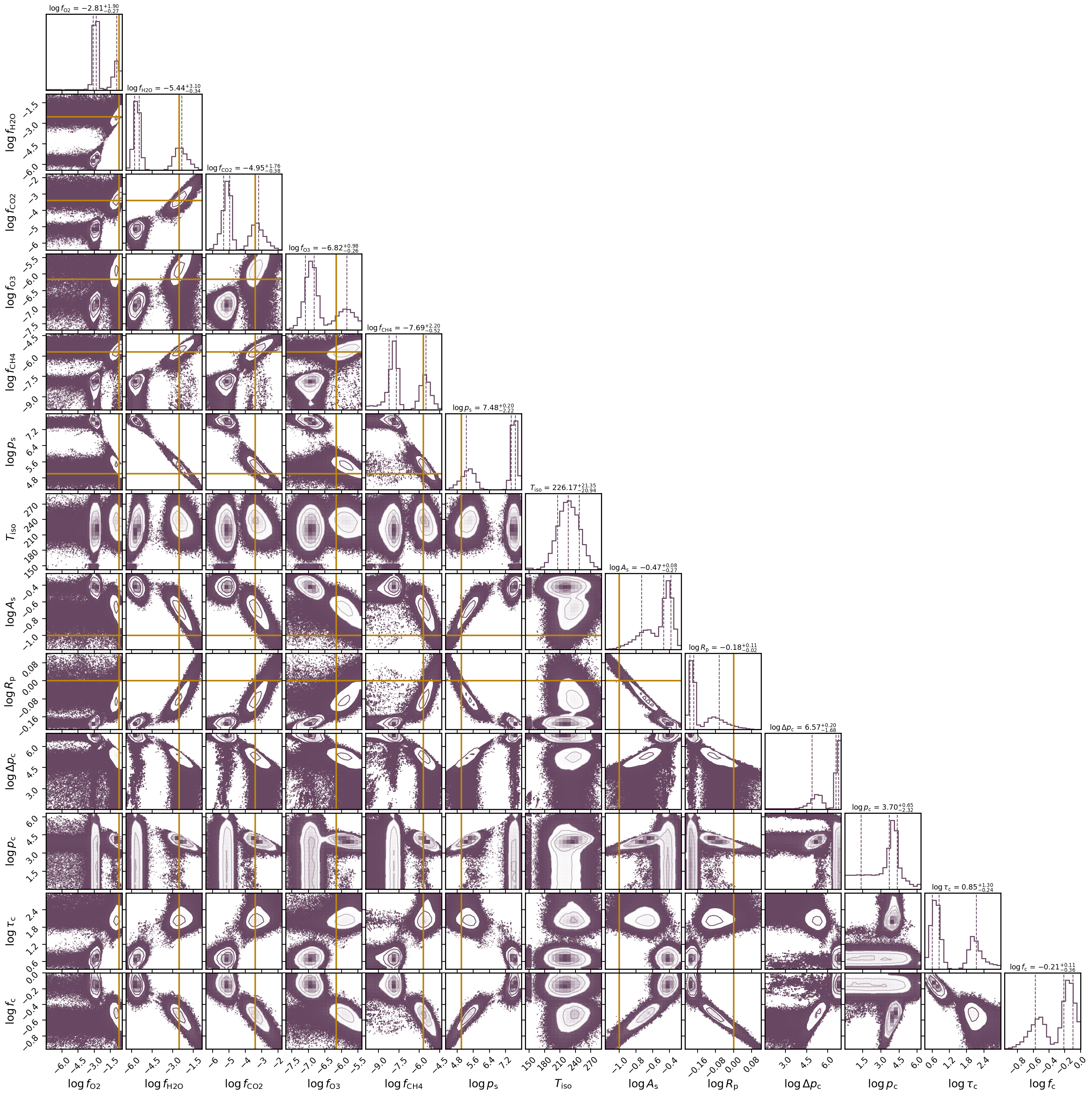}
    \caption{Same as Figure~\ref{fig:epoxi_corner_snr20} except that the underlying \textit{EPOXI} Earth observations have been degraded to a V-band SNR of 10.} 
    \label{fig:epoxi_corner_snr10}
\end{figure}

\begin{figure}
    \centering
    \includegraphics[scale=0.25]{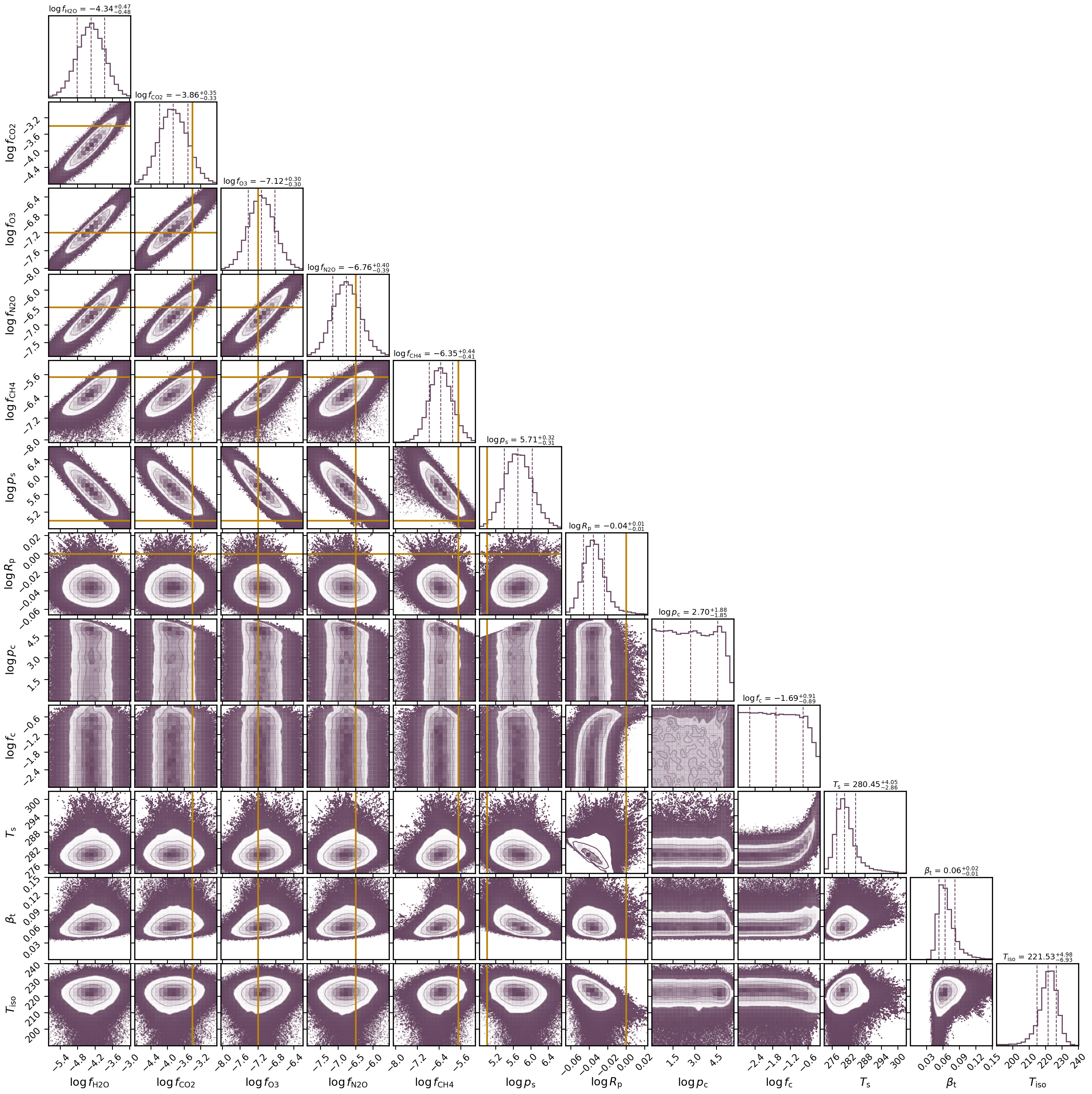}
    \caption{Visualized posterior distribution for a 12-parameters fit to the \textit{MGS}-TES Earth observations in Figure~\ref{fig:mgs_tes_data}. One-dimensional marginal distributions are shown along the diagonal with associated 16/50/84-percentile values indicated above and with the vertical dashed lines.} 
    \label{fig:mgs_tes_corner}
\end{figure}

\begin{figure}
    \centering
    \includegraphics[scale=0.25]{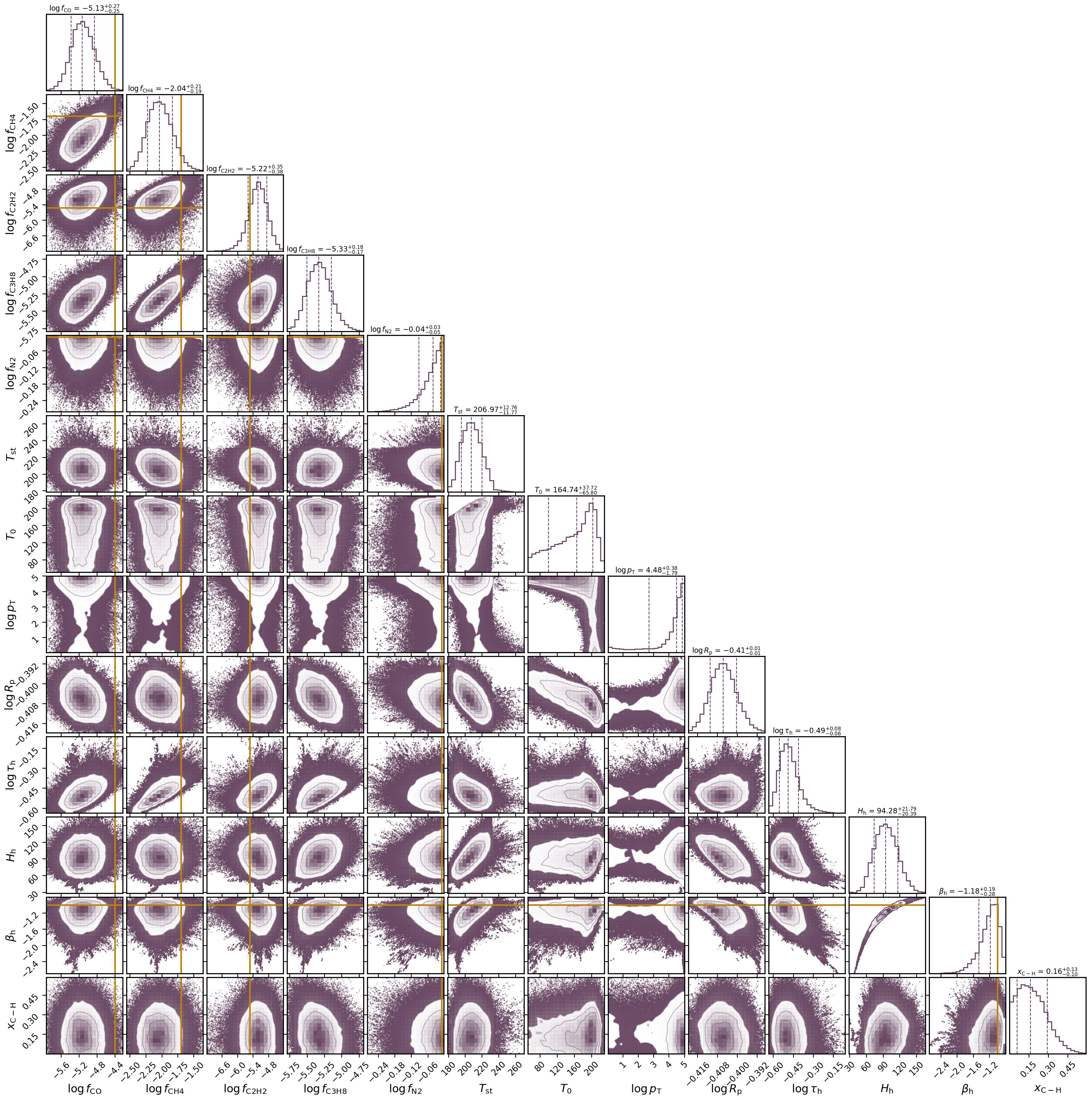}
    \caption{Visualized posterior distribution for a 13-parameters fit to the \textit{Cassini}-VIMS Titan observations (Figure~\ref{fig:cassini_data}). One-dimensional marginal distributions are shown along the diagonal with associated 16/50/84-percentile values indicated above and shown by the vertical dashed lines. Fiducial ``truth'' values from which the Figure~\ref{fig:cassini_data} observed spectrum is obtained are indicated with the vertical solid lines.} 
    \label{fig:cassini_corner}
\end{figure}

%

\clearpage

%

\begin{thebibliography}{}
\expandafter\ifx\csname natexlab\endcsname\relax\def\natexlab#1{#1}\fi
\providecommand{\url}[1]{\href{#1}{#1}}
\providecommand{\dodoi}[1]{doi:~\href{http://doi.org/#1}{\nolinkurl{#1}}}
\providecommand{\doeprint}[1]{\href{http://ascl.net/#1}{\nolinkurl{http://ascl.net/#1}}}
\providecommand{\doarXiv}[1]{\href{https://arxiv.org/abs/#1}{\nolinkurl{https://arxiv.org/abs/#1}}}

\bibitem[{{Akeson} {et~al.}(2019){Akeson}, {Armus}, {Bachelet}, {Bailey},
  {Bartusek}, {Bellini}, {Benford}, {Bennett}, {Bhattacharya}, {Bohlin},
  {Boyer}, {Bozza}, {Bryden}, {Calchi Novati}, {Carpenter}, {Casertano},
  {Choi}, {Content}, {Dayal}, {Dressler}, {Dor{\'e}}, {Fall}, {Fan}, {Fang},
  {Filippenko}, {Finkelstein}, {Foley}, {Furlanetto}, {Kalirai}, {Gaudi},
  {Gilbert}, {Girard}, {Grady}, {Greene}, {Guhathakurta}, {Heinrich},
  {Hemmati}, {Hendel}, {Henderson}, {Henning}, {Hirata}, {Ho}, {Huff},
  {Hutter}, {Jansen}, {Jha}, {Johnson}, {Jones}, {Kasdin}, {Kelly}, {Kirshner},
  {Koekemoer}, {Kruk}, {Lewis}, {Macintosh}, {Madau}, {Malhotra}, {Mandel},
  {Massara}, {Masters}, {McEnery}, {McQuinn}, {Melchior}, {Melton},
  {Mennesson}, {Peeples}, {Penny}, {Perlmutter}, {Pisani}, {Plazas}, {Poleski},
  {Postman}, {Ranc}, {Rauscher}, {Rest}, {Roberge}, {Robertson}, {Rodney},
  {Rhoads}, {Rhodes}, {Ryan}, {Sahu}, {Sand}, {Scolnic}, {Seth}, {Shvartzvald},
  {Siellez}, {Smith}, {Spergel}, {Stassun}, {Street}, {Strolger}, {Szalay},
  {Trauger}, {Troxel}, {Turnbull}, {van der Marel}, {von der Linden}, {Wang},
  {Weinberg}, {Williams}, {Windhorst}, {Wollack}, {Wu}, {Yee}, \&
  {Zimmerman}}]{akesonetal2019}
{Akeson}, R., {Armus}, L., {Bachelet}, E., {et~al.} 2019, arXiv e-prints,
  arXiv:1902.05569.
\newblock \doarXiv{1902.05569}

\bibitem[{{Alei} {et~al.}(2022){Alei}, {Konrad}, {Angerhausen}, {Grenfell},
  {Molli{\`e}re}, {Quanz}, {Rugheimer}, {Wunderlich}, \& {LIFE
  Collaboration}}]{aleietal2022}
{Alei}, E., {Konrad}, B.~S., {Angerhausen}, D., {et~al.} 2022, \aap, 665, A106,
  \dodoi{10.1051/0004-6361/202243760}

\bibitem[{{Barstow} {et~al.}(2017){Barstow}, {Aigrain}, {Irwin}, \&
  {Sing}}]{barstowetal2017}
{Barstow}, J.~K., {Aigrain}, S., {Irwin}, P.~G.~J., \& {Sing}, D.~K. 2017,
  \apj, 834, 50, \dodoi{10.3847/1538-4357/834/1/50}

\bibitem[{{Barstow} {et~al.}(2022){Barstow}, {Changeat}, {Chubb}, {Cubillos},
  {Edwards}, {MacDonald}, {Min}, \& {Waldmann}}]{barstowetal2022}
{Barstow}, J.~K., {Changeat}, Q., {Chubb}, K.~L., {et~al.} 2022, Experimental
  Astronomy, \dodoi{10.1007/s10686-021-09821-w}

\bibitem[{{Battersby} {et~al.}(2018){Battersby}, {Armus}, {Bergin}, {Kataria},
  {Meixner}, {Pope}, {Stevenson}, {Cooray}, {Leisawitz}, {Scott}, {Bauer},
  {Bradford}, {Ennico}, {Fortney}, {Kaltenegger}, {Melnick}, {Milam},
  {Narayanan}, {Padgett}, {Pontoppidan}, {Roellig}, {Sandstrom}, {Su},
  {Vieira}, {Wright}, {Zmuidzinas}, {Staguhn}, {Sheth}, {Benford}, {Mamajek},
  {Neff}, {Carey}, {Burgarella}, {De Beck}, {Gerin}, {Helmich}, {Moseley},
  {Sakon}, \& {Wiedner}}]{battersbyetal2018}
{Battersby}, C., {Armus}, L., {Bergin}, E., {et~al.} 2018, Nature Astronomy, 2,
  596, \dodoi{10.1038/s41550-018-0540-y}

\bibitem[{Bean {et~al.}(2010)Bean, Kempton, \& Homeier}]{beanetal2010}
Bean, J.~L., Kempton, E. M.-R., \& Homeier, D. 2010, Nature, 468, 669

\bibitem[{{Bellucci} {et~al.}(2009){Bellucci}, {Sicardy}, {Drossart}, {Rannou},
  {Nicholson}, {Hedman}, {Baines}, \& {Burrati}}]{belluccietal2009}
{Bellucci}, A., {Sicardy}, B., {Drossart}, P., {et~al.} 2009, \icarus, 201,
  198, \dodoi{10.1016/j.icarus.2008.12.024}

\bibitem[{{Benneke} \& {Seager}(2012)}]{benneke&seager2012}
{Benneke}, B., \& {Seager}, S. 2012, \apj, 753, 100,
  \dodoi{10.1088/0004-637X/753/2/100}

\bibitem[{{Benneke} {et~al.}(2019){Benneke}, {Wong}, {Piaulet}, {Knutson},
  {Lothringer}, {Morley}, {Crossfield}, {Gao}, {Greene}, {Dressing},
  {Dragomir}, {Howard}, {McCullough}, {Kempton}, {Fortney}, \&
  {Fraine}}]{bennekeetal2019}
{Benneke}, B., {Wong}, I., {Piaulet}, C., {et~al.} 2019, \apjl, 887, L14,
  \dodoi{10.3847/2041-8213/ab59dc}

\bibitem[{{B{\'e}tr{\'e}mieux} \&
  {Kaltenegger}(2014)}]{betremieux&kaltenegger2014}
{B{\'e}tr{\'e}mieux}, Y., \& {Kaltenegger}, L. 2014, \apj, 791, 7.
\newblock \doarXiv{1312.6625}

\bibitem[{{B{\'e}tr{\'e}mieux} \& {Swain}(2016)}]{betremieux&swain2017}
{B{\'e}tr{\'e}mieux}, Y., \& {Swain}, M.~R. 2016, ArXiv e-prints.
\newblock \doarXiv{1610.02049}

\bibitem[{Brown {et~al.}(2004)Brown, Baines, Bellucci, Bibring, Buratti,
  Capaccioni, Cerroni, Clark, Coradini, Cruikshank, {et~al.}}]{brownetal2004}
Brown, R., Baines, K., Bellucci, G., {et~al.} 2004, in The Cassini-Huygens
  Mission (Springer), 111--168

\bibitem[{{Buchner} {et~al.}(2014){Buchner}, {Georgakakis}, {Nandra}, {Hsu},
  {Rangel}, {Brightman}, {Merloni}, {Salvato}, {Donley}, \&
  {Kocevski}}]{buchneretal2014}
{Buchner}, J., {Georgakakis}, A., {Nandra}, K., {et~al.} 2014, \aap, 564, A125,
  \dodoi{10.1051/0004-6361/201322971}

\bibitem[{{Campargue} {et~al.}(2013){Campargue}, {Leshchishina}, {Wang},
  {Mondelain}, \& {Kassi}}]{campargueetal2013}
{Campargue}, A., {Leshchishina}, O., {Wang}, L., {Mondelain}, D., \& {Kassi},
  S. 2013, Journal of Molecular Spectroscopy, 291, 16,
  \dodoi{10.1016/j.jms.2013.03.001}

\bibitem[{{Christensen} \& {Pearl}(1997)}]{christensen&pearl1997}
{Christensen}, P.~R., \& {Pearl}, J.~C. 1997, \jgr, 102, 10875,
  \dodoi{10.1029/97JE00637}

\bibitem[{{Col{\'o}n} {et~al.}(2020){Col{\'o}n}, {Kreidberg}, {Welbanks},
  {Line}, {Madhusudhan}, {Beatty}, {Tamburo}, {Stevenson}, {Mandell},
  {Rodriguez}, {Barclay}, {Lopez}, {Stassun}, {Angerhausen}, {Fortney},
  {James}, {Pepper}, {Ahlers}, {Plavchan}, {Awiphan}, {Kotnik}, {McLeod},
  {Murawski}, {Chotani}, {LeBrun}, {Matzko}, {Rea}, {Vidaurri}, {Webster},
  {Williams}, {Cox}, {Tan}, \& {Gilbert}}]{colonetal2020}
{Col{\'o}n}, K.~D., {Kreidberg}, L., {Welbanks}, L., {et~al.} 2020, \aj, 160,
  280, \dodoi{10.3847/1538-3881/abc1e9}

\bibitem[{{Cours} {et~al.}(2020){Cours}, {Cordier}, {Seignovert},
  {Maltagliati}, \& {Biennier}}]{coursetal2020}
{Cours}, T., {Cordier}, D., {Seignovert}, B., {Maltagliati}, L., \& {Biennier},
  L. 2020, \icarus, 339, 113571, \dodoi{10.1016/j.icarus.2019.113571}

\bibitem[{{Cowan} {et~al.}(2015){Cowan}, {Greene}, {Angerhausen}, {Batalha},
  {Clampin}, {Col{\'o}n}, {Crossfield}, {Fortney}, {Gaudi}, {Harrington},
  {Iro}, {Lillie}, {Linsky}, {Lopez-Morales}, {Mandell}, \&
  {Stevenson}}]{cowanetal2015}
{Cowan}, N.~B., {Greene}, T., {Angerhausen}, D., {et~al.} 2015, \pasp, 127,
  311.
\newblock \doarXiv{1502.00004}

\bibitem[{{Dalba} {et~al.}(2015){Dalba}, {Muirhead}, {Fortney}, {Hedman},
  {Nicholson}, \& {Veyette}}]{dalbaetal2015}
{Dalba}, P.~A., {Muirhead}, P.~S., {Fortney}, J.~J., {et~al.} 2015, \apj, 814,
  154, \dodoi{10.1088/0004-637X/814/2/154}

\bibitem[{{Damiano} \& {Hu}(2021)}]{damiano&hu2021}
{Damiano}, M., \& {Hu}, R. 2021, \aj, 162, 200,
  \dodoi{10.3847/1538-3881/ac224d}

\bibitem[{{Dannert} {et~al.}(2022){Dannert}, {Ottiger}, {Quanz}, {Laugier},
  {Fontanet}, {Gheorghe}, {Absil}, {Dandumont}, {Defr{\`e}re}, {Gasc{\'o}n},
  {Glauser}, {Kammerer}, {Lichtenberg}, {Linz}, {Loicq}, \& {the LIFE
  collaboration}}]{dannertetal2022}
{Dannert}, F., {Ottiger}, M., {Quanz}, S.~P., {et~al.} 2022, arXiv e-prints,
  arXiv:2203.00471.
\newblock \doarXiv{2203.00471}

\bibitem[{{Esposito} {et~al.}(2004){Esposito}, {Barth}, {Colwell}, {Lawrence},
  {McClintock}, {Stewart}, {Keller}, {Korth}, {Lauche}, {Festou}, {Lane},
  {Hansen}, {Maki}, {West}, {Jahn}, {Reulke}, {Warlich}, {Shemansky}, \&
  {Yung}}]{espositoetal2004}
{Esposito}, L.~W., {Barth}, C.~A., {Colwell}, J.~E., {et~al.} 2004, \ssr, 115,
  299, \dodoi{10.1007/s11214-004-1455-8}

\bibitem[{{Feng} {et~al.}(2018){Feng}, {Robinson}, {Fortney}, {Lupu}, {Marley},
  {Lewis}, {Macintosh}, \& {Line}}]{fengetal2018}
{Feng}, Y.~K., {Robinson}, T.~D., {Fortney}, J.~J., {et~al.} 2018, \aj, 155,
  200, \dodoi{10.3847/1538-3881/aab95c}

\bibitem[{Foreman-Mackey(2016)}]{foremanmackey2016}
Foreman-Mackey, D. 2016, The Journal of Open Source Software, 24,
  \dodoi{10.21105/joss.00024}

\bibitem[{{Foreman-Mackey} {et~al.}(2013){Foreman-Mackey}, {Hogg}, {Lang}, \&
  {Goodman}}]{foremanmackeyetal2013}
{Foreman-Mackey}, D., {Hogg}, D.~W., {Lang}, D., \& {Goodman}, J. 2013, \pasp,
  125, 306, \dodoi{10.1086/670067}

\bibitem[{{Fortney} {et~al.}(2021){Fortney}, {Barstow}, \&
  {Madhusudhan}}]{fortneyetal2021}
{Fortney}, J.~J., {Barstow}, J.~K., \& {Madhusudhan}, N. 2021, in ExoFrontiers;
  Big Questions in Exoplanetary Science, ed. N.~{Madhusudhan} (IOP Publishing,
  Bristol, UK), 17--1, \dodoi{10.1088/2514-3433/abfa8fch17}

\bibitem[{{Freedman} {et~al.}(2014){Freedman}, {Lustig-Yaeger}, {Fortney},
  {Lupu}, {Marley}, \& {Lodders}}]{freedmanetal2014}
{Freedman}, R.~S., {Lustig-Yaeger}, J., {Fortney}, J.~J., {et~al.} 2014, \apjs,
  214, 25, \dodoi{10.1088/0067-0049/214/2/25}

\bibitem[{{Freedman} {et~al.}(2008){Freedman}, {Marley}, \&
  {Lodders}}]{freedmanetal2008}
{Freedman}, R.~S., {Marley}, M.~S., \& {Lodders}, K. 2008, \apjs, 174, 504,
  \dodoi{10.1086/521793}

\bibitem[{{Fujii} {et~al.}(2014){Fujii}, {Kimura}, {Dohm}, \&
  {Ohtake}}]{fujiietal2014}
{Fujii}, Y., {Kimura}, J., {Dohm}, J., \& {Ohtake}, M. 2014, Astrobiology, 14,
  753, \dodoi{10.1089/ast.2014.1165}

\bibitem[{{Gaudi} {et~al.}(2018){Gaudi}, {Seager}, {Mennesson}, {Kiessling},
  {Warfield}, {Habitable Exoplanet Observatory Science}, \& {Technology
  Definition Team}}]{gaudietal2018}
{Gaudi}, B.~S., {Seager}, S., {Mennesson}, B., {et~al.} 2018, Nature Astronomy,
  2, 600, \dodoi{10.1038/s41550-018-0549-2}

\bibitem[{{Gordon} {et~al.}(2022){Gordon}, {Rothman}, {Hargreaves}, {Hashemi},
  {Karlovets}, {Skinner}, {Conway}, {Hill}, {Kochanov}, {Tan}, {Wcis{\l}o},
  {Finenko}, {Nelson}, {Bernath}, {Birk}, {Boudon}, {Campargue}, {Chance},
  {Coustenis}, {Drouin}, {Flaud}, {Gamache}, {Hodges}, {Jacquemart}, {Mlawer},
  {Nikitin}, {Perevalov}, {Rotger}, {Tennyson}, {Toon}, {Tran}, {Tyuterev},
  {Adkins}, {Baker}, {Barbe}, {Can{\`e}}, {Cs{\'a}sz{\'a}r}, {Dudaryonok},
  {Egorov}, {Fleisher}, {Fleurbaey}, {Foltynowicz}, {Furtenbacher}, {Harrison},
  {Hartmann}, {Horneman}, {Huang}, {Karman}, {Karns}, {Kassi}, {Kleiner},
  {Kofman}, {Kwabia-Tchana}, {Lavrentieva}, {Lee}, {Long}, {Lukashevskaya},
  {Lyulin}, {Makhnev}, {Matt}, {Massie}, {Melosso}, {Mikhailenko}, {Mondelain},
  {M{\"u}ller}, {Naumenko}, {Perrin}, {Polyansky}, {Raddaoui}, {Raston},
  {Reed}, {Rey}, {Richard}, {T{\'o}bi{\'a}s}, {Sadiek}, {Schwenke},
  {Starikova}, {Sung}, {Tamassia}, {Tashkun}, {Vander Auwera}, {Vasilenko},
  {Vigasin}, {Villanueva}, {Vispoel}, {Wagner}, {Yachmenev}, \&
  {Yurchenko}}]{gordonetal2022}
{Gordon}, I.~E., {Rothman}, L.~S., {Hargreaves}, R.~J., {et~al.} 2022, \jqsrt,
  277, 107949, \dodoi{10.1016/j.jqsrt.2021.107949}

\bibitem[{{Greene} {et~al.}(2016){Greene}, {Line}, {Montero}, {Fortney},
  {Lustig-Yaeger}, \& {Luther}}]{greeneetal2016}
{Greene}, T.~P., {Line}, M.~R., {Montero}, C., {et~al.} 2016, \apj, 817, 17,
  \dodoi{10.3847/0004-637X/817/1/17}

\bibitem[{{Hapke}(1981)}]{hapke1981}
{Hapke}, B. 1981, \jgr, 86, 3039, \dodoi{10.1029/JB086iB04p03039}

\bibitem[{{Hayne} {et~al.}(2014){Hayne}, {McCord}, \& {Sotin}}]{hayneetal2014}
{Hayne}, P.~O., {McCord}, T.~B., \& {Sotin}, C. 2014, \icarus, 243, 158,
  \dodoi{10.1016/j.icarus.2014.08.045}

\bibitem[{{Heng} \& {Li}(2021)}]{heng&li2021}
{Heng}, K., \& {Li}, L. 2021, \apjl, 909, L20, \dodoi{10.3847/2041-8213/abe872}

\bibitem[{{Heng} {et~al.}(2021){Heng}, {Morris}, \& {Kitzmann}}]{hengetal2021}
{Heng}, K., {Morris}, B.~M., \& {Kitzmann}, D. 2021, Nature Astronomy, 5, 1001,
  \dodoi{10.1038/s41550-021-01444-7}

\bibitem[{{Henyey} \& {Greenstein}(1941)}]{henyey&greenstein1941}
{Henyey}, L.~G., \& {Greenstein}, J.~L. 1941, \apj, 93, 70

\bibitem[{{Horak}(1950)}]{horak1950}
{Horak}, H.~G. 1950, \apj, 112, 445, \dodoi{10.1086/145359}

\bibitem[{{Horak} \& {Little}(1965)}]{horak&little1965}
{Horak}, H.~G., \& {Little}, S.~J. 1965, \apjs, 11, 373, \dodoi{10.1086/190119}

\bibitem[{{Hubbard} {et~al.}(1993){Hubbard}, {Sicardy}, {Miles}, {Hollis},
  {Forrest}, {Nicolson}, {Appleby}, {Beisker}, {Bittner}, {Bode}, {Bruns},
  {Denzau}, {Nezel}, {Riedel}, {Struckmann}, {Arlot}, {Roques}, {Sevre},
  {Thuillot}, {Hoffmann}, {Geyer}, {Buil}, {Colas}, {Lecacheux}, {Klotz},
  {Thouvenot}, {Vidal}, {Carreira}, {Rossi}, {Blanco}, {Cristaldi}, {Nevo},
  {Reitsema}, {Brosch}, {Cernis}, {Zdanavicius}, {Wasserman}, {Hunten},
  {Gautier}, {Lellouch}, {Yelle}, {Rizk}, {Flasar}, {Porco}, {Toublanc}, \&
  {Corugedo}}]{hubbardetal1993}
{Hubbard}, W.~B., {Sicardy}, B., {Miles}, R., {et~al.} 1993, \aap, 269, 541

\bibitem[{{Hunt}(1973)}]{hunt1973}
{Hunt}, G.~e. 1973, Quarterly Journal of the Royal Meteorological Society, 099,
  346, \dodoi{10.1002/qj.49709942013}

\bibitem[{{Irwin} {et~al.}(2008){Irwin}, {Teanby}, {de Kok}, {Fletcher},
  {Howett}, {Tsang}, {Wilson}, {Calcutt}, {Nixon}, \&
  {Parrish}}]{irwinetal2008}
{Irwin}, P.~G.~J., {Teanby}, N.~A., {de Kok}, R., {et~al.} 2008, \jqsrt, 109,
  1136, \dodoi{10.1016/j.jqsrt.2007.11.006}

\bibitem[{Kaltenegger \& Traub(2009)}]{kaltenegger&traub2009}
Kaltenegger, L., \& Traub, W. 2009, The Astrophysical Journal, 698, 519

\bibitem[{{Kane} {et~al.}(2019){Kane}, {Arney}, {Crisp}, {Domagal-Goldman},
  {Glaze}, {Goldblatt}, {Grinspoon}, {Head}, {Lenardic}, {Unterborn}, {Way}, \&
  {Zahnle}}]{kaneetal2019}
{Kane}, S.~R., {Arney}, G., {Crisp}, D., {et~al.} 2019, Journal of Geophysical
  Research (Planets), 124, 2015, \dodoi{10.1029/2019JE005939}

\bibitem[{{Kane} {et~al.}(2021){Kane}, {Arney}, {Byrne}, {Dalba}, {Desch},
  {Horner}, {Izenberg}, {Mandt}, {Meadows}, \& {Quick}}]{kaneetal2021}
{Kane}, S.~R., {Arney}, G.~N., {Byrne}, P.~K., {et~al.} 2021, Journal of
  Geophysical Research (Planets), 126, e06643, \dodoi{10.1029/2020JE006643}

\bibitem[{{Karkoschka}(1998)}]{karkoschka1998}
{Karkoschka}, E. 1998, Icarus, 133, 134

\bibitem[{{Kasdin} {et~al.}(2020){Kasdin}, {Bailey}, {Mennesson}, {Zellem},
  {Ygouf}, {Rhodes}, {Luchik}, {Zhao}, {Riggs}, {Seo}, {Krist}, {Kern}, {Tang},
  {Nemati}, {Groff}, {Zimmerman}, {Macintosh}, {Turnbull}, {Debes}, {Douglas},
  \& {Lupu}}]{kasdinetal2020}
{Kasdin}, N.~J., {Bailey}, V.~P., {Mennesson}, B., {et~al.} 2020, in Society of
  Photo-Optical Instrumentation Engineers (SPIE) Conference Series, Vol. 11443,
  Society of Photo-Optical Instrumentation Engineers (SPIE) Conference Series,
  114431U, \dodoi{10.1117/12.2562997}

\bibitem[{{Keithly} \& {Savransky}(2021)}]{keithly&savransky2021}
{Keithly}, D.~R., \& {Savransky}, D. 2021, \apjl, 919, L11,
  \dodoi{10.3847/2041-8213/ac20cf}

\bibitem[{{Kitzmann} {et~al.}(2020){Kitzmann}, {Heng}, {Oreshenko}, {Grimm},
  {Apai}, {Bowler}, {Burgasser}, \& {Marley}}]{kitzmannetal2020}
{Kitzmann}, D., {Heng}, K., {Oreshenko}, M., {et~al.} 2020, \apj, 890, 174,
  \dodoi{10.3847/1538-4357/ab6d71}

\bibitem[{{Knutson} {et~al.}(2014){Knutson}, {Dragomir}, {Kreidberg},
  {Kempton}, {McCullough}, {Fortney}, {Bean}, {Gillon}, {Homeier}, \&
  {Howard}}]{knutsonetal2014b}
{Knutson}, H.~A., {Dragomir}, D., {Kreidberg}, L., {et~al.} 2014, \apj, 794,
  155, \dodoi{10.1088/0004-637X/794/2/155}

\bibitem[{{Konrad} {et~al.}(2021){Konrad}, {Alei}, {Angerhausen},
  {Carri{\'o}n-Gonz{\'a}lez}, {Fortney}, {Grenfell}, {Kitzmann},
  {Molli{\`e}re}, {Rugheimer}, {Wunderlich}, {Quanz}, \& {the LIFE
  Collaboration}}]{konradetal2021}
{Konrad}, B.~S., {Alei}, E., {Angerhausen}, D., {et~al.} 2021, arXiv e-prints,
  arXiv:2112.02054.
\newblock \doarXiv{2112.02054}

\bibitem[{{Koskinen} {et~al.}(2011){Koskinen}, {Yelle}, {Snowden}, {Lavvas},
  {Sandel}, {Capalbo}, {Benilan}, \& {West}}]{koskinenetal2011}
{Koskinen}, T.~T., {Yelle}, R.~V., {Snowden}, D.~S., {et~al.} 2011, \icarus,
  216, 507, \dodoi{10.1016/j.icarus.2011.09.022}

\bibitem[{{Kreidberg} {et~al.}(2014){Kreidberg}, {Bean}, {D{\'e}sert},
  {Benneke}, {Deming}, {Stevenson}, {Seager}, {Berta-Thompson}, {Seifahrt}, \&
  {Homeier}}]{kreidbergetal2014a}
{Kreidberg}, L., {Bean}, J.~L., {D{\'e}sert}, J.-M., {et~al.} 2014, \nat, 505,
  69, \dodoi{10.1038/nature12888}

\bibitem[{{Krissansen-Totton} {et~al.}(2018){Krissansen-Totton}, {Garland},
  {Irwin}, \& {Catling}}]{krissansentottonetal2018b}
{Krissansen-Totton}, J., {Garland}, R., {Irwin}, P., \& {Catling}, D.~C. 2018,
  \aj, 156, 114, \dodoi{10.3847/1538-3881/aad564}

\bibitem[{{Li} {et~al.}(2018){Li}, {Jiang}, {West}, {Gierasch}, {Perez-Hoyos},
  {Sanchez-Lavega}, {Fletcher}, {Fortney}, {Knowles}, {Porco}, {Baines}, {Fry},
  {Mallama}, {Achterberg}, {Simon}, {Nixon}, {Orton}, {Dyudina}, {Ewald}, \&
  {Schmude}}]{lilimingetal2018}
{Li}, L., {Jiang}, X., {West}, R.~A., {et~al.} 2018, Nature Communications, 9,
  3709, \dodoi{10.1038/s41467-018-06107-2}

\bibitem[{{Line} {et~al.}(2014){Line}, {Knutson}, {Wolf}, \&
  {Yung}}]{lineetal2014a}
{Line}, M.~R., {Knutson}, H., {Wolf}, A.~S., \& {Yung}, Y.~L. 2014, \apj, 783,
  70.
\newblock \doarXiv{1309.6663}

\bibitem[{{Line} {et~al.}(2012){Line}, {Zhang}, {Vasisht}, {Natraj}, {Chen}, \&
  {Yung}}]{lineetal2012}
{Line}, M.~R., {Zhang}, X., {Vasisht}, G., {et~al.} 2012, \apj, 749, 93.
\newblock \doarXiv{1111.2612}

\bibitem[{Line {et~al.}(2013)Line, Wolf, Zhang, Knutson, Kammer, Ellison,
  Deroo, Crisp, \& Yung}]{lineetal2013a}
Line, M.~R., Wolf, A.~S., Zhang, X., {et~al.} 2013, The Astrophysical Journal,
  775, 137

\bibitem[{{Livengood} {et~al.}(2011){Livengood}, {Deming}, {A'Hearn},
  {Charbonneau}, {Hewagama}, {Lisse}, {McFadden}, {Meadows}, {Robinson},
  {Seager}, \& {Wellnitz}}]{livengoodetal2011}
{Livengood}, T.~A., {Deming}, L.~D., {A'Hearn}, M.~F., {et~al.} 2011,
  Astrobiology, 11, 907, \dodoi{10.1089/ast.2011.0614}

\bibitem[{{Lupu} {et~al.}(2016){Lupu}, {Marley}, {Lewis}, {Line}, {Traub}, \&
  {Zahnle}}]{lupuetal2016}
{Lupu}, R.~E., {Marley}, M.~S., {Lewis}, N., {et~al.} 2016, \aj, 152, 217,
  \dodoi{10.3847/0004-6256/152/6/217}

\bibitem[{{Lustig-Yaeger} {et~al.}(2018){Lustig-Yaeger}, {Meadows}, {Tovar
  Mendoza}, {Schwieterman}, {Fujii}, {Luger}, \&
  {Robinson}}]{lustigyaegeretal2018}
{Lustig-Yaeger}, J., {Meadows}, V.~S., {Tovar Mendoza}, G., {et~al.} 2018, \aj,
  156, 301, \dodoi{10.3847/1538-3881/aaed3a}

\bibitem[{{Macdonald} \& {Cowan}(2019)}]{macdonald&cowan2019}
{Macdonald}, E. J.~R., \& {Cowan}, N.~B. 2019, \mnras, 489, 196,
  \dodoi{10.1093/mnras/stz2047}

\bibitem[{{MacDonald} \& {Lewis}(2021)}]{macdonald&lewis2021}
{MacDonald}, R.~J., \& {Lewis}, N.~K. 2021, arXiv e-prints, arXiv:2111.05862.
\newblock \doarXiv{2111.05862}

\bibitem[{{MacDonald} \& {Madhusudhan}(2017)}]{macdonald&madhusudhan2017}
{MacDonald}, R.~J., \& {Madhusudhan}, N. 2017, \mnras, 469, 1979,
  \dodoi{10.1093/mnras/stx804}

\bibitem[{{Madhusudhan}(2018)}]{madhusudhan2018}
{Madhusudhan}, N. 2018, in Handbook of Exoplanets, ed. H.~J. {Deeg} \& J.~A.
  {Belmonte} (Springer, Cham), 104, \dodoi{10.1007/978-3-319-55333-7\_104}

\bibitem[{{Madhusudhan} \& {Burrows}(2012)}]{madhusudhan&burrows2012}
{Madhusudhan}, N., \& {Burrows}, A. 2012, \apj, 747, 25,
  \dodoi{10.1088/0004-637X/747/1/25}

\bibitem[{{Madhusudhan} \& {Seager}(2009)}]{madhusudhan&seager2009}
{Madhusudhan}, N., \& {Seager}, S. 2009, \apj, 707, 24.
\newblock \doarXiv{0910.1347}

\bibitem[{{Maltagliati} {et~al.}(2015){Maltagliati}, {B{\'e}zard}, {Vinatier},
  {Hedman}, {Lellouch}, {Nicholson}, {Sotin}, {de Kok}, \&
  {Sicardy}}]{maltagliatietal2015}
{Maltagliati}, L., {B{\'e}zard}, B., {Vinatier}, S., {et~al.} 2015, \icarus,
  248, 1, \dodoi{10.1016/j.icarus.2014.10.004}

\bibitem[{{Mansfield} {et~al.}(2022){Mansfield}, {Wiser}, {Stevenson}, {Smith},
  {Line}, {Bean}, {Fortney}, {Parmentier}, {Kempton}, {Arcangeli},
  {D{\'e}sert}, {Kilpatrick}, {Kreidberg}, \& {Malik}}]{mansfieldetal2022}
{Mansfield}, M., {Wiser}, L., {Stevenson}, K.~B., {et~al.} 2022, arXiv
  e-prints, arXiv:2203.01463.
\newblock \doarXiv{2203.01463}

\bibitem[{{Marley} {et~al.}(2014){Marley}, {Lupu}, {Lewis}, {Line}, {Morley},
  \& {Fortney}}]{marleyetal2014}
{Marley}, M., {Lupu}, R., {Lewis}, N., {et~al.} 2014, ArXiv e-prints.
\newblock \doarXiv{1412.8440}

\bibitem[{{Mayorga} {et~al.}(2020){Mayorga}, {Charbonneau}, \&
  {Thorngren}}]{mayorgaetal2020}
{Mayorga}, L.~C., {Charbonneau}, D., \& {Thorngren}, D.~P. 2020, \aj, 160, 238,
  \dodoi{10.3847/1538-3881/abb8df}

\bibitem[{{Mayorga} {et~al.}(2016){Mayorga}, {Jackiewicz}, {Rages}, {West},
  {Knowles}, {Lewis}, \& {Marley}}]{mayorgaetal2016}
{Mayorga}, L.~C., {Jackiewicz}, J., {Rages}, K., {et~al.} 2016, \aj, 152, 209,
  \dodoi{10.3847/0004-6256/152/6/209}

\bibitem[{{Mayorga} {et~al.}(2021){Mayorga}, {Lustig-Yaeger}, {May}, {Sotzen},
  {Gonzalez-Quiles}, {Kilpatrick}, {Martin}, {Mandt}, {Stevenson}, \&
  {Izenberg}}]{mayorgaetal2021}
{Mayorga}, L.~C., {Lustig-Yaeger}, J., {May}, E.~M., {et~al.} 2021, \psj, 2,
  140, \dodoi{10.3847/PSJ/ac0c85}

\bibitem[{{McClatchey} {et~al.}(1972){McClatchey}, {Fenn}, {Selby}, {Volz}, \&
  {Garing}}]{mcclatcheyetal1972}
{McClatchey}, R.~A., {Fenn}, R.~W., {Selby}, J.~E.~A., {Volz}, F.~E., \&
  {Garing}, J.~S. 1972, {Optical Properties of the Atmosphere (Third Edition)},
  Tech. rep., Air Force Cambridge Research Labs

\bibitem[{{Meadows} \& {Crisp}(1996)}]{meadows&crisp1996}
{Meadows}, V.~S., \& {Crisp}, D. 1996, \jgr, 101, 4595,
  \dodoi{10.1029/95JE03567}

\bibitem[{{Min} {et~al.}(2020){Min}, {Ormel}, {Chubb}, {Helling}, \&
  {Kawashima}}]{minetal2020}
{Min}, M., {Ormel}, C.~W., {Chubb}, K., {Helling}, C., \& {Kawashima}, Y. 2020,
  \aap, 642, A28, \dodoi{10.1051/0004-6361/201937377}

\bibitem[{{Misra} {et~al.}(2014){Misra}, {Meadows}, \& {Crisp}}]{misraetal2014}
{Misra}, A., {Meadows}, V., \& {Crisp}, D. 2014, \apj, 792, 61.
\newblock \doarXiv{1407.3265}

\bibitem[{{Molli{\`e}re} {et~al.}(2019){Molli{\`e}re}, {Wardenier}, {van
  Boekel}, {Henning}, {Molaverdikhani}, \& {Snellen}}]{molliereetal2019}
{Molli{\`e}re}, P., {Wardenier}, J.~P., {van Boekel}, R., {et~al.} 2019, \aap,
  627, A67, \dodoi{10.1051/0004-6361/201935470}

\bibitem[{{Morley} {et~al.}(2016){Morley}, {Knutson}, {Line}, {Fortney},
  {Thorngren}, {Marley}, {Teal}, \& {Lupu}}]{morleyetal2016}
{Morley}, C.~V., {Knutson}, H., {Line}, M., {et~al.} 2016, ArXiv e-prints.
\newblock \doarXiv{1610.07632}

\bibitem[{{Nayak} {et~al.}(2017){Nayak}, {Lupu}, {Marley}, {Fortney},
  {Robinson}, \& {Lewis}}]{nayaketal2017}
{Nayak}, M., {Lupu}, R., {Marley}, M.~S., {et~al.} 2017, \pasp, 129, 034401,
  \dodoi{10.1088/1538-3873/129/973/034401}

\bibitem[{{Nixon} \& {Madhusudhan}(2022)}]{nixon&madhusudhan2022}
{Nixon}, M.~C., \& {Madhusudhan}, N. 2022, arXiv e-prints, arXiv:2201.03532.
\newblock \doarXiv{2201.03532}

\bibitem[{{Pendleton} \& {Allamandola}(2002)}]{pendleton&allamandola2002}
{Pendleton}, Y.~J., \& {Allamandola}, L.~J. 2002, \apjs, 138, 75,
  \dodoi{10.1086/322999}

\bibitem[{{Piette} {et~al.}(2022){Piette}, {Madhusudhan}, \&
  {Mandell}}]{pietteetal2022}
{Piette}, A. A.~A., {Madhusudhan}, N., \& {Mandell}, A.~M. 2022, \mnras, 511,
  2565, \dodoi{10.1093/mnras/stab3612}

\bibitem[{Pont {et~al.}(2008)Pont, Knutson, Gilliland, Moutou, \&
  Charbonneau}]{pontetal2008}
Pont, F., Knutson, H., Gilliland, R., Moutou, C., \& Charbonneau, D. 2008,
  Monthly Notices of the Royal Astronomical Society, 385, 109

\bibitem[{{Porco} {et~al.}(2004){Porco}, {West}, {Squyres}, {McEwen}, {Thomas},
  {Murray}, {Del Genio}, {Ingersoll}, {Johnson}, {Neukum}, {Veverka}, {Dones},
  {Brahic}, {Burns}, {Haemmerle}, {Knowles}, {Dawson}, {Roatsch}, {Beurle}, \&
  {Owen}}]{porcoetal2004}
{Porco}, C.~C., {West}, R.~A., {Squyres}, S., {et~al.} 2004, \ssr, 115, 363,
  \dodoi{10.1007/s11214-004-1456-7}

\bibitem[{{Quanz} {et~al.}(2021{\natexlab{a}}){Quanz}, {Absil}, {Benz},
  {Bonfils}, {Berger}, {Defr{\`e}re}, {van Dishoeck}, {Ehrenreich}, {Fortney},
  {Glauser}, {Grenfell}, {Janson}, {Kraus}, {Krause}, {Labadie}, {Lacour},
  {Line}, {Linz}, {Loicq}, {Miguel}, {Pall{\'e}}, {Queloz}, {Rauer}, {Ribas},
  {Rugheimer}, {Selsis}, {Snellen}, {Sozzetti}, {Stapelfeldt}, {Udry}, \&
  {Wyatt}}]{quanzetal2021}
{Quanz}, S.~P., {Absil}, O., {Benz}, W., {et~al.} 2021{\natexlab{a}},
  Experimental Astronomy, \dodoi{10.1007/s10686-021-09791-z}

\bibitem[{{Quanz} {et~al.}(2021{\natexlab{b}}){Quanz}, {Ottiger}, {Fontanet},
  {Kammerer}, {Menti}, {Dannert}, {Gheorghe}, {Absil}, {Airapetian}, {Alei},
  {Allart}, {Angerhausen}, {Blumenthal}, {Buchhave}, {Cabrera},
  {Carri{\'o}n-Gonz{\'a}lez}, {Chauvin}, {Danchi}, {Dandumont}, {Defr{\`e}re},
  {Dorn}, {Ehrenreich}, {Ertel}, {Fridlund}, {Garc{\'\i}a Mu{\~n}oz},
  {Gasc{\'o}n}, {Girard}, {Glauser}, {Grenfell}, {Guidi}, {Hagelberg},
  {Helled}, {Ireland}, {Kopparapu}, {Korth}, {Kozakis}, {Kraus}, {L{\'e}ger},
  {Leedj{\"a}rv}, {Lichtenberg}, {Lillo-Box}, {Linz}, {Liseau}, {Loicq},
  {Mahendra}, {Malbet}, {Mathew}, {Mennesson}, {Meyer}, {Mishra},
  {Molaverdikhani}, {Noack}, {Oza}, {Pall{\'e}}, {Parviainen}, {Quirrenbach},
  {Rauer}, {Ribas}, {Rice}, {Romagnolo}, {Rugheimer}, {Schwieterman},
  {Serabyn}, {Sharma}, {Stassun}, {Szul{\'a}gyi}, {Wang}, {Wunderlich},
  {Wyatt}, \& {the LIFE collaboration}}]{quanzetal2021b}
{Quanz}, S.~P., {Ottiger}, M., {Fontanet}, E., {et~al.} 2021{\natexlab{b}},
  arXiv e-prints, arXiv:2101.07500.
\newblock \doarXiv{2101.07500}

\bibitem[{{Roberge} \& {Moustakas}(2018)}]{robergeetal2018}
{Roberge}, A., \& {Moustakas}, L.~A. 2018, Nature Astronomy, 2, 605,
  \dodoi{10.1038/s41550-018-0543-8}

\bibitem[{{Roberge} {et~al.}(2017){Roberge}, {Rizzo}, {Lincowski}, {Arney},
  {Stark}, {Robinson}, {Snyder}, {Pueyo}, {Zimmerman}, {Jansen}, {Nesvold},
  {Meadows}, \& {Turnbull}}]{robergeetal2017}
{Roberge}, A., {Rizzo}, M.~J., {Lincowski}, A.~P., {et~al.} 2017, \pasp, 129,
  124401, \dodoi{10.1088/1538-3873/aa8fc4}

\bibitem[{{Robinson}(2017)}]{robinson2017}
{Robinson}, T.~D. 2017, \apj, 836, 236, \dodoi{10.3847/1538-4357/aa5ea8}

\bibitem[{{Robinson} \& {Catling}(2014)}]{robinson&catling2014}
{Robinson}, T.~D., \& {Catling}, D.~C. 2014, Nature Geoscience, 7, 12,
  \dodoi{10.1038/ngeo2020}

\bibitem[{{Robinson} \& {Crisp}(2018)}]{robinson&crisp2018}
{Robinson}, T.~D., \& {Crisp}, D. 2018, \jqsrt, 211, 78,
  \dodoi{10.1016/j.jqsrt.2018.03.002}

\bibitem[{{Robinson} {et~al.}(2017){Robinson}, {Fortney}, \&
  {Hubbard}}]{robinsonetal2017}
{Robinson}, T.~D., {Fortney}, J.~J., \& {Hubbard}, W.~B. 2017, \apj, 850, 128,
  \dodoi{10.3847/1538-4357/aa951e}

\bibitem[{{Robinson} {et~al.}(2014){Robinson}, {Maltagliati}, {Marley}, \&
  {Fortney}}]{robinsonetal2014a}
{Robinson}, T.~D., {Maltagliati}, L., {Marley}, M.~S., \& {Fortney}, J.~J.
  2014, Proceedings of the National Academy of Science, 111, 9042,
  \dodoi{10.1073/pnas.1403473111}

\bibitem[{{Robinson} {et~al.}(2010){Robinson}, {Meadows}, \&
  {Crisp}}]{robinsonetal2010}
{Robinson}, T.~D., {Meadows}, V.~S., \& {Crisp}, D. 2010, \apjl, 721, L67,
  \dodoi{10.1088/2041-8205/721/1/L67}

\bibitem[{Robinson \& Reinhard(2020)}]{robinson&reinhard2020}
Robinson, T.~D., \& Reinhard, C.~T. 2020, Earth as an Exoplanet (University of
  Arizona Press), 379--416.
\newblock \url{http://www.jstor.org/stable/j.ctv105bb62.21}

\bibitem[{{Robinson} {et~al.}(2011){Robinson}, {Meadows}, {Crisp}, {Deming},
  {A'Hearn}, {Charbonneau}, {Livengood}, {Seager}, {Barry}, {Hearty},
  {Hewagama}, {Lisse}, {McFadden}, \& {Wellnitz}}]{robinsonetal2011}
{Robinson}, T.~D., {Meadows}, V.~S., {Crisp}, D., {et~al.} 2011, Astrobiology,
  11, 393, \dodoi{10.1089/ast.2011.0642}

\bibitem[{{Sidis} \& {Sari}(2010)}]{sidis&sari2010}
{Sidis}, O., \& {Sari}, R. 2010, The Astrophysical Journal, 720, 904

\bibitem[{{Sing} {et~al.}(2009){Sing}, {D{\'e}sert}, {Lecavelier Des Etangs},
  {Ballester}, {Vidal-Madjar}, {Parmentier}, {Hebrard}, \&
  {Henry}}]{singetal2009}
{Sing}, D.~K., {D{\'e}sert}, J.~M., {Lecavelier Des Etangs}, A., {et~al.} 2009,
  \aap, 505, 891, \dodoi{10.1051/0004-6361/200912776}

\bibitem[{{Smith} {et~al.}(2020){Smith}, {Feng}, {Fortney}, {Robinson},
  {Marley}, {Lupu}, \& {Lewis}}]{smithetal2020}
{Smith}, A. J.~R.~W., {Feng}, Y.~K., {Fortney}, J.~J., {et~al.} 2020, \aj, 159,
  36, \dodoi{10.3847/1538-3881/ab5a8a}

\bibitem[{{Sobolev}(1975)}]{sobolev1975}
{Sobolev}, V.~V. 1975, {Light scattering in planetary atmospheres} (Pergamon
  Press, Oxford)

\bibitem[{{Spurr} \& {Natraj}(2011)}]{spurr&natraj2011}
{Spurr}, R., \& {Natraj}, V. 2011, \jqsrt, 112, 2630,
  \dodoi{10.1016/j.jqsrt.2011.06.014}

\bibitem[{{Stam}(2008)}]{stam2008}
{Stam}, D.~M. 2008, \aap, 482, 989, \dodoi{10.1051/0004-6361:20078358}

\bibitem[{{Stevenson} {et~al.}(2014){Stevenson}, {D{\'e}sert}, {Line}, {Bean},
  {Fortney}, {Showman}, {Kataria}, {Kreidberg}, {McCullough}, {Henry},
  {Charbonneau}, {Burrows}, {Seager}, {Madhusudhan}, {Williamson}, \&
  {Homeier}}]{stevensonetal2014}
{Stevenson}, K.~B., {D{\'e}sert}, J.-M., {Line}, M.~R., {et~al.} 2014, Science,
  346, 838, \dodoi{10.1126/science.1256758}

\bibitem[{{Swain} {et~al.}(2008){Swain}, {Vasisht}, \&
  {Tinetti}}]{swainetal2008}
{Swain}, M.~R., {Vasisht}, G., \& {Tinetti}, G. 2008, \nat, 452, 329

\bibitem[{{Tinetti} {et~al.}(2005){Tinetti}, {Meadows}, {Crisp}, {Fong },
  {Velusamy}, \& {Snively}}]{tinettietal2005}
{Tinetti}, G., {Meadows}, V.~S., {Crisp}, D., {et~al.} 2005, Astrobiology, 5,
  461

\bibitem[{{Tinetti} {et~al.}(2006){Tinetti}, {Meadows}, {Crisp}, {Fong},
  {Fishbein}, {Turnbull}, \& {Bibring}}]{tinettietal2006a}
---. 2006, Astrobiology, 6, 34, \dodoi{10.1089/ast.2006.6.34}

\bibitem[{Tinetti {et~al.}(2007)Tinetti, Vidal-Madjar, Liang, Beaulieu, Yung,
  Carey, Barber, Tennyson, Ribas, Allard, {et~al.}}]{tinettietal2007}
Tinetti, G., Vidal-Madjar, A., Liang, M.-C., {et~al.} 2007, Nature, 448, 169

\bibitem[{{Tinetti} {et~al.}(2016){Tinetti}, {Drossart}, {Eccleston},
  {Hartogh}, {Heske}, {Leconte}, {Micela}, {Ollivier}, {Pilbratt}, {Puig},
  {Turrini}, {Vandenbussche}, {Wolkenberg}, {Pascale}, {Beaulieu}, {G{\"u}del},
  {Min}, {Rataj}, {Ray}, {Ribas}, {Barstow}, {Bowles}, {Coustenis}, {Coud{\'e}
  du Foresto}, {Decin}, {Encrenaz}, {Forget}, {Friswell}, {Griffin}, {Lagage},
  {Malaguti}, {Moneti}, {Morales}, {Pace}, {Rocchetto}, {Sarkar}, {Selsis},
  {Taylor}, {Tennyson}, {Venot}, {Waldmann}, {Wright}, {Zingales}, \&
  {Zapatero-Osorio}}]{tinettietal2016}
{Tinetti}, G., {Drossart}, P., {Eccleston}, P., {et~al.} 2016, in Society of
  Photo-Optical Instrumentation Engineers (SPIE) Conference Series, Vol. 9904,
  Space Telescopes and Instrumentation 2016: Optical, Infrared, and Millimeter
  Wave, ed. H.~A. {MacEwen}, G.~G. {Fazio}, M.~{Lystrup}, N.~{Batalha},
  N.~{Siegler}, \& E.~C. {Tong}, 99041X, \dodoi{10.1117/12.2232370}

\bibitem[{{Tomasko} {et~al.}(2008){Tomasko}, {Doose}, {Engel}, {Dafoe}, {West},
  {Lemmon}, {Karkoschka}, \& {See}}]{tomaskoetal2008}
{Tomasko}, M.~G., {Doose}, L., {Engel}, S., {et~al.} 2008, \planss, 56, 669,
  \dodoi{10.1016/j.pss.2007.11.019}

\bibitem[{Toon {et~al.}(1989)Toon, McKay, Ackerman, \&
  Santhanam}]{toonetal1989}
Toon, O.~B., McKay, C., Ackerman, T., \& Santhanam, K. 1989, Journal of
  Geophysical Research: Atmospheres (1984--2012), 94, 16287,
  \dodoi{10.1029/JD094iD13p16287}

\bibitem[{{Tremblay} {et~al.}(2020){Tremblay}, {Line}, {Stevenson}, {Kataria},
  {Zellem}, {Fortney}, \& {Morley}}]{tremblayetal2020}
{Tremblay}, L., {Line}, M.~R., {Stevenson}, K., {et~al.} 2020, \aj, 159, 117,
  \dodoi{10.3847/1538-3881/ab64dd}

\bibitem[{{Tribbett} {et~al.}(2021){Tribbett}, {Robinson}, \&
  {Koskinen}}]{tribbettetal2021}
{Tribbett}, P.~D., {Robinson}, T.~D., \& {Koskinen}, T.~T. 2021, \psj, 2, 109,
  \dodoi{10.3847/PSJ/abf92d}

\bibitem[{{Tsiaras} {et~al.}(2019){Tsiaras}, {Waldmann}, {Tinetti}, {Tennyson},
  \& {Yurchenko}}]{tsiarasetal2019}
{Tsiaras}, A., {Waldmann}, I.~P., {Tinetti}, G., {Tennyson}, J., \&
  {Yurchenko}, S.~N. 2019, Nature Astronomy, 3, 1086,
  \dodoi{10.1038/s41550-019-0878-9}

\bibitem[{{Vinatier} {et~al.}(2015){Vinatier}, {B{\'e}zard}, {Lebonnois},
  {Teanby}, {Achterberg}, {Gorius}, {Mamoutkine}, {Guandique}, {Jolly},
  {Jennings}, \& {Flasar}}]{vinatieretal2015}
{Vinatier}, S., {B{\'e}zard}, B., {Lebonnois}, S., {et~al.} 2015, \icarus, 250,
  95, \dodoi{10.1016/j.icarus.2014.11.019}

\bibitem[{{von Paris} {et~al.}(2013){von Paris}, {Hedelt}, {Selsis},
  {Schreier}, \& {Trautmann}}]{vonparisetal2013}
{von Paris}, P., {Hedelt}, P., {Selsis}, F., {Schreier}, F., \& {Trautmann}, T.
  2013, \aap, 551, A120, \dodoi{10.1051/0004-6361/201220009}

\bibitem[{{Webber} {et~al.}(2015){Webber}, {Lewis}, {Marley}, {Morley},
  {Fortney}, \& {Cahoy}}]{webberetal2015}
{Webber}, M.~W., {Lewis}, N.~K., {Marley}, M., {et~al.} 2015, \apj, 804, 94,
  \dodoi{10.1088/0004-637X/804/2/94}

\bibitem[{{Zhang} {et~al.}(2019){Zhang}, {Chachan}, {Kempton}, \&
  {Knutson}}]{zhangetal2019}
{Zhang}, M., {Chachan}, Y., {Kempton}, E. M.~R., \& {Knutson}, H.~A. 2019,
  \pasp, 131, 034501, \dodoi{10.1088/1538-3873/aaf5ad}

\bibitem[{{Zugger} {et~al.}(2010){Zugger}, {Kasting}, {Williams}, {Kane}, \&
  {Philbrick}}]{zuggeretal2010}
{Zugger}, M.~E., {Kasting}, J.~F., {Williams}, D.~M., {Kane}, T.~J., \&
  {Philbrick}, C.~R. 2010, \apj, 723, 1168,
  \dodoi{10.1088/0004-637X/723/2/1168}

\end{thebibliography}

%

\end{document}